\documentclass[fleqn,usenatbib]{mnras}

\usepackage{newtxtext,newtxmath}
\usepackage{hyperref}
\usepackage{siunitx}
\usepackage{amsmath}
\usepackage{booktabs}
\let\oldAA\AA
\renewcommand{\AA}{\text{\normalfont\oldAA}}
\usepackage[T1]{fontenc}
\usepackage{xcolor}
\DeclareRobustCommand{\VAN}[3]{#2}
\let\VANthebibliography\thebibliography
\def\thebibliography{\DeclareRobustCommand{\VAN}[3]{##3}\VANthebibliography}

\usepackage{graphicx}
\usepackage{grffile}% Including figure files
\title[Cosmic BBH Evolution in Star Clusters]{The Cosmic Evolution of Binary Black Holes in Young, Globular and Nuclear Star Clusters: Rates, Masses, Spins and Mixing Fractions}

\author[M. Mapelli et al.]{Michela Mapelli$^{1,2,3}$\thanks{E-mail: \href{michela.mapelli@unipd.it}{michela.mapelli@unipd.it}},
Yann Bouffanais$^{1,2}$,
Filippo Santoliquido$^{1,2}$,
Manuel Arca Sedda$^{4}$,
\newauthor{M. Celeste Artale$^{1,2,5}$}
\\
$^{1}$Physics and Astronomy Department Galileo Galilei, University of Padova, Vicolo dell'Osservatorio 3, I--35122, Padova, Italy\\
$^{2}$INFN--Padova, Via Marzolo 8, I--35131 Padova, Italy\\
$^{3}$INAF--Osservatorio Astronomico di Padova, Vicolo dell'Osservatorio 5, I--35122, Padova, Italy\\
$^{4}$Astronomisches Rechen-Institut, Zentr\"um f\"ur Astronomie, Universit\"at Heidelberg, M\"onchofstr. 12-14, Heidelberg, Germany\\
$^{5}$Institut f{\"u}r  Astro- und Teilchenphysik, Universit{\"a}t Innsbruck, Technikerstrasse 25/8, A-6020, Innsbruck, {\"O}sterreich
}

\date{Accepted XXX. Received YYY; in original form ZZZ}

\pubyear{2021}

\begin{document}
\label{firstpage}
\pagerange{\pageref{firstpage}--\pageref{lastpage}}
\maketitle

\begin{abstract}
The growing population of binary black holes (BBHs) observed by gravitational wave detectors is a potential Rosetta stone for understanding their formation channels. Here, we use an upgraded version of our  semi-analytic codes {\sc fastcluster} and {\sc cosmo$\mathcal{R}$ate} to investigate the cosmic evolution of four different BBH populations: isolated BBHs and dynamically formed BBHs in nuclear star clusters (NSCs), globular clusters (GCs), and young star clusters (YSCs).  With our approach, we can study different channels assuming the same stellar and binary input physics. We find that the merger rate density of BBHs in GCs and NSCs is barely affected by stellar metallicity ($Z$), while the rate of  isolated BBHs changes wildly with $Z$. BBHs in YSCs behave in an intermediate way between isolated and GC/NSC BBHs. The local merger rate density of Nth-generation black holes (BHs), obtained by summing up  hierarchical mergers in GCs, NSCs and YSCs, ranges from $\sim{1}$ to $\sim{4}$ Gpc$^{-3}$ yr$^{-1}$ and is mostly sensitive to the spin parameter.  We find that the mass function of primary BHs evolves with redshift in GCs and NSCs, becoming more top-heavy at higher $z$. In contrast, the primary BH mass function almost does not change with redshift in YSCs and in the field. This signature of the BH mass function has relevant implications for Einstein Telescope and Cosmic Explorer. Finally, our analysis suggests that multiple channels contribute to the BBH population of the second gravitational-wave transient catalog. 

\end{abstract}


\section{Introduction}

A variety of formation channels have been proposed for binary black holes (BBHs;see, e.g., \citealt{mapelli2021review} for a recent review): BBH mergers can be the outcome of isolated binary evolution via common envelope \citep{tutukov1973,bethe1998,portegieszwart1998,belczynski2002,belczynski2008,belczynski2016,eldridge2016,dvorkin2016,dvorkin2018,stevenson2017,mapelli2017,mapelli2019,kruckow2018,spera2019,tanikawa2020,belczynski2020,klencki2021,olejak2021}, stable mass transfer \citep{giacobbo2018,neijssel2019,bavera2020,gallegosgarcia2021,shao2021} or chemically homogeneous evolution \citep{marchant2016,mandel2016,demink2016,dubuisson2020,riley2021}. Alternatively, BBHs can form dynamically  in triples \citep[e.g.,][]{antonini2017,silsbee2017,arcasedda2018b,fragione2020,vigna2021}, multiples \citep[e.g.,][]{fragione2019b,liu2019,liu2021,hamers2020}, young star clusters (YSCs, \citealt[][]{banerjee2010,mapelli2016,banerjee2017,banerjee2020,dicarlo2019,dicarlo2020a,kumamoto2019,kumamoto2020}),  globular clusters (GCs, \citealt[][]{portegieszwart2000,tanikawa2013,samsing2014,rodriguez2016,askar2017,fragionekocsis2018,choksi2019,hong2018,kamlah2021}), and nuclear star clusters (NSCs, \citealt[][]{antonini2016,petrovich2017,antonini2019,arcasedda2020,arcasedda2020b,fragione2020b}). Furthermore, gas torques in AGN discs  trigger  the formation of BBHs and speed up their mergers \citep[e.g.,][]{bartos2017,stone2017,mckernan2018,yang2019,tagawa2020,ishibashi2020}. Finally, primordial black holes (BHs), born from gravitational collapses in the early Universe, might also pair up and merge via gravitational wave (GW) emission \citep[e.g.,][]{carr1974,carr2016, sasaki2016,alihaimoud2017,clesse2017,deluca2021}.

One of the key signatures of the dynamical scenario is the formation of massive BHs via hierarchical merger chains \citep{miller2002,giersz2015,fishbach2017,gerosa2017,rodriguez2019,arcasedda2021b,mapelli2021,gerosa2021review}: the remnant of a BBH merger is a single object at birth, but, if it is inside a dense stellar environment, it may pair up dynamically with other BHs and merge again. The merger remnant has a distinctive feature, which is a large spin magnitude $\chi\sim{0.7}$,  mostly inherited from pre-merger orbital angular momentum \citep{jimenez-forteza2017,gerosa2021review}. The efficiency of hierarchical mergers is hampered by relativistic kicks, that the merger remnant suffers at birth because of radiation
of linear momentum through beamed GW emission \citep{fitchett1983,favata2004,campanelli2007,lousto2011}.  %asymmetries in linear momentum transport by gravitational waves.
 The magnitude of the relativistic kick is generally comparable to (or higher than) the escape velocity of a massive star cluster, and can lead to the ejection of the merger remnant, interrupting the hierarchical chain \citep{holley-bockelmann2008,moody2009}.

Advanced LIGO \citep{LIGOdetector} and Virgo \citep{VIRGOdetector} observed more than 50 BBH mergers to date \citep{abbottO3a,abbottGWTC-2.1}. Population analyses on these BBHs moderately support the co-existence of multiple formation channels %, because a single evolutionary pathway struggles to produce the variety of properties of the observed BBHs 
\citep{abbottO3popandrate,callister2021,zevin2021,wong2021,bouffanais2021}. %, but see \citealt{roulet2021} for a different result).
Moreover, GW190521 \citep{abbottGW190521,abbottGW190521astro}, and possibly GW190403\_051519 and GW190426\_190642 \citep{abbottGWTC-2.1} challenge current models of massive star evolution, hosting BHs in the pair-instability mass gap \citep{belczynski2016pair,woosley2017,spera2017,marchant2019,stevenson2019}. On the one hand, an isolated formation channel cannot be ruled out for these events, because the boundaries of the mass gap are affected by large uncertainties \citep[e.g.,][]{farmer2019,farmer2020,mapelli2020,farrell2020,belczynski2020b,vink2021,costa2021,tanikawa2020}. On the other hand, dynamics can partially fill the pair-instability mass gap via hierarchical BH mergers \citep{rodriguez2019,anagnostou2020,fragione2020,kimball2020,mapelli2021,arcasedda2021,liu2021,gerosa2021} or stellar collisions \citep{dicarlo2019,dicarlo2020a,dicarlo2020b,kremer2020,renzo2020b,gonzalez2021}.

Several studies performed a multi-channel analysis, trying to constrain the relative contribution of each formation scenario to the observed BBH population \citep[e.g.,][]{zevin2017,stevenson2017,mandel2018,bouffanais2019,bouffanais2021,zevin2021,wong2021}.  To compare different channels self-consistently, the models would need to have the same underlying stellar/binary evolution models, and the same physical assumptions. %, making the model prediction self-consistent between each other which is fundamental requirement for a proper model selection analysis.

  Comparing catalogs of BBHs simulated with different input assumptions might lead to biased results: %the differences among the considered BBH catalogs might be due to the adopted numerical approach rather than to the intrinsic physical differences among channels \citep[e.g.,][]{belczynski2021}.
  for example, if the assumed initial BH mass function is different for different models, the results of the multi-channel comparison will be conditioned by this discrepancy in the initial conditions.

  The only way to avoid such bias is to simulate different dynamical channels with the same input physics, starting from the same underlying initial assumptions (e.g., the same BH mass function). This is a challenging task, because  models of different formation channels are generally produced with different numerical techniques, which encode dramatically different input physics. For example,  BBHs in NSCs are generally studied with semi-analytical models \citep[e.g.,][]{antonini2016,arcasedda2020b}, BBHs in GCs are often modelled with hybrid Monte Carlo simulations \citep[e.g.,][]{rodriguez2016,askar2017}, BBHs  in YSCs with direct N-body simulations \citep[e.g.,][]{banerjee2010,ziosi2014,fujii2014} and isolated BBHs with population-synthesis simulations \citep[e.g.,][]{belczynski2016,mapelli2017,eldridge2016}, run with different codes and assumptions.

%Several studies performed a multi-channel analysis, trying to constrain the relative contribution of each formation scenario to the observed BBH population \citep[e.g.,][]{zevin2017,stevenson2017,mandel2018,bouffanais2019,bouffanais2021,zevin2021,wong2021}. Models of different formation channels are generally produced with different numerical techniques: For example,  BBHs in NSCs are generally studied with semi-analytical models \citep[e.g.,][]{antonini2016,arcasedda2020b}, BBHs in GCs are often modelled with hybrid Monte Carlo simulations \citep[e.g.,][]{rodriguez2016,askar2017}, BBHs  in YSCs with direct N-body simulations \citep[e.g.,][]{banerjee2010,ziosi2014,fujii2014} and isolated BBHs with population-synthesis simulations \citep[e.g.,][]{belczynski2016,mapelli2017,eldridge2016}, run with different codes and assumptions.
%
%At present, this cannot be done with direct N-body simulations, because of their high computational cost, especially for the most massive and long-lived star clusters \citep{wang2020}.

The  purpose of this work is to compare the merger rate and other BBH properties (mass and spin distribution) we obtain for different channels, by adopting the same input physics (e.g., the initial BH mass function) and the same numerical code for all the considered scenarios.  
We use the semi-analytic dynamical code {\sc fastcluster} \citep{mapelli2021}, which can handle isolated BBHs and dynamical BBHs in YSCs, GCs and NSCs within the same numerical framework. {\sc fastcluster} overcomes the numerical challenge of simulating BBHs in massive and long-lived star clusters by integrating the effect of dynamical hardening and GW emission with a fast semi-analytic approach, calibrated on direct N-body models. 
 Finally, we derive the mixing fraction of each channel, by running Bayesian hierarchical inference on the public data of the second GW transient catalog (GWTC-2, \citealt{abbottO3a}).

\section{Methods}

\subsection{Isolated BBHs}

Isolated BBHs %are BBHs that %form from unperturbed binary stars %
form and evolve in the field; 
they are not perturbed by dynamical interactions. To generate masses, delay times\footnote{The delay time is the time between the formation of a binary star and the merger of the final BBH.} and spin orientations  of isolated BBHs, we use the population-synthesis code {\sc mobse} \citep{giacobbo2018,giacobbo2018b}. 
%This ensures that the underlying BH mass spectrum is the same for both isolated BBHs and dynamical BBHs. For isolated BBHs, we use {\sc mobse}  to generate BBH delay times (i.e., the time between the formation of a binary star and the merger of the final BBH) and spin orientations. 

{\sc mobse} is an upgraded and custom version of {\sc bse} \citep{hurley2002}. It implements up-to-date models for stellar winds \citep{vink2001,graefener2008,chen2015}, core-collapse supernovae \citep[SNe,][]{fryer2012}, pair-instability SNe \citep{mapelli2020} and SN kicks \citep{giacobbo2020}. For more details, we refer to \cite{giacobbo2018} and \cite{giacobbo2018b}. BHs with mass up to $\approx{65}$ M$_\odot$ can form from metal-poor stars in {\sc mobse}, but only BHs with mass up to $\approx{45}$ M$_\odot$ merge within a Hubble time in isolated binary systems (see, e.g., Figure~11 of \citealt{giacobbo2018b}). The main reason of this difference is that tight isolated binary stars, which are the progenitors of isolated BBH mergers, evolve via mass transfer or common envelope. These are dissipative processes and lead to the complete removal of stellar envelopes, leaving behind naked He cores. The maximum mass of a BH that forms from a naked He core is $\approx{45}$ M$_\odot$ in {\sc mobse} models.  In contrast, single stars and stars in detached binary systems can retain a portion of their hydrogen-rich envelope until their final collapse, producing BHs with mass up to $\approx{65}$ M$_\odot$.

 Several authors have studied the origin of BH spin magnitudes, either in single or binary stars \citep[e.g.,][]{fuller2019,fullerma2019,belczynski2020,qin2018,qin2019,bavera2020b,olejak2021b}. The main uncertainties come from the theory of angular momentum transport in massive stars and hamper the predictive power of current models. BBHs in GW events provide mild support for relatively low spins \citep{abbottO3popandrate}. Given the uncertainties of the models, in this work we adopt a phenomenological approach for 1g BH spins, and draw dimensionless spin magnitudes $\chi$ from a Maxwellian distribution truncated at $\chi=1$ \citep{bouffanais2021}, with root mean square $\sigma_{\chi}=0.1$ (fiducial case) or $\sigma_{\chi}=0.01$ (low-spin case). In particular, the case with $\sigma_{\chi}=0.1$ is reminiscent of the spins inferred from GWTC-2 \citep[see Figure 10 of ][]{abbottO3popandrate}, while the case with  $\sigma_{\chi}=0.01$ matches the models by \cite{fullerma2019}, which predict vanishingly small BH spins. 
%AGGIUNGERE UNA NOTA SULLE MASSE DEI BH O LA MASS FUNCTION (FIGURA)?

\subsection{First generation (1g) BBHs in star clusters}
\label{sec:firstgen} 

To generate catalogs of BBH mergers in dynamical environments (YSCs, GCs and NSCs), we use the semi-analytic code {\sc fastcluster} \citep{mapelli2021}. %Figure~\ref{fig:FASTCLUSTER_scheme} is a flow chart  of {\sc fastcluster}. 
Here below, we summarize the main features of this code and  refer to \cite{mapelli2021} for more details. %See~\ref{sec:appendix}
{\sc fastcluster} takes into account two classes of BBHs: original and dynamical BBHs. The former originate from binary stars that are already present in the initial conditions (hereafter, original binaries), while the latter are dynamically assembled. Both original and dynamical BBHs evolve inside their parent star cluster and are affected by dynamical encounters.

A dynamical BBH forms in a timescale 
\begin{equation}\label{eq:tdyn}
    t_{\rm dyn}=\max{\left[t_{\rm SN},\,{}t_{\rm DF}+\min{(t_{\rm 3bb},t_{\rm 12})}\right]},
\end{equation}
where $t_{\rm SN}$ is the time of the core-collapse SN explosion or direct collapse, $t_{\rm DF}$ is the dynamical friction timescale  \citep{chandrasekhar1943}, $t_{\rm 3bb}$ is the timescale for dynamical formation of a BBH via three-body encounters \citep{goodman1993,lee1995} and $t_{\rm 12}$ is the timescale for dynamical formation of a BBH via exchange into an existing binary star \citep{millerlauburg2009}. For the aforementioned timescales, we use the following approximations:
\begin{eqnarray}
  t_{\rm DF}=\frac{3}{4\left(2\,{}\pi{}\right)^{1/2}\,{}G^2\ln{\Lambda{}}}\,{}\frac{\sigma^3}{m_{\rm BH}\,{}\rho{}},\nonumber\\
  t_{\rm 3bb}=125\,{}{\rm Myr}\,{}\left(\frac{10^6\,{}{\rm M}_\odot\,{}{\rm pc}^{-3}}{\rho_{\rm c}}\right)^2\,{}\left(\zeta{}^{-1}\,{}\frac{\sigma_{\rm 1D}}{30\,{}{\rm km}\,{}{\rm s}^{-1}}\right)^9\,{}\left(\frac{20\,{}{\rm M}_\odot}{m_{\rm BH}}\right)^5,\nonumber\\
   t_{\rm 12}=3\,{}{\rm Gyr}\,{}\left(\frac{0.01}{f_{\rm bin}}\right)\,{}\left(\frac{10^6\,{}{\rm M}_\odot\,{}{\rm pc}^{-3}}{\rho_{\rm c}}\right)\,{}\left(\frac{\sigma}{50\,{}{\rm km}\,{}{\rm s}^{-1}}\right) \nonumber\\
  \times{}\,{}\left(\frac{12\,{}{\rm M}_\odot}{m_{\rm BH}+2\,{}m_\ast}\right)\,{}\left(\frac{1\,{}{\rm AU}}{a_{\rm hard}}\right),
% \label{eq:tdyn}
  \end{eqnarray}
  where $G$ is the gravity constant, $m_{\rm BH}$ is the mass of the BH, $\sigma{}$ is the 3D velocity dispersion, $\rho{}$ is the mass density at the half-mass radius, $\ln{}\Lambda{}\sim{}10$ is the Coulomb logarithm, %After a time $t_{\rm DF}$, the BH has sunk to the core of the cluster and can acquire a companion by three-body encounters or by exchange.
  $\rho_{\rm c}$ is the central density of the star cluster, $\sigma{}_{\rm 1D}=\sigma{}/\sqrt{3}$ is the one-dimensional velocity dispersion at the half-mass radius (assuming an isotropic distribution of stellar velocities) and $\zeta{}\leq{1}$ accounts for deviations from equipartition of a BH subsystem (here we assume that there is equipartition, \citealt{spitzer1969}).
%\begin{equation}\label{eq:zeta}
%\zeta{}=\frac{m_\ast{}\,{}\sigma{}^2}{m_{\rm BH}\,{}\sigma_{\rm BH}^2}.
%\end{equation} In eq.~\ref{eq:zeta}, $\sigma_{\rm BH}$ is the velocity dispersion associated with massive BHs of mass $m_{\rm BH}$. In case of equipartition, $\zeta{}=1$. If the system is not in equipartition (i.e., Spitzer instability takes place, \citealt{spitzer1969}), then $\zeta{}<1$. Here, we assume $\zeta{}=1$.  The strong dependence of eq.~\ref{eq:t3bb} on $\sigma_{\rm 1D}$ and $m_{\rm BH}$ makes it critical for the formation of BBHs in dense stellar systems. Here, we use the formalism discussed in \cite{oleary2006} and \cite{morscher2015}, which has been adopted in hybrid Monte Carlo simulations  and compares well with direct N-body simulations of globular clusters \citep{rodriguez2016compare}.
%Finally, the timescale for the dynamical exchange of a BH into a binary star is \citep{millerlauburg2009}
%\begin{eqnarray}\label{eq:t12}
%\end{eqnarray}
Furthermore, $f_{\rm bin}$ is the binary fraction, $m_\ast$ is the average mass of a star in the cluster and $a_{\rm hard}=G\,{}m_\ast/\sigma{}^2$ is the minimum semi-major axis of a hard binary system. Equation~\ref{eq:tdyn} indicates that a dynamical BBH forms only after the primary BH had enough time to sink to the cluster core by dynamical friction and acquire a companion via either three-body or exchange interactions.

The masses of both original and dynamical BBHs are generated from the population-synthesis code {\sc mobse} \citep{giacobbo2018,giacobbo2018b}. {\sc fastcluster} can take any other possible initial conditions for BH masses. However, this choice ensures that the underlying BH mass spectrum is the same for isolated, original and dynamical BBHs. The main difference between original and dynamical BBHs is that the masses of original BBHs are taken from isolated BBH simulations (they are the same as isolated BBHs), while the masses of dynamical BBHs are extracted from the distribution of single BHs. The secondary component mass of  a dynamical BBH is extracted from a distribution  $p(m_2)\propto{}(m_1+m_2)^4$, where $m_1$ and $m_2$ are the primary and secondary component, respectively \citep{oleary2016}.

Consistently with isolated BBHs, BH spin magnitudes are randomly sampled from a Maxwellian distribution with root mean square $\sigma_{\chi}=0.1$ (fiducial case) or $\sigma_{\chi}=0.01$ (low-spin case). We randomly draw spin directions isotropic over the sphere, because dynamics resets any spin alignments.

The semi-major axis $a$ and the eccentricity $e$ at the time of BBH formation are calculated with {\sc mobse} in the case of original BBHs and are drawn from the following probability distributions in the case of dynamical BBHs \citep{heggie1975}:
\begin{eqnarray}
p(a)\propto{}a^{-1}\quad{}\quad{}a\in[1,\,{}10^3]\,{}{\rm R}_\odot\nonumber\\
p(e)=2\,{}e\quad{}\quad{}e\in[0,\,{}1).
\end{eqnarray}

At the beginning of the integration, we check if a (dynamical or original) BBH is hard, i.e. if its binding energy $E_{\rm b}$ satisfies the following relationship \citep{heggie1975}:
\begin{equation}
    E_{\rm b}=\frac{G\,{}m_1\,{}m_2}{2\,{}a}\geq{}\frac{1}{2}m_\ast{}\,{}\sigma^2.
\end{equation}
If the binary is hard, we integrate its orbital evolution. Otherwise, we assume it breaks via dynamical encounters.

%%%%%%%%%%%%%%%%%%%%%%%%%%%%FIGURE%%%%%%%%%%%%%%%%%%%%%%%%%%%%%%%%%%%%%%%%%%%%%%%%%%%
\begin{figure}
  \begin{center}
    \includegraphics[width = 0.53 \textwidth]{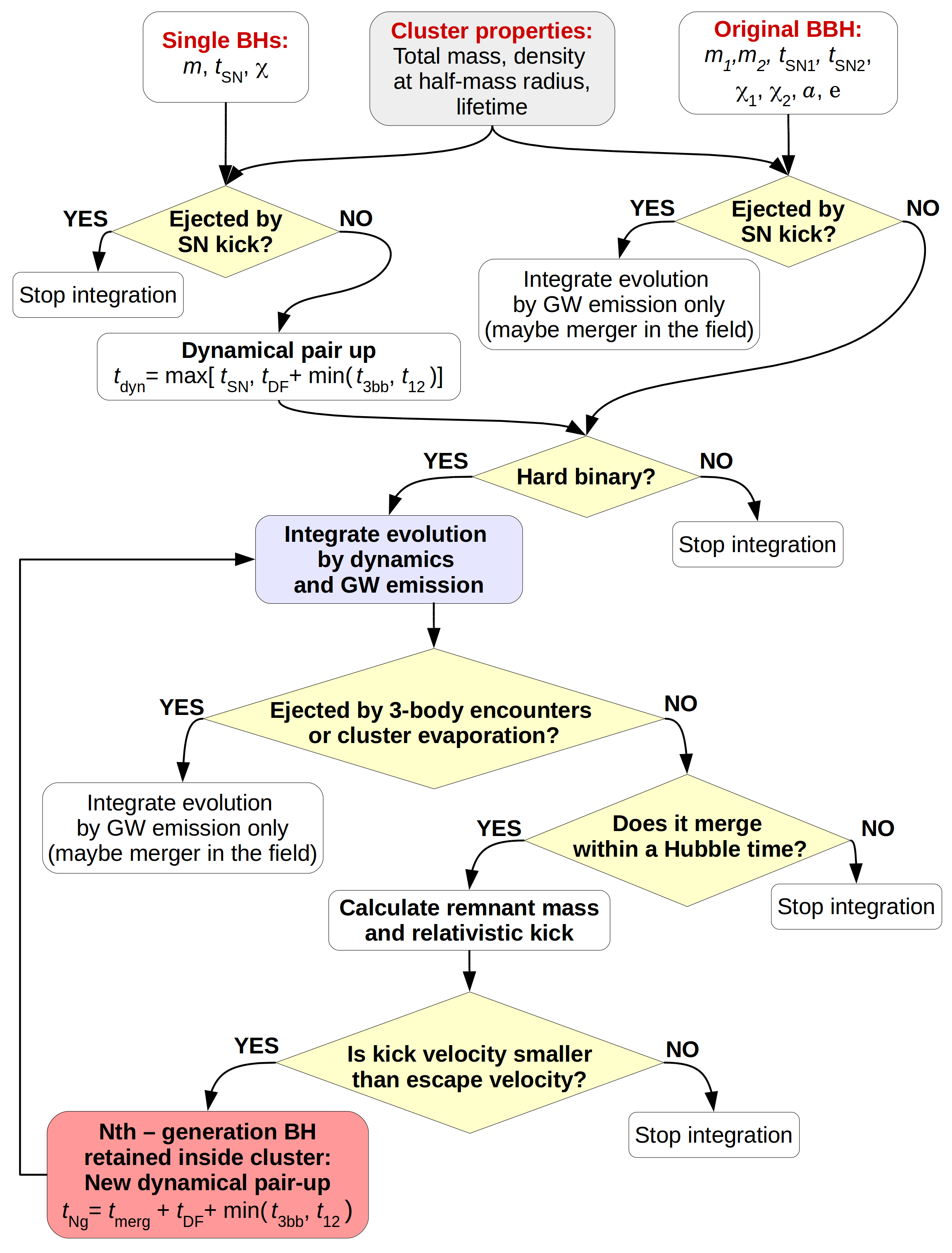}
    \end{center}
  \caption{Flow chart of {\sc fastcluster}. \label{fig:FASTCLUSTER_scheme}}
\end{figure}
%%%%%%%%%%%%%%%%%%%%%%%%%%%%%%%%%%%%%%%%%%%%%%%%%%%%%%%%%%%%%%%%%%%%%%%%%%%%%%%%%%%%%%

\subsection{Orbital evolution}\label{eq:orbev}

When a BBH is hard and is inside its parent star cluster, the evolution of its semi-major axis $a$ and eccentricity $e$ can be described as \citep{mapelli2021review}:
\begin{eqnarray}\label{eq:mapelli2018}
   \frac{{\rm d}a}{{\rm d}t}=-2\,{}\pi{}\,{}\xi{}\,{}\frac{G\,{}\rho{}_{\rm c}}{\sigma}\,{}a^2-\frac{64}{5}\,{} \frac{G^3 \,{} m_1 \,{} m_2 \,{} (m_1+m_2)}{c^5 \,{} a^3\,{} (1-e^2)^{7/2}}\,{}f_1(e) \nonumber\\
  \frac{{\rm d}e}{{\rm d}t}=2\,{}\pi{}\,{}\xi{}\,{}\kappa{}\,{}\frac{G\,{}\rho{}_{\rm c}}{\sigma}\,{}a-\frac{304}{15}\,{} e \frac{ G^3 \,{} m_1 \,{} m_2 \,{} (m_1+m_2)}{c^5 \,{}a^4 \,{}  (1-e^2)^{5/2}}\,{}f_2(e),\nonumber\\
\end{eqnarray}
where $c$ is the speed of light and \citep{peters1964}
\begin{eqnarray}
  f_1(e)=\left(1+\frac{73}{24}\,{}e^2+\frac{37}{96}\,{} e^4\right) \nonumber\\
f_2(e)=\left(1+\frac{121}{304} \,{} e^2\right).
\end{eqnarray}
In  the system of ordinary differential equations~\ref{eq:mapelli2018}, $\xi$ and $\kappa$ are two dimensionless parameters, calibrated with direct N-body simulations \citep{hills1983,quinlan1996,miller2002,sesana2006}. Here, we assume $\xi=3$ \citep{quinlan1996} and $\kappa=0.1$ \citep{sesana2006}. Equations~\ref{eq:mapelli2018} are  composed of two terms.  The first terms  in both equations  ($\frac{{\rm d}a}{{\rm d}t}\propto{}-a^{2}$ and $\frac{{\rm d}e}{{\rm d}t}\propto{}a$) describe the dynamical hardening and the evolution of eccentricity via Newtonian dynamical scatterings;  the second terms ($\frac{{\rm d}a}{{\rm d}t}\propto{}-a^{-3}$ and $\frac{{\rm d}e}{{\rm d}t}\propto{}-a^{-4}$) describe hardening and circularization via GW emission \citep{peters1964}.

{\sc fastcluster} integrates   the system of equations~\ref{eq:mapelli2018} until one of the following conditions is satisfied: (i) the BBH is ejected from the cluster, (ii) the BBH merges, (iii) the star cluster dies by evaporation, or (iv)  we reach the Hubble time (which one of these four cases happens first). If the BBH is ejected from the cluster, {\sc fastcluster} integrates only the second terms of eqs.~\ref{eq:mapelli2018} (hardening and circularization by GW emission) until either the BBH merges in the field or a Hubble time has elapsed.

A binary is assumed to be ejected from the cluster when $a_{\rm ej}>a_{\rm GW}$ \citep{baibhav2020} with
\begin{eqnarray}\label{eq:aej}
a_{\rm ej}=\frac{2\,{}\xi{}\,{}m_\ast{}^2}{(m_1+m_2)^3}\,{}\frac{G\,{}m_1\,{}m_2}{v_{\rm esc}^2}\nonumber{}\\
a_{\rm GW}=\left[\frac{32\,{}G^2}{5\,{}\pi{}\,{}\xi{}\,{}c^5}\,{}\frac{\sigma{}\,{}m_1\,{}m_2\,{}(m_1+m_2)}{\rho{}_{\rm c}\,{}(1-e^2)^{7/2}}\,{}f_1(e)\right]^{1/5}.
\end{eqnarray}
The former of the two eqs.~\ref{eq:aej} describes the semi-major axis below which the BBH is ejected by dynamical recoil, while the latter describes the maximum semi-major axis for the regime of efficient orbital decay via GW emission. 

\subsection{Nth generation (Ng) dynamical BBHs}

If the BBH merges in less than a Hubble time,  we estimate the mass and spin of the merger remnant using the fitting formulas by \cite{jimenez-forteza2017}. If the BBH merges inside its parent star cluster, we also calculate the relativistic kick magnitude $v_{\rm K}$ using the fit by \cite{lousto2012}. We assume that the merger remnant remains inside its parent cluster if the relativistic kick magnitude $v_{\rm K}<v_{\rm esc}$, where $v_{\rm esc}$ is the escape velocity from the star cluster. Otherwise, the merger remnant is ejected from the parent cluster and cannot participate in any further hierarchical mergers.

Even when the merger remnant remains inside its parent cluster, the kick sends it  in the cluster's halo, where the stellar density is orders of magnitude lower with respect to the core \citep{spitzer1987}.  %far away from the cluster core.
The BH must sink back to the core via dynamical friction before it can acquire new companions via three-body encounters or exchanges. We then calculate the timescale $t_{\rm Ng}$ for the merger remnant to pair up dynamically with a new companion BH as
\begin{equation}\label{eq:tNg}
    t_{\rm Ng}=t_{\rm merg}+t_{\rm DF}+\min{(t_{\rm 3bb},\,{}t_{\rm 12})}.
\end{equation}
In the above equation, $t_{\rm merg}=t_{\rm dyn}+t_{\rm GW}$ is the delay time of the first generation (1g) BBH, where $t_{\rm dyn}$ is defined in eq.~\ref{eq:tdyn}, while $t_{\rm GW}$ is the time elapsed from the formation of the BBH to its merger, according to eqs.~\ref{eq:mapelli2018}. If $t_{\rm Ng}$ is shorter than the Hubble time, we start the loop again by integrating the second generation (2g) BBH with eqs~\ref{eq:mapelli2018}. We iterate the hierarchical merger chain until the merger remnant is ejected from the cluster, or the cluster evaporates, or we reach the Hubble time. Figure~\ref{fig:FASTCLUSTER_scheme} is a flow chart  of {\sc fastcluster}.

\subsection{Properties of star clusters}\label{sec:clusters}
 
  We consider three different flavours %kinds 
  of star clusters: NSCs, GCs and YSCs. Each star cluster is uniquely defined by its lifetime $t_{\rm SC}$, total mass $M_{\rm tot}$, binary fraction $f_{\rm bin}$ and half-mass density $\rho{}$. We assume $t_{\rm SC}=13.6$, 13.6 and 1 Gyr for NSCs, GCs \citep{gratton1997,gratton2003,vandenberg2013} and YSCs  \citep{portegieszwart2010}, respectively. Furthermore, we assume $f_{\rm bin}=0.01$, 0.1 and 1 in NSCs \citep{antonini2016}, GCs \citep{jibregman2015} and YSCs \citep{sana2012}, respectively. We draw the total masses  from a log-normal distribution with mean $\langle{}\log_{10}{M_{\rm tot}/{\rm M}_\odot}\rangle{}=6.18,\,{}5.6$ and 4.3 for NSCs, GCs and YSCs, respectively. We assume a fiducial standard deviation $\sigma_{\rm M}=0.4$ for all star cluster flavours.% We also consider the cases in which $\sigma_{\rm M}=0.2,$ and 0.6. 
  We draw the  density at the half-mass radius from a log-normal distribution with mean $\langle{}\log_{10}{\rho{}/({\rm M}_\odot\,{}{\rm pc}^{-3})}\rangle{}=5,$ 3.7 and 3.3 for NSCs, GCs and YSCs, respectively. We assume a fiducial standard deviation $\sigma_\rho=0.4$ for all star cluster flavours.  The values of $M_{\rm tot}$ and $\rho{}$ are inferred from the observations reported in \cite{neumayer2020} for NSCs and GCs (see also \citealt{harris1996,georgiev2016}) and from \cite{portegieszwart2010} for YSCs.  For each star cluster, we assume a core density $\rho_{\rm c}=20\,{}\rho$.  We derive the escape velocity from $M_{\rm tot}$ and $\rho$ \citep{georgiev2009a,georgiev2009b,fragione2020} using the following relationship
\begin{equation}\label{eq:vesc}
  v_{\rm esc}=40\,{}{\rm km}\,{}{\rm s}^{-1}\,{}\left(\frac{M_{\rm tot}}{10^5\,{}{\rm M}_\odot}\right)^{1/3}\,{}\left(\frac{\rho}{10^5\,{}{\rm M}_\odot\,{}{\rm pc}^{-3}}\right)^{1/6}.
\end{equation}
 Equation~\ref{eq:vesc} results in a distribution of escape velocities fairly consistent with the observational sample reported in Figure~1 of \cite{antonini2016} for GCs and NSCs. In the initial conditions, we generate each star cluster by randomly drawing a value of $M_{\rm TOT}$ and $\rho$ from the aforementioned distributions. We simulate only one BBH per each randomly drawn star cluster, in order to better sample the parameter space of BBHs and possible host clusters. Here, we do not consider NSCs that host a supermassive BH. In such clusters, most of the binaries inside the influence radius of the supermassive BH are soft. % and are disrupted over a timescale \citep{binney1987}:
%    \begin{equation}
%      t_{\rm ev}=\frac{(m_1+m_2)\,{}\sigma{}}{16\,{}\sqrt{\pi}\,{}G\,{}m_\ast{}\,{}\rho{}_{\rm c}\,{}a\,{}\ln{\Lambda}}.
%    \end{equation}
We refer to \cite{arcasedda2020b} for a detailed treatment of this case. We assume, for the sake of simplicity, that the star cluster properties do not evolve in time. We will add the evolution of the star cluster in a follow-up study.

%\section{Methods: BBH merger rate}

\subsection{BBH merger rate  density}\label{sec:MRD}

The BBH merger rate  density  per each channel $i$ can be estimated as
\begin{eqnarray}\label{eq:cosmorate}
  \mathcal{R}_i(z) = %\frac{\rm d\quad{}\quad{}}{{\rm d}t(z)}
  \int_{z_{\rm max}}^{z}\psi_i(z')\,{}\frac{{\rm d}t(z')}{{\rm d}z'} \,{}   
   \left[\int_{Z_{\rm min}(z')}^{Z_{\rm max}(z')}\eta{}_i(Z)\,{}\mathcal{F}_i(z',z, Z)\,{}{\rm d}Z\right] \,{}{\rm d}z',
\end{eqnarray}
where $t(z')$ is the look-back time at redshift $z'$  and ${\rm d}t(z')/{\rm d}z'=(1+z')^{-1}\,{}H(z')^{-1}$, with $H(z')=H_0\,{}\left[(1+z')^3\,{}\Omega_{\rm M}+\Omega_\Lambda\right]^{1/2}$. Furthermore,  $\psi_i(z')$ is the  formation rate density  at redshift $z'$ for the $i-$th channel, where $i=$~NSCs, GCs, YSCs or field, $Z_{\rm min}(z')$ and $Z_{\rm max}(z')$ are the minimum and maximum metallicity of stars formed at redshift $z'$, $\eta{}_i(Z)$ is the merger efficiency at metallicity $Z$, and $\mathcal{F}_i(z', z, Z)$ is the %fraction of
 merger rate of BBHs belonging to a given channel $i$ that form at redshift $z'$ from stars with metallicity $Z$ and merge at redshift $z$, normalized to all BBHs belonging to the same channel $i$ that form from stars with metallicity $Z$. To calculate the look-back time we take the cosmological parameters ($H_{0}$, $\Omega_{\rm M}$ and $\Omega_{\Lambda}$)  from \cite{planck2016}. 

\subsubsection{Formation rate density}

\begin{figure}
  \begin{center}
    \includegraphics[width = 0.45 \textwidth]{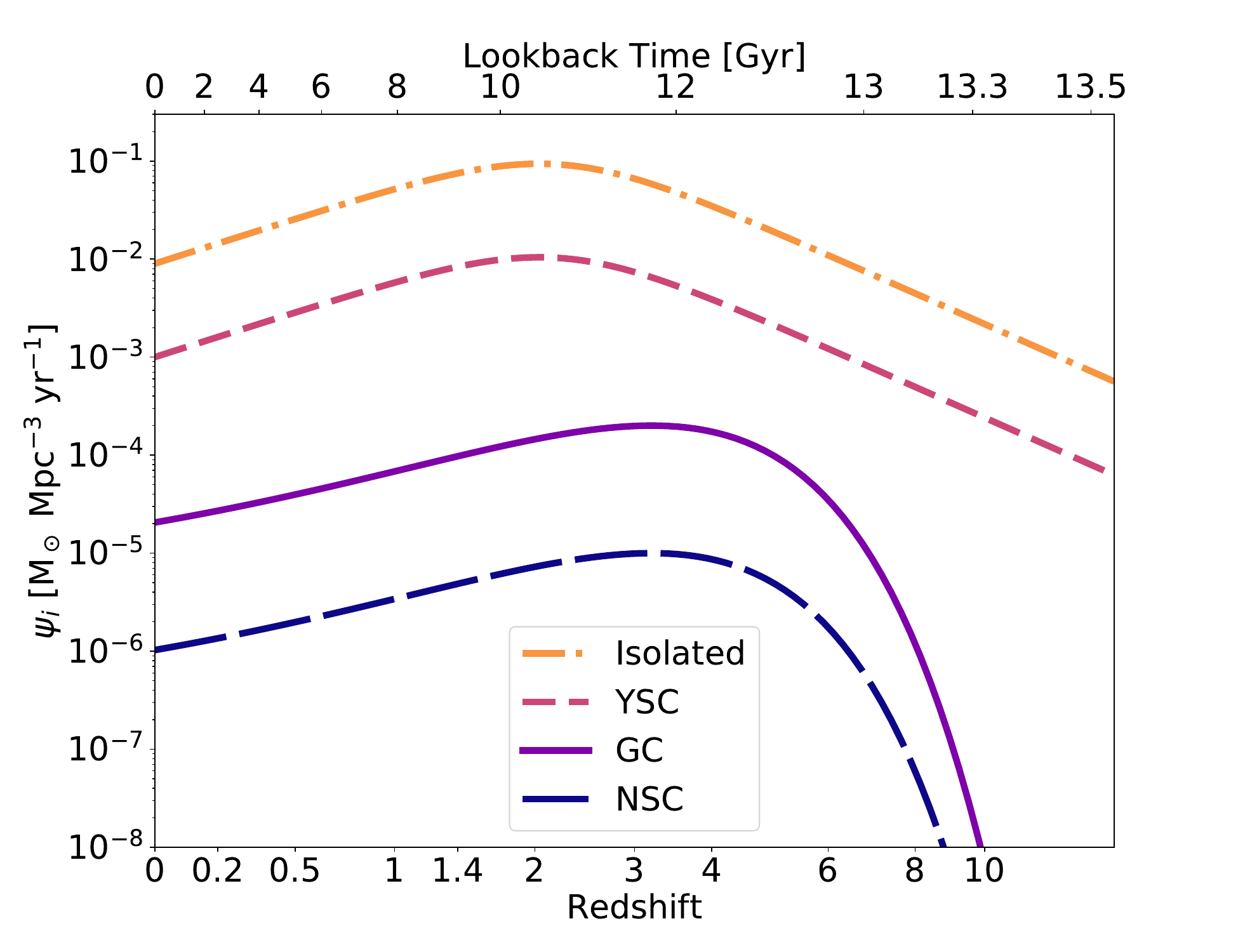}
    \end{center}
  \caption{Star formation rate density as a function of redshift for isolated stars (orange dot-dashed line), YSCs (magenta short-dashed line), GCs (violet solid line) and NSCs (blue long-dashed line). \label{fig:SFR}}
\end{figure}

In our fiducial model, we define $\psi_i(z)$ as follows. For the formation rate of GCs as a function of redshift we assume a Gaussian distribution
\begin{equation}\label{eq:GCs}
\psi_{\rm GC}(z)=\mathcal{B}_{\rm GC}\,{}\exp{\left[-(z-z_{\rm GC})^2/(2\,{}\sigma_{\rm GC}^2)\right]}, 
\end{equation}
where, in the fiducial model, $z_{\rm GC}=3.2$ is the redshift where the formation rate of GCs is maximum, $\sigma_{\rm GC}=1.5$ is the standard deviation of the distribution and $\mathcal{B}_{\rm GC}$ is the normalization factor. This distribution is reminiscent of the one estimated by \cite{el-badry2019} (see also \citealt{rodriguezloeb2018}). In particular, the fiducial normalization we adopt, $\mathcal{B}_{\rm GC}=2\times{}10^{-4}\,{}{\rm M}_{\odot}\,{}{\rm Mpc}^{-3}\,{}{\rm yr}^{-1}$, is consistent with both \cite{el-badry2019} and \cite{reina-campos2019}. The peak redshift $z_{\rm GC}=3.2$ is not taken from  \cite{el-badry2019}, who report $z_{\rm GC}=4$, but rather is calibrated on the distribution of the ages of Galactic GCs, which peaks at $z=3.2$ \citep{gratton1997,gratton2003,vandenberg2013}. %The normalization is consistent with both \cite{el-badry2019} and \cite{reina-campos2019}. 
In Section~\ref{sec:uncertainties}, we will discuss the impact of these parameters on the merger rate. If we assume that none of our GCs dies by evaporation, eq.~\ref{eq:GCs} yields a density of GCs in the local Universe  $n_{\rm GC}\approx{4}$ Mpc$^{-3}$. This is higher than the observed value ($n_{\rm GC}\approx{2.5}$ Mpc$^{-3}$, \citealt{portegieszwart2000}), but our estimate of $n_{\rm GC}$ must be regarded as an upper limit because we assume that all GCs, even the least massive, survive to redshift zero. Fig.~\ref{fig:SFR} shows the formation rate density as a function of redshift for the four channels considered here. 

The uncertainty on the formation rate of NSCs is even higher. According to several models \citep{tremaine1975,capuzzo1993,capuzzo2008,antonini2012}, NSCs form from the merger of GCs sinking to the centre of their host galaxies by dynamical friction. Thus, %as a reasonable guess, 
for NSCs we adopt  the same functional form as for GC  formation history, but we reduce the normalization:
\begin{equation}\label{eq:NSCs}
\psi_{\rm NSC}(z)=\mathcal{B}_{\rm NSC}\,{}\exp{\left[-(z-z_{\rm NSC})^2/(2\,{}\sigma_{\rm NSC}^2)\right]}, 
\end{equation}
where, in the fiducial model, $z_{\rm NSC}=3.2$ and  $\sigma_{\rm NSC}=1.5$ for analogy with GCs. This formalism is subject to large uncertainties, because of the scarce observational constraints. In Section~\ref{sec:uncertainties}, we will comment on these uncertainties. In our fiducial model, the normalization of eq.~\ref{eq:NSCs} is $\mathcal{B}_{\rm NSC}=10^{-5}\,{}{\rm M}_{\odot}\,{}{\rm Mpc}^{-3}\,{}{\rm yr}^{-1}$, and  was chosen so that we obtain a NSC density in the local Universe comparable with the observed one. If we assume that all NSCs survive to redshift zero (which is reasonable for NSCs) and integrate eq.~\ref{eq:NSCs} over cosmic time, we find a current density of NSCs $n_{\rm NSC}\approx{0.06}$ Mpc$^{-3}$. For comparison, if we take the density of galaxies with stellar mass $>10^7$ M$_\odot$ from observations \citep{conselice2016} and  assume that all such galaxies have a  NSC, we expect a current NSC density $n_{\rm NSC}\approx{0.05-0.1}$ Mpc$^{-3}$, which is the same order of magnitude as our estimate. %which compares nicely with our estimate. 
%Gives a current density of NSCs ~ 0.06 Mpc$^-3$ Roughly consistent with estimates from observations: if we assume that all galaxies with stellar mass $>10^7$ Msun have a NSC Their density from observations (Conselice et al. 2016, ApJ, 830, 83) is ~0.05-0.1 Mpc$^-3$

Modelling the redshift evolution of YSCs is a somewhat easier task, because YSCs are expected to trace the total cosmic star formation rate density \citep{lada2003, portegieszwart2010}. Hence, we assume
\begin{equation}\label{eq:YSCs}
\psi_{\rm YSC}(z)=\mathcal{B}_{\rm YSC}(z)\,{}\psi{}(z),
\end{equation}
where
\begin{equation}\label{eq:madau}
\psi{}(z)=0.01\,{}\frac{(1+z)^{2.6}}{1+[(1+z)/3.2]^{6.2}}~\text{M}_\odot\,{}\text{Mpc}^{-3}\,{}\text{yr}^{-1}
\end{equation}
is the fit to the total cosmic star formation rate density by \cite{madau2017} and $\mathcal{B}_{\rm YSC}(z)$ is the fraction of the cosmic star formation rate density that happens in YSCs. In our fiducial model, we adopt 
\begin{equation}\label{eq:BYSC}
\mathcal{B}_{\rm YSC}(z)=\max{\left\{0,\min{\left[0.1,\,{}1-\frac{\psi_{\rm NSC}(z)}{\psi{}(z)}-\frac{\psi_{\rm GC}(z)}{\psi{}(z)}\right]}\right\}}.
\end{equation}
In the above equation, we impose that $\mathcal{B}_{\rm YSC}(z)$ cannot take unphysical negative values, that $(\psi_{\rm NSC}+\psi_{\rm GC}+\psi_{\rm YSC})\leq{}\psi{}$ (i.e., the sum of star formation rate density in NSCs, GCs and YSCs cannot be higher than the total star formation rate density in the Universe at a given redshift) and that $\psi_{\rm YSC}(z)\leq{}0.1\,{}\psi{}(z)$.  Actually, for any reasonable values of $\psi{}(z)$, $\psi_{\rm NSC}(z)$ and $\psi_{\rm GC}(z)$ (see Figure~\ref{fig:SFR}), eq.~\ref{eq:BYSC} is equivalent to 
assume that YSCs represent $\sim{10}$\% of the total cosmic star formation rate \citep{kruijssen2014}. %,  as a reasonable guess from \cite{kruijssen2014}. 
%and impose that the total star formation rate density in our model at a given redshift cannot be higher than $\psi{}(z)$.

Finally, the star formation rate in the field will be equal to the remaining portion of the total cosmic star formation rate density:
\begin{equation}\label{eq:field}
\psi_{\rm iso}(z)=\mathcal{B}_{\rm iso}(z)\,{}\psi{}(z),
\end{equation}
where 
\begin{equation}
\mathcal{B}_{\rm iso}(z)=\max{\left[0,\,{}1-\frac{\psi_{\rm NSC}(z)}{\psi{}(z)}-\frac{\psi_{\rm GC}(z)}{\psi{}(z)}-\frac{\psi_{\rm YSC}(z)}{\psi{}(z)}\right]}.
\end{equation}
In the above equation, we assume that all the star formation rate that does not take place in star clusters goes into isolated binary formation, %at any given redshift,
  and impose that the total star formation rate density in our model at a given redshift cannot be higher than $\psi{}(z)$. Actually, the star formation rate in the field is always dominant over the other channels in our fiducial model, as shown in Fig.~\ref{fig:SFR}. 

%%%%%%%%%%%%%%%%%%%%%%%%%%%%FIGURE%%%%%%%%%%%%%%%%%%%%%%%%%%%%%%%%%%%%%%%%%%%%%%%%%%%
%%%%%%%%%%%%%%%%%%%%%%%%%%%%%%%%%%%%%%%%%%%%%%%%%%%%%%%%%%%%%%%%%%%%%%%%%%%%%%%%%%%%%%

\subsubsection{Merger efficiency}

The merger efficiency is the total number of BBHs of a given population that merge within a Hubble time divided by the total initial stellar mass of that population \citep{giacobbo2018,klencki2018}. For  isolated BBHs, this is simply
\begin{equation}\label{eq:eta_field}
\eta_{\rm field}(Z) = \frac{\mathcal{N}_{\text{TOT}}(Z)}{M_\ast{}(Z)},
\end{equation}
where $N_{\rm TOT}(Z)$ is the number of BBH mergers for a given metallicity $Z$ and $M_\ast(Z)$ is the total initial stellar mass of the population, assuming a Kroupa mass function between 0.1 and 150 M$_\odot$ \citep{kroupa2001}.

For  dynamical and original BBHs, the calculation of $\eta{}(Z)$ is less straightforward, because {\sc fastcluster} does not integrate the entire BH population of a star cluster, but only a sub-set, in order to sample the parameter space more efficiently (see Section~\ref{sec:clusters}). We thus estimate the merger efficiency in star clusters as
\begin{equation}\label{eq:eta_dyn}
\eta_{\rm SC}(Z) = \frac{\mathcal{N}_{\rm merg,\,{}sim }(Z)}{\mathcal{N}_{\rm sim}(Z)}\,{}\frac{\mathcal{N}_{\rm BH}(Z)}{M_\ast{}(Z)},
\end{equation}
where $\mathcal{N}_{\rm merg,\,{}sim }(Z)$ is the number of  BHs simulated with {\sc fastcluster} that merge within a Hubble time for a given metallicity $Z$,  $\mathcal{N}_{\rm sim}(Z)$ is the number of BHs simulated with {\sc fastcluster} for a given metallicity $Z$, $\mathcal{N}_{\rm BH}(Z)$ is the total number of BHs associated with a given metallicity (including the BHs we did not simulate with {\sc fastcluster}) and   $M_\ast{}(Z)$ is the total initial stellar mass for a given metallicity $Z$.  $\mathcal{N}_{\rm merg,\,{}sim }(Z)$ and $\mathcal{N}_{\rm sim }(Z)$  are directly extracted from the simulations. We calculate $M_\ast{}(Z)=\sum{}M_{\rm TOT}(Z)$, i.e. the sum of the initial total mass of all simulated star clusters with a given $Z$. We derive $\mathcal{N}_{\rm BH}(Z)$ as the number of BHs we expect from a stellar population following a Kroupa mass function between 0.1 and 150~M$_\odot$, assuming that all stars with zero-age main sequence mass $\ge{}20$~M$_\odot$ are BH progenitors\footnote{This threshold should be regarded as an approximation. As already shown by several authors, the transition between neutron star and BH progenitors depends not only on the zero-age main sequence mass, but rather  on a plethora of additional factors and is still highly uncertain \citep[e.g.,][]{oconnor2011,ugliano2012,pejcha2015,sukhbold2016,ertl2020, patton2020}.} \citep{heger2003}. In our definition, $\mathcal{N}_{\rm merg,\,{}sim }(Z)$ includes even Nth generation (Ng) mergers, while $\mathcal{N}_{\rm sim }(Z)$ counts only 1g BHs. Hence, the ratio $\mathcal{N}_{\rm merg,\,{}sim }(Z)/\mathcal{N}_{\rm sim}(Z)$ can be $>1$ if hierarchical mergers are extremely efficient.

%%%%%%%%%%%%%%%%%%%%%%%%%%%%%%%%TABLE%%%%%%%%%%%%%%%%%%%%%%%%%%%%%%%%%%%
\begin{table}[h]
	\begin{center}
	\caption{Model properties.\label{tab:table1}}
	\begin{tabular}{lcccccc}
          		\toprule
  Model & Channel & SN model & $\alpha$ & $\sigma_{\chi}$ & $\sigma_{\rm Z}$ & $f_{\rm orig}$\\
  \midrule
A02 & Isolated & delayed & 1 & 0.1 & 0.2 & 1\\
A02 & YSC      & delayed & 1 & 0.1 & 0.2 & 0.6\\
A02  & GC       & delayed & 1 & 0.1 & 0.2 & 0.1\\
A02 & NSC      & delayed & 1 & 0.1 & 0.2 & 0.01\\ %\vspace{0.1cm}\\
A03 & Isolated & delayed & 1 & 0.1 & 0.3 & 1 \\
A03 & YSC      & delayed & 1 & 0.1 & 0.3 & 0.6 \\
A03  & GC       & delayed & 1 & 0.1 & 0.3 & 0.1 \\
A03 & NSC      & delayed & 1 & 0.1 & 0.3 & 0.01\\ %\vspace{0.1cm}\\
A04 & Isolated & delayed & 1 & 0.1 & 0.4 & 1 \\
A04 & YSC      & delayed & 1 & 0.1 & 0.4 & 0.6\\
A04  & GC       & delayed & 1 & 0.1 & 0.4 & 0.1 \\
A04 & NSC      & delayed & 1 & 0.1 & 0.4 & 0.01 \vspace{0.1cm}\\

B02 & Isolated & rapid & 1 & 0.1 & 0.2 & 1 \\
B02 & YSC      & rapid & 1 & 0.1 & 0.2 & 0.6 \\
B02  & GC       & rapid & 1 & 0.1 & 0.2 & 0.1 \\
B02 & NSC      & rapid & 1 & 0.1 & 0.2 & 0.01\\ %\vspace{0.1cm}\\
B03 & Isolated & rapid & 1 & 0.1 & 0.3 & 1\\
B03 & YSC      & rapid & 1 & 0.1 & 0.3 & 0.6\\
B03  & GC       & rapid & 1 & 0.1 & 0.3 & 0.1\\
B03 & NSC      & rapid & 1 & 0.1 & 0.3 & 0.01\\ %\vspace{0.1cm}\\
B04 & Isolated & rapid & 1 & 0.1 & 0.4 & 1 \\
B04 & YSC      & rapid & 1 & 0.1 & 0.4 & 0.6\\
B04  & GC       & rapid & 1 & 0.1 & 0.4 & 0.1\\
B04 & NSC      & rapid & 1 & 0.1 & 0.4 & 0.01\vspace{0.1cm}\\

C02 & Isolated & delayed & 1 & 0.01 & 0.2 &1\\
C02 & YSC      & delayed & 1 & 0.01 & 0.2 &0.6\\
C02  & GC       & delayed & 1 & 0.01 & 0.2 &0.1\\
C02 & NSC      & delayed & 1 & 0.01 & 0.2 &0.01\\ %\vspace{0.1cm}\\
C03 & Isolated & delayed & 1 & 0.01 & 0.3 &1\\
C03 & YSC      & delayed & 1 & 0.01 & 0.3 &0.6\\
C03  & GC       & delayed & 1 & 0.01 & 0.3 &0.1\\
C03 & NSC      & delayed & 1 & 0.01 & 0.3 &0.01\\ %\vspace{0.1cm}\\
C04 & Isolated & delayed & 1 & 0.01 & 0.4 &1\\
C04 & YSC      & delayed & 1 & 0.01 & 0.4 &0.6\\
C04  & GC       & delayed & 1 & 0.01 & 0.4 &0.1\\
C04 & NSC      & delayed & 1 & 0.01 & 0.4 &0.01\vspace{0.1cm}\\

D02 & Isolated & delayed & 5 & 0.1 & 0.2 &1\\
D02 & YSC      & delayed & 5 & 0.1 & 0.2 &0.6\\
D02  & GC       & delayed & 5 & 0.1 & 0.2 &0.1\\
D02 & NSC      & delayed & 5 & 0.1 & 0.2 &0.01\\ %\vspace{0.1cm}\\
D03 & Isolated & delayed & 5 & 0.1 & 0.3 &1\\
D03 & YSC      & delayed & 5 & 0.1 & 0.3 &0.6\\
D03  & GC       & delayed & 5 & 0.1 & 0.3 &0.1\\
D03 & NSC      & delayed & 5 & 0.1 & 0.3 &0.01\\ %\vspace{0.1cm}\\
D04 & Isolated & delayed & 5 & 0.1 & 0.4 &1\\
D04 & YSC      & delayed & 5 & 0.1 & 0.4 &0.6\\
D04  & GC       & delayed & 5 & 0.1 & 0.4 &0.1\\
D04 & NSC      & delayed & 5 & 0.1 & 0.4 &0.01\\
		\bottomrule
	\end{tabular}
	\end{center}
%	\shiftleft{
	\footnotesize{Column 1: Name of the model, composed of a letter (A, B, C and D) followed  by a number indicating the metallicity spread (02, 03 and 04 indicate $\sigma_{\rm Z}=0.2,$ 0.3 and 0.4, respectively); column 2: formation channel (isolated, YSC, GC or NSC); column 3: core-collapse SN model (delayed or rapid); column 4: parameter $\alpha$ of common envelope for isolated binaries and original binaries; column 5: spin parameter $\sigma_{\chi}=0.1$ or 0.01; column 6: metallicity spread $\sigma_{\rm Z}=0.2, $ 0.3, 0.4; column 7 ($f_{\rm orig}$): original BBH fraction (in the isolated channel every binary is original).}
\end{table}

\subsubsection{Metallicity evolution}

For the metallicity evolution, we adopt a formalism similar to the one described by \cite{bouffanais2021}, namely we use the fit to the mass-weighted metallicity evolution given by \cite{madau2017}:
\begin{equation}\label{eq:met}
    \log{\langle{}Z/{\rm Z}_\odot\rangle{}}=0.153-0.074\,{}z^{1.34}
\end{equation}
To describe the spread around the mass-weighted metallicity, we assume that metallicities are distributed according to a log-normal distribution: %\micmap{with mean value $\log{\langle{}Z/{\rm Z}_\odot\rangle{}}$} and standard deviation $\sigma{}_Z$:
\begin{equation}
\label{eq:pdf}
p(z', Z) = \frac{1}{\sqrt{2 \pi\,{}\sigma_{\rm Z}^2}}\,{} \exp\left\{{-\,{} \frac{\left[\log{(Z(z')/{\rm Z}_\odot)} - {\langle{}\log{Z(z')/Z_\odot}\rangle{}}\right]^2}{2\,{}\sigma_{\rm Z}^2}}\right\},
\end{equation}
where %\citep{bavera2020}
\begin{equation}
   \langle{}\log{Z(z')/Z_\odot}\rangle{}=\log{\langle{}Z(z')/Z_\odot\rangle{}}-\frac{{\ln(10)}\,{}\sigma_{\rm Z}^2}{2}.
    \label{eq:average_Z}
\end{equation}

The standard deviation $\sigma{}_Z$ is highly uncertain. Here, we probe different values of $\sigma{}_Z=0.2,$ 0.3 and 0.4. Equation~\ref{eq:pdf} allows us to estimate 
the term $\mathcal{F}_i(z',z,Z)$ of eq.~\ref{eq:cosmorate}:
\begin{equation}\label{eq:Fz}
\mathcal{F}_i(z',z,Z)=\frac{\dot{\mathcal{N}}_i(z',z,Z)}{\mathcal{N}_{\text{TOT\,{}i}}(Z)}\,{}p(z', Z),
\end{equation}
where $\dot{\mathcal{N}}_i(z',z,Z)$ is the total number of BBHs of channel $i$ that form at redshift $z'$ with metallicity $Z$ and merge at redshift $z$ per unit time, while $\mathcal{N}_{\text{TOT,\,{}i}}(Z)$ is the total number of BBH mergers of channel $i$ with progenitor's metallicity $Z$. We use the same metallicity formalism for all the considered channels.

\subsubsection{Fraction of original and dynamical BBHs}\label{sec:fractions}

With {\sc fastcluster}, we evaluate original BBHs (i.e., BBHs that form from a binary star but then evolve dynamically in a star cluster) and dynamical BBHs (i.e., BBHs that form via three-body encounters or exchanges), separately. In order to estimate the total BBH merger rate, we need to know the percentage of original and dynamical BBHs. Ideally, the mixing fraction between  original and dynamical BBHs can be obtained by running Bayesian inference on GWTC-2. However, this would significantly increase the number of dimensions of our multi-channel analysis (see the next section); hence, we prefer to assume some physically motivated guess for the fraction of original BBHs.

In NSCs, the fraction of original binaries surviving dynamical interactions is expected to be of the order of $\sim{0.01}$, because most binary systems are soft in such extreme environment \citep{antonini2016}. Hence, we assume that the fraction of original BBHs in NSCs is $\sim{0.01}$, analogous to the total surviving binary fraction. %In the core of GCs, the fraction of binary systems is usually estimated to be $\sim{0.1}$, with large fluctuations from cluster to cluster \citep{sollima2007,milone2012}. 
We also assume that the fraction of original BBHs in GCs is $\sim{0.1}$, corresponding to the typical binary fraction measured in the core of GCs, with large fluctuations from cluster to cluster \citep{sollima2007,milone2012}. %This is probably an upper limit because a fraction of the binary systems observed in GCs are the result of exchanges.  %%This percentage supposedly includes not only original but also exchanged systems; but since observations mostly probe  low-mass binary systems, which are less likely to originate from a previous exchange, we can assume the order of magnitude is correct for original binary systems. Hence, we assume that the fraction of original BBHs in GCs is $\sim{0.1}$.
For YSCs we use the recent results by \cite{dicarlo2020b} and \cite{rastello2021}. Based on direct N-body simulations of YSCs, they find that the percentage of original BBH mergers is $\approx{60}$\%, with large fluctuations depending on metallicity. In Section~\ref{sec:uncertainties}, we will comment on the impact of these assumptions about the original BBH merger fraction. Finally, in the isolated BBH channel, each BBH is original by definition. The only difference between isolated BBHs and original binaries in YSCs/GCs/NSCs is that the latter are perturbed by dynamical encounters, while the former are unperturbed.

%The merger efficiency $\eta{}(Z)$ is estimated as the number of BBHs that merge within a Hubble time in a coeval population of star with initial mass $M_\ast$ and metallicity $Z$, divided by $M_\ast$. See \cite{santoliquido2020} for further details.

%\subsection{Star cluster formation rates}
%Two words on {\sc cosmo$\mathcal{R}$ate}...

\subsection{Description of runs}

For the isolated BBH channel, we ran $1.44\times{}10^8$ 
%$7.2\times{}10^8$ 
massive isolated binary systems with {\sc mobse}, considering twelve different metallicities ($Z=0.0002,$ 0.0004, 0.0008, 0.0012, 0.0016, 0.002, 0.004, 0.006, 0.008, 0.012, 0.016, $0.02$), two different SN models (rapid and delayed model, from \citealt{fryer2012}) and two %different 
values for the parameter $\alpha{}$ of common envelope ($\alpha=1,$ 5). 
%In particular, we have simulated $10^7$ binaries per each metallicity comprised between $Z = 0.0002$ and 0.002, and $2\times{}10^7$ binaries per each metallicity $Z\ge{}0.004$, since higher metallicities are associated with lower BBH merger efficiency (e.g. \citealt{giacobbo2018b,klencki2018}). Thus, we have simulated $1.8\times{}10^8$ isolated binaries per each SN and $alpha$ parameter. 
The zero-age main-sequence masses of the primary component of each binary star are distributed according to a Kroupa \citep{kroupa2001} initial mass function in the range $[5,\,{}150]\,{}{\rm M}_\odot$. The orbital periods, eccentricities and mass ratios of binaries are drawn from \cite{sana2012}. In particular, we derive the mass ratio $q$ as $\mathcal{F}(q) \propto q^{-0.1}$ with $q\in [0.1,\,{}1]$, the orbital period $P$ from $\mathcal{F}(\Pi) \propto \Pi^{-0.55}$ with $\Pi = \log{(P/\text{day})} \in [0.15,\,{} 5.5]$ and the eccentricity $e$ from $\mathcal{F}(e) \propto e^{-0.42}~~\text{with}~~ 0\leq e \leq 0.9$.

For the dynamical channels, we ran 288 different realizations of our models with {\sc fastcluster}, half of them for original binaries and the other half for dynamical binaries. Each of these 288 realizations consists of $10^6$ BBH systems. We consider three %different 
families of star clusters (NSCs, GCs and YSCs), twelve %different 
metallicities (the same as for the isolated BBHs), two %different 
values of the spin magnitude parameter ($\sigma_\chi=0.01$ and 0.1), two %different 
core-collapse SN models (rapid and delayed model, from \citealt{fryer2012}) and two %different 
values for the parameter $\alpha{}$ of common envelope ($\alpha=1,$ 5).  The properties of the star clusters are the same as described in Section~\ref{sec:clusters}.

For each of the isolated and dynamical models, we ran the {\sc cosmo${\mathcal{R}}$ate} code, in order to derive the merger rate of each specific channel. We considered three %different 
values of the metallicity spread $\sigma{}_{\rm Z}=0.2$, 0.3 and 0.4. Table~\ref{tab:table1} summarizes the details of each resulting model. Each model presented in Table~\ref{tab:table1} includes the 12 simulated progenitor metallicities, mixed according to the formalism of {\sc cosmo${\mathcal{R}}$ate} (Section~\ref{sec:MRD}). Furthermore, each star cluster model in Table~\ref{tab:table1} includes both dynamical and original BBHs, mixed according to the fractions described in Section~\ref{sec:fractions}.  In Section~\ref{sec:uncertainties}, we will consider additional models with respect to the ones summarized in Table~\ref{tab:table1}, to discuss the main uncertainties related to the formation rate of each channel, to the proportion between original and dynamical BBHs and to the properties of the considered star clusters.

%%%%%%%%%%%%%%%%%%%%%%%%%%%%%%%%TABLE%%%%%%%%%%%%%%%%%%%%%%%%%%%%%%%%%%%
\begin{table}
	\begin{center}
	\caption{BBH merger rate density at redshift $z=0$. %and maximum BBH merger rate density.
	\label{tab:table2}}
	\begin{tabular}{lccc}
          		\toprule
  Model & Channel & $\mathcal{R}(0)$ & $\mathcal{R}_{\rm Ng}(0)$\\ % & $\mathcal{R}_{\rm max}$ & $\mathcal{R}_{\rm Ng,\,{}max}$\\
  \midrule
A02 & Isolated & 5.14 & -- \\
A02 & YSC & 2.35 & 0.07 \\
A02 & GC  & 3.64 & 0.82 \\
A02 & NSC & 1.31 & 0.47\\
A03 & Isolated & 17.53 & -- \\
A03 & YSC & 4.40 & 0.11\\
A03 & GC  & 4.58 & 0.98\\
A03 & NSC & 1.41 & 0.51 \\
A04 & Isolated & 60.67 & -- \\
A04 & YSC & 8.72 & 0.16\\
A04 & GC & 5.59 & 1.14\\
A04 & NSC & 1.50 & 0.54
\vspace{0.1cm}\\

B02 & Isolated & 7.41 & -- \\
B02 & YSC & 3.10 & 0.08\\
B02 & GC  & 5.62 & 1.27\\
B02 & NSC & 2.08 & 0.76 \\
B03 & Isolated & 24.66 & -- \\
B03 & YSC & 6.07 & 0.14\\
B03  & GC & 6.97 & 1.50\\
B03 & NSC & 2.15 & 0.79\\
B04 & Isolated & 77.75 & -- \\
B04 & YSC & 11.93 & 0.23\\
B04  & GC & 8.47 & 1.74\\
B04 & NSC & 2.22 & 0.81
\vspace{0.1cm}\\

C02 & Isolated & 5.14 & --\\
C02 & YSC & 2.74 & 0.37\\
C02 & GC  & 4.58 & 1.75\\
C02 & NSC & 1.50 & 0.66\\
C03 & Isolated & 17.53 & --\\
C03 & YSC &4.98 & 0.61\\
C03  & GC & 5.74 & 2.14\\
C03 & NSC & 1.62 & 0.71\\
C04 & Isolated & 60.67 & -- \\
C04 & YSC & 9.55 & 0.94\\
C04  & GC & 6.98 & 2.53\\
C04 & NSC & 1.72 & 0.76
\vspace{0.1cm}\\

D02 & Isolated & 4.41 & --\\
D02 & YSC & 2.38 & 0.08\\
D02 & GC  & 3.73 & 0.83\\
D02 & NSC & 1.32 & 0.47 \\
D03 & Isolated & 13.23 & --\\
D03 & YSC & 4.39 & 0.11\\
D03  & GC & 4.74 & 1.00\\
D03 & NSC & 1.43 & 0.51\\
D04 & Isolated & 45.85 & --\\
D04 & YSC & 8.43 & 0.17\\
D04  & GC & 5.82 & 1.17\\
D04 & NSC & 1.52 & 0.54\\

		\bottomrule
	\end{tabular}
	\end{center}
%	\shiftleft{
	\footnotesize{Column 1: Model name; column 2: formation channel; column 3, $\mathcal{R}(0)$: merger rate density of BBHs at $z=0$ in units of Gpc$^{-3}$ yr$^{-1}$; column 4, $\mathcal{R}_{\rm Ng}(0)$: merger rate density of Nth generation (Ng) BBHs with ${\rm N}>1$ at $z=0$, in units of Gpc$^{-3}$ yr$^{-1}$.}
	%; column 5 ($\mathcal{R}_{\rm max}$): maximum merger rate density of BBHs over all redshifts; column 6 ($\mathcal{R}_{\rm max}$): maximum merger rate density of Nth generation (Ng) BBHs over all redshifts. }
\end{table}
\begin{figure*}
  \begin{center}
    \includegraphics[width = 0.9 \textwidth]{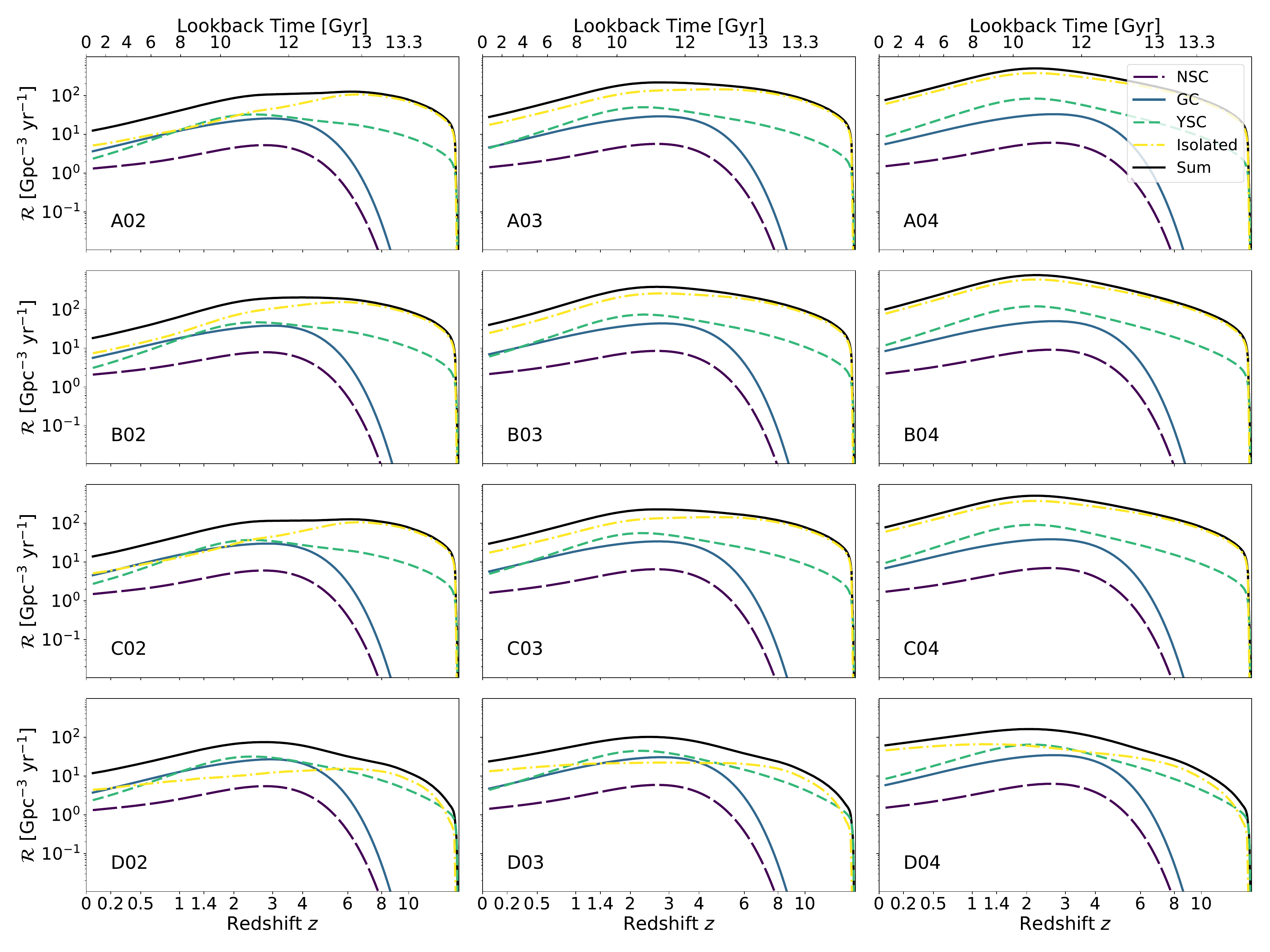}
    \end{center}
  \caption{BBH merger rate density $\mathcal{R}(z)$ as a function of redshift, in the comoving frame, for all the models listed in Table~\ref{tab:table1}. From left to right, the upper row shows models A02, A03 and A04, the second row models B02, B03 and B04, the third row models C02, C03 and C04 and the lower row models D02, D03 and D04. In all the panels, yellow dot-dashed line: isolated BBHs; light-blue short-dashed line: BBHs in YSCs; blue solid  line: BBHs in GCs; dark-blue long-dashed line: BBHs in NSCs; black solid  line: total merger rate density. \label{fig:rate}}
\end{figure*}
%%%%%%%%%%%%%%%%%%%%%%%%%%%%%%%%%%%%%%%%%%%%%%%%%%%%%%%%%%%%%%%%%%%%%%%%%%%%%%%%%%%%%%

%%%%%%%%%%%%%%%%%%%%%%%%%%%%FIGURE%%%%%%%%%%%%%%%%%%%%%%%%%%%%%%%%%%%%%%%%%%%%%%%%%%%
\begin{figure*}
  \begin{center}
    \includegraphics[width = 0.9 \textwidth]{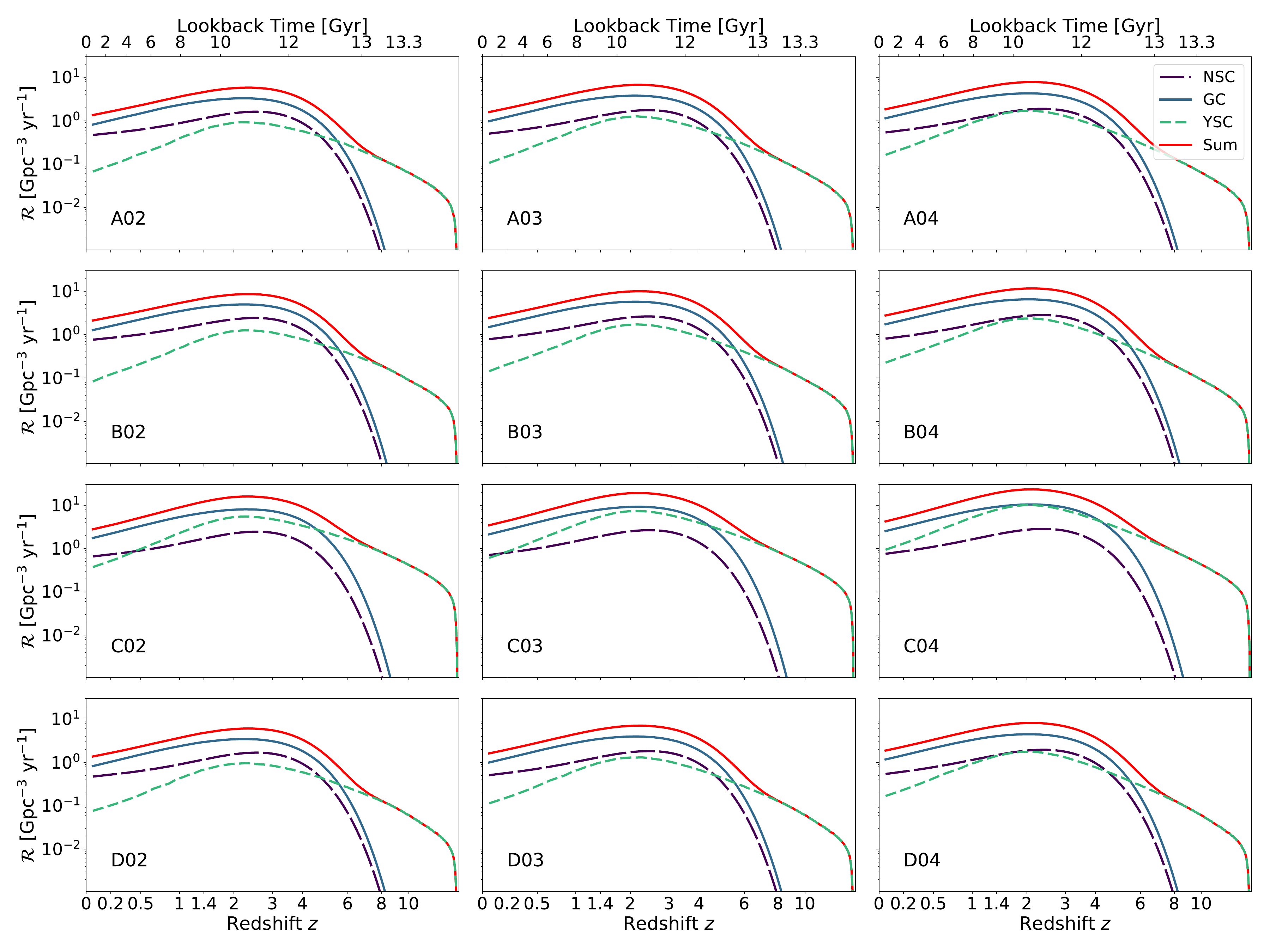}
    \end{center}
  \caption{Merger rate density of Nth generation (Ng) BBHs $\mathcal{R}_{\rm Ng}(z)$ as a function of redshift, in the comoving frame, for all the models listed in Table~\ref{tab:table1}. The order of the panels is the same as in Fig.~\ref{fig:rate}. In all the panels, light-blue short-dashed line: BBHs in YSCs; blue solid  line: BBHs in GCs; dark-blue long-dashed line: BBHs in NSCs; red solid  line: total merger rate density of Ng BBHs. \label{fig:rate_ng}}
\end{figure*}
%%%%%%%%%%%%%%%%%%%%%%%%%%%%%%%%%%%%%%%%%%%%%%%%%%%%%%%%%%%%%%%%%%%%%%%%%%%%%%%%%%%%%%

\subsection{Bayesian inference and mixing fractions}

To compare our models against GW events in the first (O1), second (O2) and in the first part of the third observing run (O3a) of the LIGO--Virgo collaboration (LVC), we use a hierarchical Bayesian approach. Given a number $N_{\rm obs}$ of GW observations, $\mathcal{H}=\lbrace h^{k} \rbrace_{k=1}^{N_{\rm obs}}$, described by an ensemble of parameters $\theta$, the posterior distribution of the hyper-parameters $\lambda{}$ associated with the models is described as an in-homogeneous Poisson distribution \citep{loredo2004,mandel2018}

\begin{eqnarray}\label{eq:post_hier_model}
p(\lambda{}, N_\lambda | \mathcal{H}) = \text{e}^{-\mu_{\lambda}}\,{}  \pi(\lambda{}, N_\lambda{}) \prod_{k=1}^{N_{\rm obs}}  N_{\lambda} \int_{\theta} \mathcal{L}^{k}(h^k | \theta) \,{}p(\theta | \lambda )\,{}{\rm d}\theta{}, %\nonumber{}\\
\end{eqnarray}
where $\theta$ are the GW parameters, $N_{\lambda}$ is the number of events predicted by the astrophysical model, $\mu_{\lambda}$ is the predicted number of detections associated with the model and the GW detector, $\pi{}(\lambda{},N_\lambda{})$ is the prior distribution on $\lambda$ and $N_\lambda$, and $\mathcal{L}^{k}(\lbrace h\rbrace^k | \theta)$ is the likelihood of the $k-$th detection. The predicted number of detections is given by $\mu{}(\lambda{})=N_\lambda\,{}\beta{}(\lambda{})$, where 
\begin{equation}\label{eq:beta}
\beta{}(\lambda{})=\int_\theta p(\theta{}|\lambda{})\,{}p_{\rm det}(\theta{})\,{}{\rm d}\theta    
\end{equation} 
is the detection efficiency of the model. In eq.~\ref{eq:beta}, $p_{\rm det}(\theta{})$ is the probability of detecting a source with parameters $\theta$ and can be inferred by computing the optimal signal-to-noise ratio and comparing it to a detection threshold, as described, e.g., in \cite{bouffanais2021}. The values for the event's log-likelihood are derived from the posterior and prior samples released by the LVC, such that the integral in eq.~\ref{eq:post_hier_model} is approximated with a Monte Carlo approach as 
\begin{equation}\label{eq:approx_integral_likeli}
\mathcal{I}^{k} = \int_{\theta}\mathcal{L}^{k}(h^k | \theta) \,{}p(\theta | \lambda )\,{}{\rm d}\theta{}\sim{}\frac{1}{N_s^k}\,{}\sum_{i=1}^{N_s^k}\frac{p(\theta^k_i | \lambda{})}{\pi^k(\theta_i^k)},
\end{equation}
where $\theta_i^k$ is the $i-$th posterior sample for the $k-$th detection and $N_s^k$ is the total number of posterior samples for the $k-$th detection. Both the model and prior distributions are estimated with Gaussian kernel density estimation.

%Finally, we can also choose to neglect the information coming from the number of sources predicted by the model when estimating the posterior distribution. By doing so, we can have some insights on the impact of the rate on the analysis. In practice, this can be done by 
In our analysis, we further marginalise eq.~\ref{eq:post_hier_model} over $N_{\lambda}$ using a prior $\pi(N_{\lambda}) \sim 1 / N_{\lambda}$ \citep{fishbach2018}, which yields the following expression
\begin{eqnarray}\label{eq:post_hier_model_marg} 
p(\lambda| \mathcal{H}) \sim \pi(\lambda) \prod_{k=1}^{N_{\rm obs}}  \dfrac{\mathcal{I}^k}{\beta(\lambda)} ,
%p(\lambda| \mathcal{H}) \sim \pi(\lambda) \prod_{k=1}^{N_{\rm obs}}  \left[ \dfrac{\int \mathcal{L}^{k}(h^{k} | \theta) \,{} p(\theta | \lambda)  \,{} \text{d} \theta}{\beta(\lambda)} \right],}
\end{eqnarray}
where the integral $\mathcal{I}^k$ can be approximated in the same way as in eq.~\ref{eq:approx_integral_likeli} and $\beta(\lambda)$ is given by eq.~\ref{eq:beta}. We make this choice to neglect the information coming from the number of sources predicted by the model when estimating the posterior distribution. By doing this assumption, our analysis is not affected by the large uncertainties on the rates (see, e.g., Section~\ref{sec:uncertainties}). More details on this procedure are described in \cite{mandel2018} and \cite{bouffanais2021}.

In our analysis, our model distribution is the sum of the contributions from multiple channels (isolated BBHs, dynamical BBHs in YSCs, GCs and NSCs) weighted by mixing fraction hyper-parameters as
\begin{eqnarray}\label{eq:mixfrac}
p(\theta{}|f_{\rm iso},\,{}f_{\rm YSC},\,{}f_{\rm GC},\,{}f_{\rm NSC},\lambda{})=f_{\rm iso}\,{}p(\theta{}|{\rm iso}, \lambda{})\nonumber\\
+f_{\rm YSC}\,{}p(\theta{}|{\rm YSC},{}\lambda{})
+f_{\rm GC}\,{}p(\theta{}|{\rm GC},{}\lambda{})+f_{\rm NSC}\,{}p(\theta{}|{\rm NSC},{}\lambda{}),
\end{eqnarray}
where $f_{\rm iso}$, $f_{\rm YSC}$, $f_{\rm GC}$ and $f_{\rm NSC}$ are the mixing fractions of BBHs from isolated binary stars, YSCs, GCs and NSCs, defined so that $f_{\rm iso} + f_{\rm YSC} + f_{\rm GC} + f_{\rm NSC} = 1$.  Based on this definition, the mixing fraction for each channel approximately is the fraction of merger events associated with that specific channel. Since we decided to  neglect the information coming from the number of sources predicted by the model when estimating the posterior distribution (eq.~\ref{eq:post_hier_model_marg}), the mixing fractions are sensitive to the properties of the sources (masses, spins and redshift at which the merger occurs) but do not depend on the merger rate density we estimated for each channel.  
 Furthermore, this definition of the mixing fraction assumes that all GWTC-2 events originate from the four channels we considered here. In future work, we will extend our analysis by including the other possible channels (e.g., primordial BHs, AGN discs, triples and multiples) we have not considered here.

In our  analysis,  we do not consider all GWTC-2 event candidates \citep{abbottO3a}  but only the 45 BBHs analyzed in \cite{abbottO3popandrate}, which represent a sub-sample with false alarm rate $<1$~yr$^{-1}$. 
For these 45 BBHs, we  use the GWTC-2 posterior samples  for $\theta=\{\mathcal{M},\,{}q,\,{}\chi_{\rm eff},z\}$, where $\mathcal{M}=(m_1\,{}m_2)^{3/5}(m_1+m_2)^{-1/5}$ is the chirp mass, $q=m_2/m_1$ is the mass ratio, and $\chi_{\rm eff}$ is the effective spin:
\begin{equation}\label{eq:chieff}
\chi_{\rm eff}= \frac{(m_1\,{}\vec{\chi}_1+m_2\,{}\vec{\chi}_2)}{m_1+m_2}\cdot{}\frac{\vec{L}}{L},
\end{equation}
where $\vec{L}$ is the BBH orbital angular momentum,  while 
$\vec{\chi}_1$ and $\vec{\chi}_2$
%$\vec{\chi}_1\equiv{}c\,{}\vec{S}_1/(G\,{}m_1^2)$ and $\vec{\chi}_2\equiv{}c\,{}\vec{S}_2/(G\,{}m_2^2)$ 
are the dimensionless spin vectors. We used a Metropolis-Hastings algorithm to generate samples from the posterior of eq. \ref{eq:post_hier_model_marg}. We ran chains of $10^{7}$ iterations for each set of hyper-parameters, and then trimmed the chains using auto-correlation length.

\section{Results}\label{sec:results}

%%%%%%%%%%%%%%%%%%%%%%%%%%%%FIGURE%%%%%%%%%%%%%%%%%%%%%%%%%%%%%%%%%%%%%%%%%%%%%%%%%%%
\begin{figure*}
  \begin{center}
    \includegraphics[width = 0.8 \textwidth]{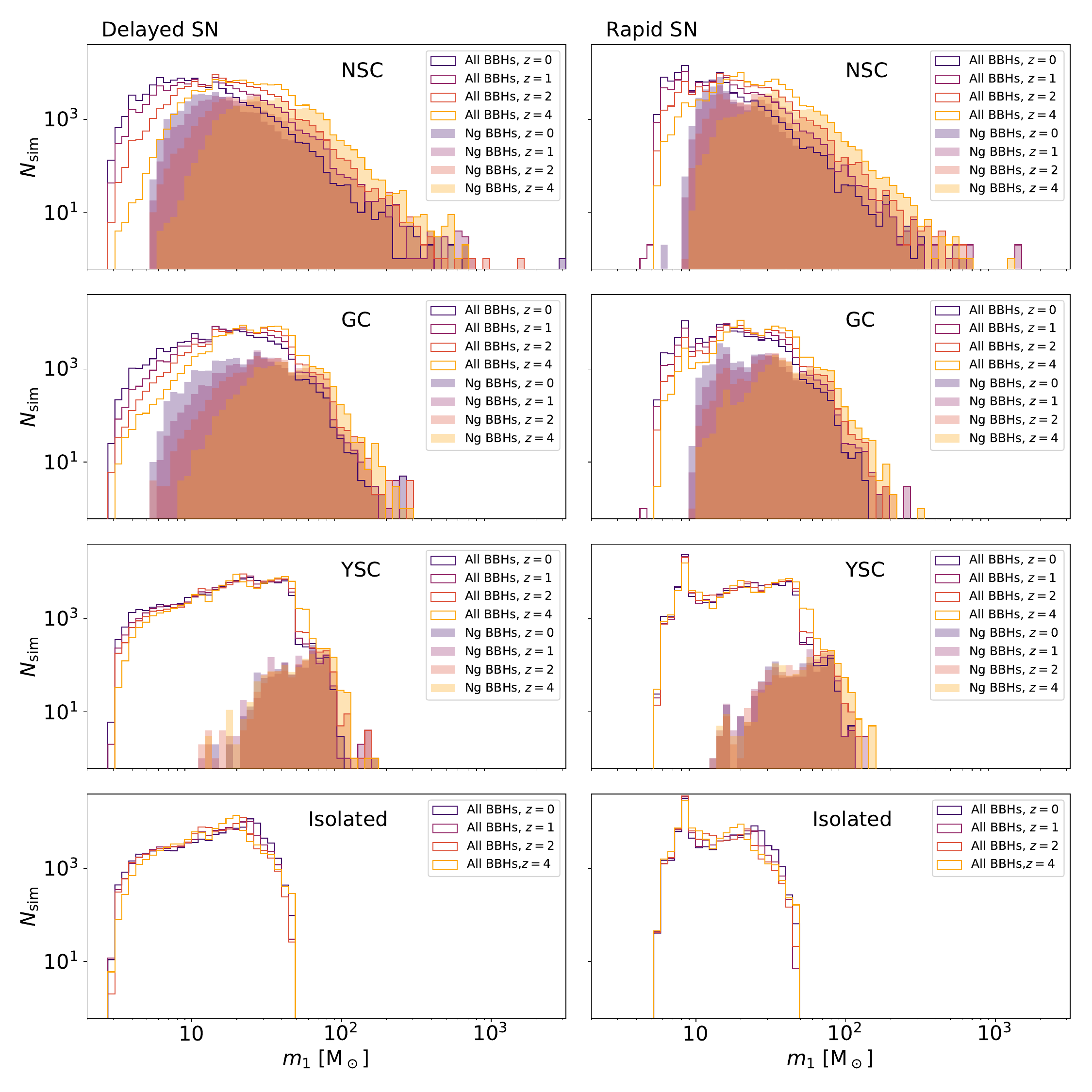}
    \end{center}
  \caption{Distribution of primary BH masses in BBH mergers. From top to bottom: NSCs, GCs, YSCs and isolated BBHs. Left-hand column: model A03 (exploiting the delayed SN model); right-hand column: model B03 (with the rapid SN model). Unfilled histograms: all BBH mergers; filled histograms: Ng BBH mergers (with ${\rm N}>1$). Blue, purple, pink and orange histograms: BBHs merging at $z=0,$ 1, 2 and 4, respectively. We show the same number of simulated BBHs per each channel and per each redshift. \label{fig:masses}}
\end{figure*}
%%%%%%%%%%%%%%%%%%%%%%%%%%%%%%%%%%%%%%%%%%%%%%%%%%%%%%%%%%%%%%%%%%%%%%%%%%%%%%%%%%%%%%

%%%%%%%%%%%%%%%%%%%%%%%%%%%%FIGURE%%%%%%%%%%%%%%%%%%%%%%%%%%%%%%%%%%%%%%%%%%%%%%%%%%%
\begin{figure*}
  \begin{center}
    \includegraphics[width = 0.9 \textwidth]{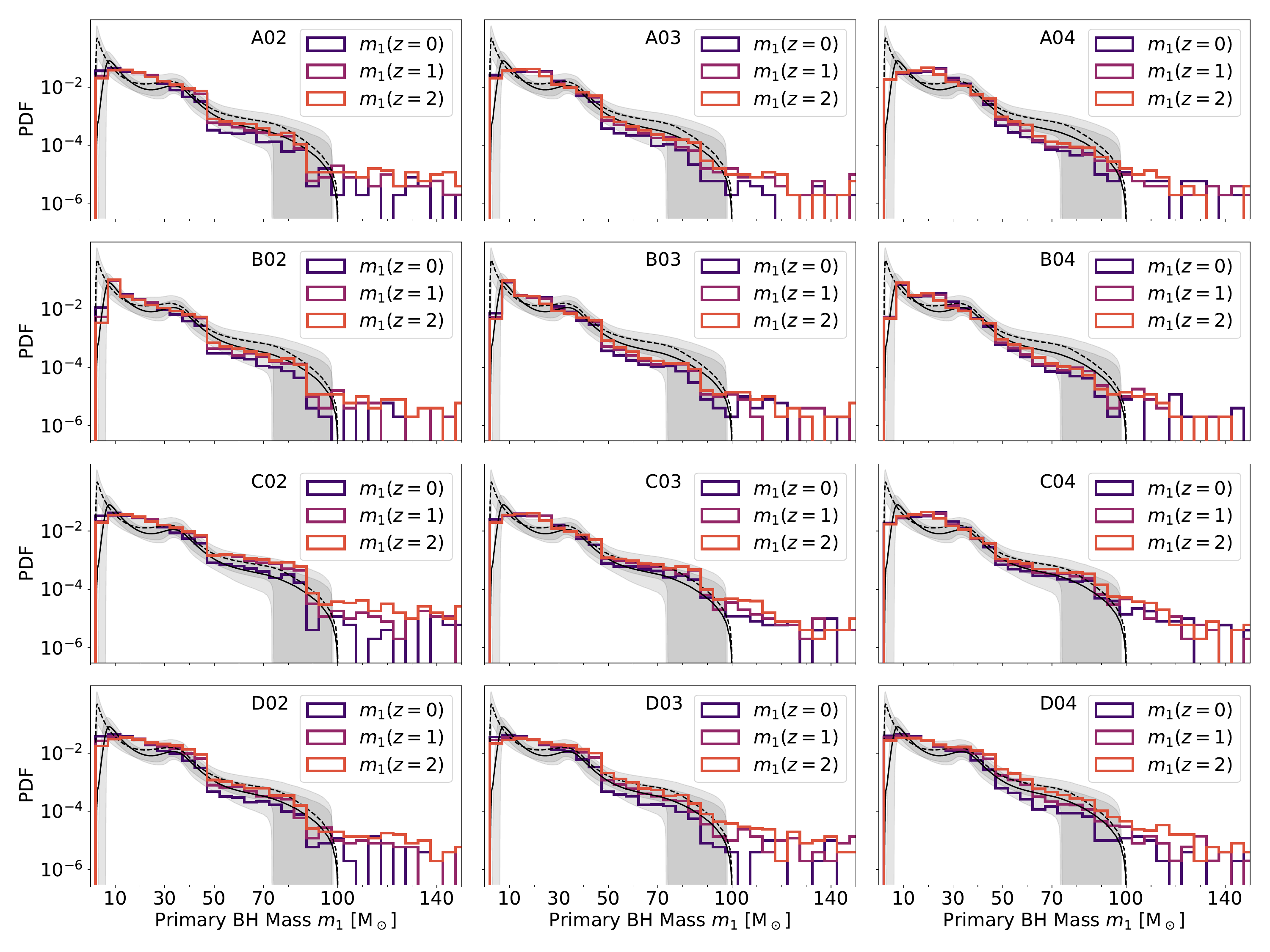}
    \end{center}
  \caption{Probability distribution function of primary BH masses ($m_1$) of BBHs merging at redshift $z=0$ (blue), $z=1$ (purple) and $z=2$ (pink). In each panel, we have put together different channels (isolated, YSC, GC and NSC) based on their merger rate, to obtain a synthetic Universe. We truncate the plots at 150 M$_\odot$ to improve the readability of this Figure, but there are several BHs with even higher masses (see Fig.~\ref{fig:masses}). The order of the panels is the same as in Fig.~\ref{fig:rate}. The black solid (dashed) line is the median value of the {\sc power law + peak} model applied to GWTC-2 BBHs excluding (including) GW190814 \protect{\citep{abbottO3popandrate}}. The shaded gray areas are the corresponding 90\% credible intervals. We arbitrarily re-scaled the {\sc power law + peak} model on the $y$ axis.  \label{fig:mix_mass}}
\end{figure*}
%%%%%%%%%%%%%%%%%%%%%%%%%%%%%%%%%%%%%%%%%%%%%%%%%%%%%%%%%%%%%%%%%%%%%%%%%%%%%%%%%%%%%%

%%%%%%%%%%%%%%%%%%%%%%%%%%%%FIGURE%%%%%%%%%%%%%%%%%%%%%%%%%%%%%%%%%%%%%%%%%%%%%%%%%%%
\begin{figure*}
  \begin{center}
    \includegraphics[width = 0.75 \textwidth]{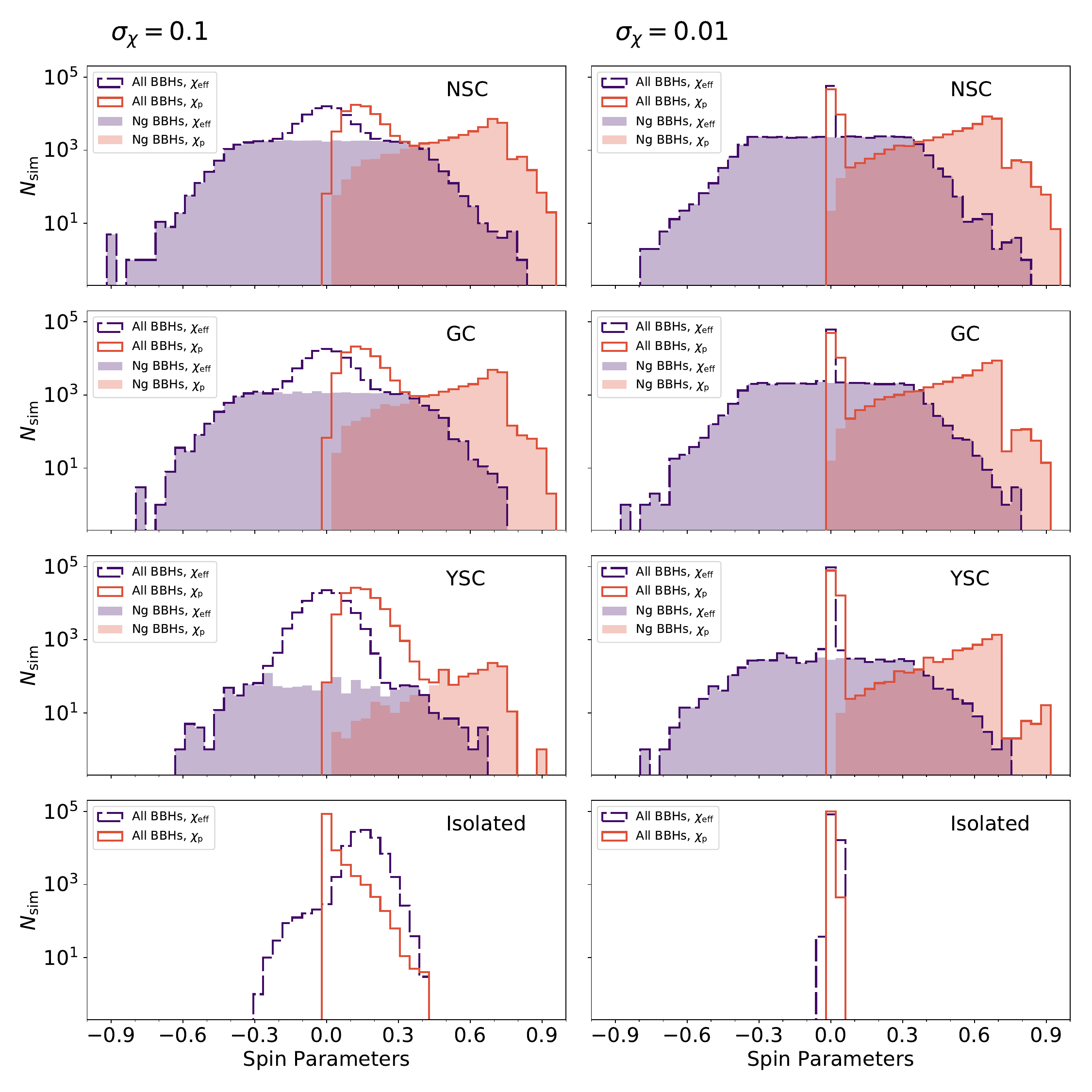}
    \end{center}
  \caption{Distribution of effective spins ($\chi_{\rm eff}$, blue) and precessing spins ($\chi_{\rm p}$, red) in BBH mergers at redshift $z=0$. From top to bottom: NSCs, GCs, YSCs and isolated BBHs. Left-hand column: model A03 ($\sigma_{\chi}=0.1$); right-hand column: model C03 ($\sigma{}_\chi{}=0.01$). Unfilled histograms: all BBH mergers; filled histograms: Ng BBH mergers (with ${\rm N}>1$). We show the same number of simulated BBHs per each channel. \label{fig:spins}}
\end{figure*}
%%%%%%%%%%%%%%%%%%%%%%%%%%%%%%%%%%%%%%%%%%%%%%%%%%%%%%%%%%%%%%%%%%%%%%%%%%%%%%%%%%%%%%

%%%%%%%%%%%%%%%%%%%%%%%%%%%%FIGURE%%%%%%%%%%%%%%%%%%%%%%%%%%%%%%%%%%%%%%%%%%%%%%%%%%%
\begin{figure*}
  \begin{center}
    \includegraphics[width = 0.9 \textwidth]{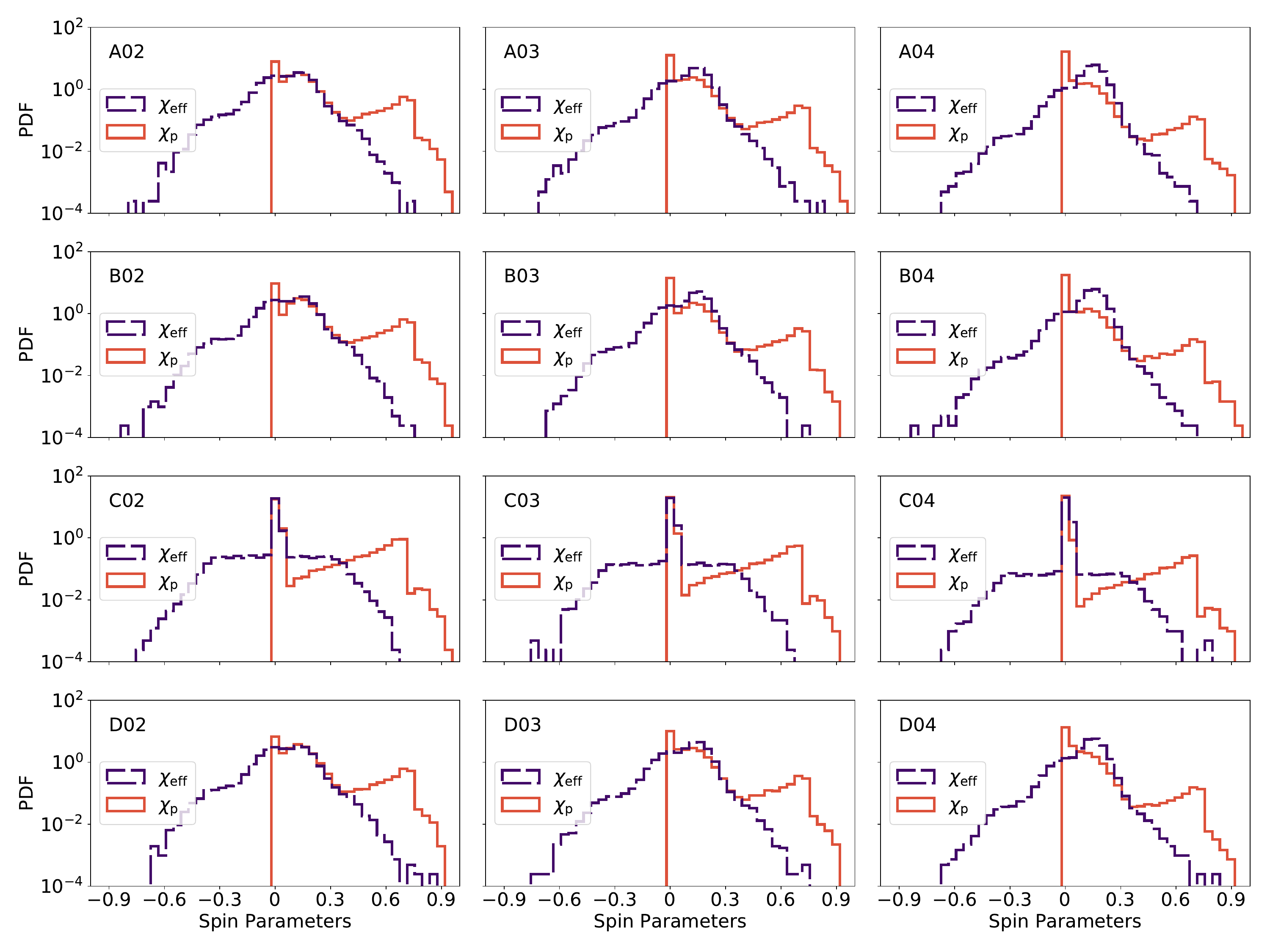}
    \end{center}
  \caption{Probability distribution function of effective ($\chi_{\rm eff}$, blue dashed line) and precessing spin ($\chi_{\rm p}$, red solid line) of BBHs merging at $z=0$. In each panel, we have put together different channels (isolated, YSC, GC and NSC) based on their merger rate, to obtain a synthetic Universe, as already done in Fig.~\ref{fig:mix_mass}. The order of the panels is the same as in Fig.~\ref{fig:rate}. \label{fig:mix_spin}}
\end{figure*}
%%%%%%%%%%%%%%%%%%%%%%%%%%%%%%%%%%%%%%%%%%%%%%%%%%%%%%%%%%%%%%%%%%%%%%%%%%%%%%%%%%%%%%

\subsection{Multi-channel rates}
%CALCOLARE MMAX VS MCLUSTER

Figure~\ref{fig:rate} shows the BBH merger rate density  as a function of redshift, while Table~\ref{tab:table2} shows the BBH merger rate density at $z=0$, for all of our models. The BBH merger rate density evolution of NSCs and GCs  are only weakly affected by the metallicity spread $\sigma_{\rm Z}$, because BBHs efficiently pair up and harden in these massive star clusters, regardless of progenitor's metallicity. In contrast, the merger rate density of isolated BBHs is dramatically affected by progenitor's metallicity. As already discussed in previous  works \citep[e.g.,][]{chruslinska2019,santoliquido2021,mandel2021}, there is about one order of magnitude difference in the BBH merger rate if we assume $\sigma_{\rm Z}=0.2$ or $0.4$. The main reason of this difference is that a larger value of $\sigma_{\rm Z}$ allows the formation of a larger fraction of metal-poor stars at low-redshift. In isolated binaries, the merger efficiency of BBHs born from metal-poor stars is three--four orders of magnitude higher than that of BBHs born from metal-rich stars. The behaviour of YSCs is intermediate between the field and GCs/NSCs. 

The impact of the core-collapse SN model is nearly the same for all considered channels: the local BBH merger rate is $\approx{40-60\%}$ higher if we assume the rapid instead of the delayed SN model. This happens because the minimum mass of 1g BHs is higher in the rapid ($m_{\rm min}=5$ M$_\odot$) than in the delayed model ($m_{\rm min}=3$ M$_\odot$), leading to a shorter GW decay time. Actually, the difference between the two SN models tends to be slightly higher for GC and NSC BBHs than for isolated BBHs,  because larger BH masses favour their retention inside the parent star cluster after the SN kick. 

The merger rate density of isolated BBHs is $\approx{16-32}$\% higher if we assume $\alpha_{\rm CE}=1$ than if we assume $\alpha_{\rm CE}=5$, as already discussed by \cite{giacobbo2020}. In contrast, the BBH merger rate density of YSCs, GCs and NSCs is almost unaffected by the choice of the common envelope parameter.

The merger rate density of isolated BBHs and YSC BBHs extends to higher redshift with respect to GC BBHs and NSC BBHs, but this is probably a mere effect of the extrapolation of the  fitting formula for the star formation rate density  to redshift $z\gtrsim{10}$, where we do not have measurements \citep{madau2017}. Furthermore, we do not model population III stars in this work (see, e.g. \citealt{kinugawa2016,hartwig2016,belczynski2017,liubromm2020,ng2021,tanikawa2021} for an accurate modeling of BBHs from metal-free stars).

The total local merger rate density (i.e., the sum of the merger rate densities of the four channels) is within the 90\% credible interval of the value inferred by the LVC [$\mathcal{R}(0)=19.3_{-9}^{+15}$ Gpc$^{-3}$ yr$^{-1}$ if GW190814 is not considered a BBH and if we allow the merger rate to evolve with redshift, \citealt{abbottO3popandrate}] for the models A02, A03, B02, C02, C03, D02 and D03, while it is too high in the other models. In particular, the models  with $\sigma_{\rm Z}=0.4$ (A04, B04, C04 and D04) always produce a total local merger rate $\mathcal{R}>60$ Gpc$^{-3}$ yr$^{-1}$, which is a factor of $>3$ higher than the median value inferred by the LVC. %However, such a comparison is not self-consistent, because our models have a different mass function with respect to the {\sc{power law + peak}} model assumed by the LVC. We will derive the expected number of detections from our models in Section~\ref{sec:bayes}. 

Figure~\ref{fig:rate_ng} shows the merger rate density evolution of Ng BHs only, with ${\rm N}>1$. The fourth column of Table~\ref{tab:table2} reports the local merger rate density  of Ng BBHs. GCs give the main contribution to the merger rate of Ng BBHs in all our models [$\mathcal{R}_{\rm Ng}(0)\approx{0.8-2.5}$ Gpc$^{-3}$ yr$^{-1}$], followed by NSCs [$\mathcal{R}_{\rm Ng}(0)\approx{0.5-0.8}$ Gpc$^{-3}$ yr$^{-1}$] and YSCs [$\mathcal{R}_{\rm Ng}(0)\approx{0.1-0.9}$ Gpc$^{-3}$ yr$^{-1}$]. In models  A02--A04, the merger rate of Ng BHs is $\approx{36}\%$, $\approx{22}\%$ and $\approx{2-3}\%$ of the total merger rate in NSCs, GCs and YSCs, respectively.

%* if we consider 2g only GCs win but NSCs are extremely close

%* effect we already discussed in MM 2021: if sigmax=0.01 both GCs and YSCs are boosted but not NSCs, because in NSCs the escape velocity is already higher

Models C02--CO4 have higher values of $\mathcal{R}_{\rm Ng}$ with respect to the other models, because lower spin magnitudes are associated with lower relativistic kicks and hence favour the merger of Ng BHs. The  spin magnitude parameter $\sigma_{\chi}$ has a stronger impact on the rate of Ng BBH mergers in YSCs than in GCs and especially NSCs. For example, the local Ng BBH rate in YSCs is a factor of $\approx{6}$ higher in model C03 ($\sigma_{\chi}=0.01$) with respect to model A03 ($\sigma_{\chi}=0.1$). In the case of GCs and NSCs, the difference between the C03 and A03 models is equal to a factor of $\approx{2}$ and $\approx{1.4}$, respectively. This trend is a consequence of the different escape velocities of YSCs, GCs and NSCs: in our models, NSCs have escape velocities of the order of 100 km s$^{-1}$, hence they retain a large fraction of the merger remnants even if $\sigma_{\chi}=0.1$; in contrast, GCs and especially YSCs have lower escape velocities and lose most of their BH merger remnants if $\sigma_{\chi}=0.1$. If we lower $\sigma_{\chi}$ to 0.01, even YSCs can efficiently retain their BH merger remnants. Accounting for all these uncertainties, the total merger rate of Ng BBHs in the local Universe ranges from $\approx{1}$ Gpc$^{-3}$ yr$^{-1}$ (D02) to $\approx{4}$ Gpc$^{-3}$ yr$^{-1}$ (C04).
%* despite these differences, global second generation rate between ~2 and ~ 5 (chi=0.01, sigmZ=0.4)

\subsection{BBH mass}

%* masses remarkably dependent on SN model
Figure~\ref{fig:masses} shows the distribution of the primary BH masses in the four considered channels at different redshifts, for models A03 
%(left-hand panel, 
(delayed SN model) and B03  %(right-hand panel, 
(rapid SN model). The overall primary BH mass distribution strongly depends on the core-collapse SN model by construction: while the delayed SN model allows the formation of BHs with mass as low as 3 M$_\odot$, the rapid SN model prevents the formation of BHs with mass $<5$ M$_\odot$.

%* YSCs similar to field (because dynamics less efficient)
The mass function of primary BHs in YSCs is similar to the distribution of primary BHs in isolated binary systems, but while the latter has a sharp truncation at $\approx{50}$ M$_\odot$, the former has a tail up to $\approx{200}$ M$_\odot$ because of Ng systems.

The contribution of dynamically formed BBHs and Ng BBHs is more important for NSCs and GCs, which are the most dynamically active systems. However,  BHs with mass $>100$ M$_\odot$ are extremely rare even in GCs and NSCs. As already discussed in \cite{mapelli2021}, NSCs are the channel with the largest number of low-mass primary BHs. This happens because single BHs that receive a SN kick higher than the escape velocity  leave their parent star cluster and cannot pair up dynamically. Since the natal kick in our models is higher for  less massive BHs, this strongly suppresses the formation of light BBHs  in YSCs and GCs, which have relatively low escape velocity, while NSCs are able to retain even the least massive BHs.  

%* impact of second generation larger in  NSCs and GCs than YSCs

The mass distribution of isolated BBHs and YSC BBHs does not show any strong dependence on the merger redshift. In contrast, the mass distribution of BBHs in GCs and especially NSCs shows a relevant trend: low-mass BBH mergers are more common at low redshift than at high redshift. This is a consequence of dynamics: more massive BHs %have smaller values of $t_{\rm dyn}$ (eq.~\ref{eq:tdyn}) and 
dynamically pair up  on a shorter timescale than lighter BHs (eq.~\ref{eq:tdyn}). Moreover, the timescale for GW decay is shorter for more massive systems than for lighter ones  (eqs.~\ref{eq:mapelli2018}).  

 Previous work has shown that the mass distribution of isolated BBHs might even have an opposite trend with respect to dynamical BBHs: low-mass isolated BBHs might have a shorter delay time than massive isolated BBHs as a consequence of common envelope \citep{mapelli2019,vanson2021}. This might result in a dearth of massive isolated BBH mergers at high redshift. In the bottom panel of Fig.~\ref{fig:masses}, we do see a very weak trend, with the peak of the massive isolated primary BHs shifting from $m_1\sim{25-30}$ M$_\odot$ at $z=4$ to $m_1\sim{20}$ M$_\odot$ at $z=0$. However, this shift is much weaker than the opposite trend for dynamically formed BBHs. Furthermore, the delay time of isolated BBHs is drastically affected by a number of factors, such as the common envelope parameter $\alpha$, the accretion efficiency and the stellar metallicity (see, e.g., Figure A2 of \citealt{bouffanais2021b}). 

%More important dependence for NSCs and GCs: lighter BHs merge less efficiently at high z because they do not have time (in isolation this is compensated by binary evolution common envelope)

%Figure~\ref{fig:mix_mass} shows a realization of the BH mass distribution we obtain at $z=0$ when we put together BBHs from various channels according to their merger rate. In other words, this is the entire population of BBH mergers at redshift $z=0$ in our synthetic Universe, without including  observation biases. The most notable difference is between the rapid and delayed model, the former displaying a stronger peak at primary mass $m_1\approx{10}$ M$_\odot$ with respect to the latter. The tail of high-mass BHs ($\geq{}50$ M$_\odot$) is more populated in models C02--C04 with respect to the other models, because the low-spin models have a larger percentage of Ng BBHs. 

Figure~\ref{fig:mix_mass} shows a realization of the primary BH mass distribution we obtain by putting together BBHs from various channels according to their merger rate. In other words, this is the entire population of BBH mergers at $z=0,$ 1 and 2 in our synthetic Universe, without including  observation biases. The most notable difference is between the rapid and delayed model, the former displaying a stronger peak at primary mass $m_1\approx{10}$ M$_\odot$ with respect to the latter. The tail of high-mass BHs ($\geq{}50$ M$_\odot$) is more populated in models C02--C04 with respect to the other models, because the low-spin models have a higher percentage of Ng BBHs.

Going from models with $\sigma_{\rm Z}=0.2$ to models with $\sigma_{\rm Z}=0.4$, the contribution of primary BHs with mass $m_{\rm BH}\sim{20}$ M$_\odot$ becomes more and more important, because the isolated BBH channel (which has the largest population of BHs with mass $m_{\rm BH}\sim{20}$ M$_\odot$, Fig.~\ref{fig:masses}) is associated with a higher merger rate for larger values of $\sigma_{\rm Z}$.  This happens because a larger value of $\sigma_{\rm Z}$ allows the formation of a larger fraction of metal-poor stars at low redshift, which results in a higher merger efficiency for isolated BBHs at $z\sim{0}$. 
%Actually, the increase of the peak at ~20 Msun is just a consequence of the different weight of the isolated channel with respect to the other channels when we move from sigma_Z=0.2 to sigma_Z=0.4. As the referee says, at sigma_Z=0.4 we have the formation of more metal-poor stars at redshift~0, which have higher merger efficiency.

In Figure~\ref{fig:mix_mass}, we also visually compare our synthetic populations with the {\sc power law + peak} model from Fig.~8 of \cite{abbottO3popandrate}. Our populations 
%visually 
match the {\sc power law + peak} model, the main difference being the number of primary BHs with mass $\sim{20}$~M$_\odot$, which is higher in our models, especially if we adopt the delayed model and $\sigma_{\rm Z}=0.4$. At the high-mass end, our low spin models C02--C04 better match the {\sc power law + peak} model than the other runs, but all of our synthetic populations are within the 90\% credible interval of the phenomenological model by \cite{abbottO3popandrate}.

Figure~\ref{fig:mix_mass} compares the population of BBHs at redshift 0, 1 and 2. The fraction of BBHs with primary mass $\ge{}30$ M$_\odot$ increases with redshift, because of the %relative 
contribution of GCs and NSCs to the overall BBH population (Fig.~\ref{fig:masses}). % increases from $z=0$ to $z=3$. 
This dependence on redshift is more evident in models with $\sigma_{\rm Z}=0.2$ and $\alpha=5$, in which the contribution of isolated BBHs is quenched.  From GWTC-2 data, there is no clear evidence that the mass of BBH mergers evolves with redshift \citep{abbottO3a,abbottO3popandrate}, but some recent analysis suggests a possible weak trend under several assumptions \citep{fishbach2021a,fishbach2021b}. In our models, we also predict a weak trend, driven by BBHs in GCs and NSCs. %Future observing runs of the LVC and next-generation 

%and The 30-50 and higher mass tail are more populated at $z=1,$ 2 than at redshift zero + Maya

\subsection{BBH spins}

%Spin magnitudes of 1g BBHs are toy models in our simulations, while spin orientations are 
Figure~\ref{fig:spins} shows the distribution of effective ($\chi_{\rm eff}$,  eq.~\ref{eq:chieff}) and precessing spins ($\chi_{\rm p}$) for our BBH mergers at redshift $z=0$. We calculated  $\chi_{\rm p}$ according to the following definition:
\begin{eqnarray}\label{eq:chip}
%\chi_{\rm eff}= \frac{(m_1\,{}\vec{\chi}_1+m_2\,{}\vec{\chi}_2)}{m_1+m_2}\cdot{}\frac{\vec{L}}{L},
%\nonumber{}\\
%\chi_{\rm p}=\frac{c}{B_1\,{}G\,{}m_1^2}\,{}\max{(B_1\,{}S_{1\perp{}},\,{}B_2\,{}S_{2\,{}\perp})},
\chi_{\rm p}=\max{\left[\chi_{\rm 1\perp},\,{}\frac{q\,{}(4\,{}q+3)}{4+3\,{}q}\,{}\chi_{\rm 2\perp}\right]},
\end{eqnarray}
where  $\chi_{1\perp{}}$ and $\chi{}_{2\perp{}}$ are the components of the dimensionless spin vectors ($\vec{\chi}_1$ and $\vec{\chi}_2$) perpendicular to the orbital angular momentum. %Finally $\vec{\chi}_1\equiv{}c\,{}\vec{S}_1/(G\,{}m_1^2)$ and $\vec{\chi}_2\equiv{}c\,{}\vec{S}_2/(G\,{}m_2^2)$.

The distributions of $\chi_{\rm eff}$ and $\chi_{\rm p}$ are nearly independent of redshift, but this is not surprising, because we derive the magnitudes of 1g BBHs from a toy model which does not depend on either redshift or mass. %We compare a model with low-spin (C03 with $\sigma{}_\chi=0.01$) with our fiducial spin case (A03 with $\sigma{}_\chi=0.1$). 
In the dynamical channels $\chi_{\rm eff}$ is symmetric around zero, because we assume isotropic spin orientation, while the isolated channel has a strong preference for positive $\chi_{\rm eff}$ because of angular momentum alignment during mass transfer and tidal evolution. Ng mergers extend the distribution of $\chi_{\rm eff}$ to very low and very high values with respect to 1g mergers.

The distribution of $\chi_{\rm p}$ for isolated BBHs has a strong peak at zero, because of the preferential alignment, while the distribution of $\chi_{\rm p}$ for dynamical BBHs has two peaks. The position of the primary peak depends on the choice of $\sigma_{\chi}$, while the secondary peak is at  $\chi_{\rm p}\approx{0.7}$ and is completely determined by Ng BBHs. %The relative importance of the secondary peak with respect to the first one

Figure~\ref{fig:mix_spin} shows the distribution of $\chi_{\rm eff}$ and $\chi_{\rm p}$ we obtain by putting together BBHs formed via different channels according to their merger rate at $z=0$. In the fiducial spin case ($\sigma_{\chi}=0.1$), the distribution of $\chi_{\rm eff}$ becomes more asymmetric if we assume a larger value of $\sigma_{\rm Z}$, because the contribution of isolated BBHs to the total merger population increases for larger metallicity spreads. In the low-spin case ($\sigma_{\chi}=0.01$), this dependence on $\sigma_{\rm Z}$ is not visible because all 1g BBHs have vanishingly small spins.

In the fiducial spin case ($\sigma_{\chi}=0.1$), the total distribution of $\chi_{\rm p}$ shows three peaks: a first narrow peak at zero because of isolated BBHs, a second broader peak at $\chi_{\rm p}\sim{0.1-0.2}$ because of 1g dynamical BBHs and a third peak at  $\chi_{\rm p}\sim{0.7}$ because of Ng mergers. The peak at $\chi_{\rm p}\sim{0.1-0.2}$ is an effect of our choice of $\sigma_\chi$. In the low-spin case ($\sigma_{\chi}=0.01$), $\chi_{\rm p}$ has only two peaks: one at zero (due to both isolated BBHs and 1g dynamical BBHs) and the other at 0.7 (Ng BBHs).

\subsection{Mixing fractions}\label{sec:bayes}
%%%%%%%%%%%%%%%%%%%%%%%%%%%%FIGURE%%%%%%%%%%%%%%%%%%%%%%%%%%%%%%%%%%%%%%%%%%%%%%%%%%%
\begin{figure*}
  \begin{center}
    \includegraphics[width = 0.9 \textwidth]{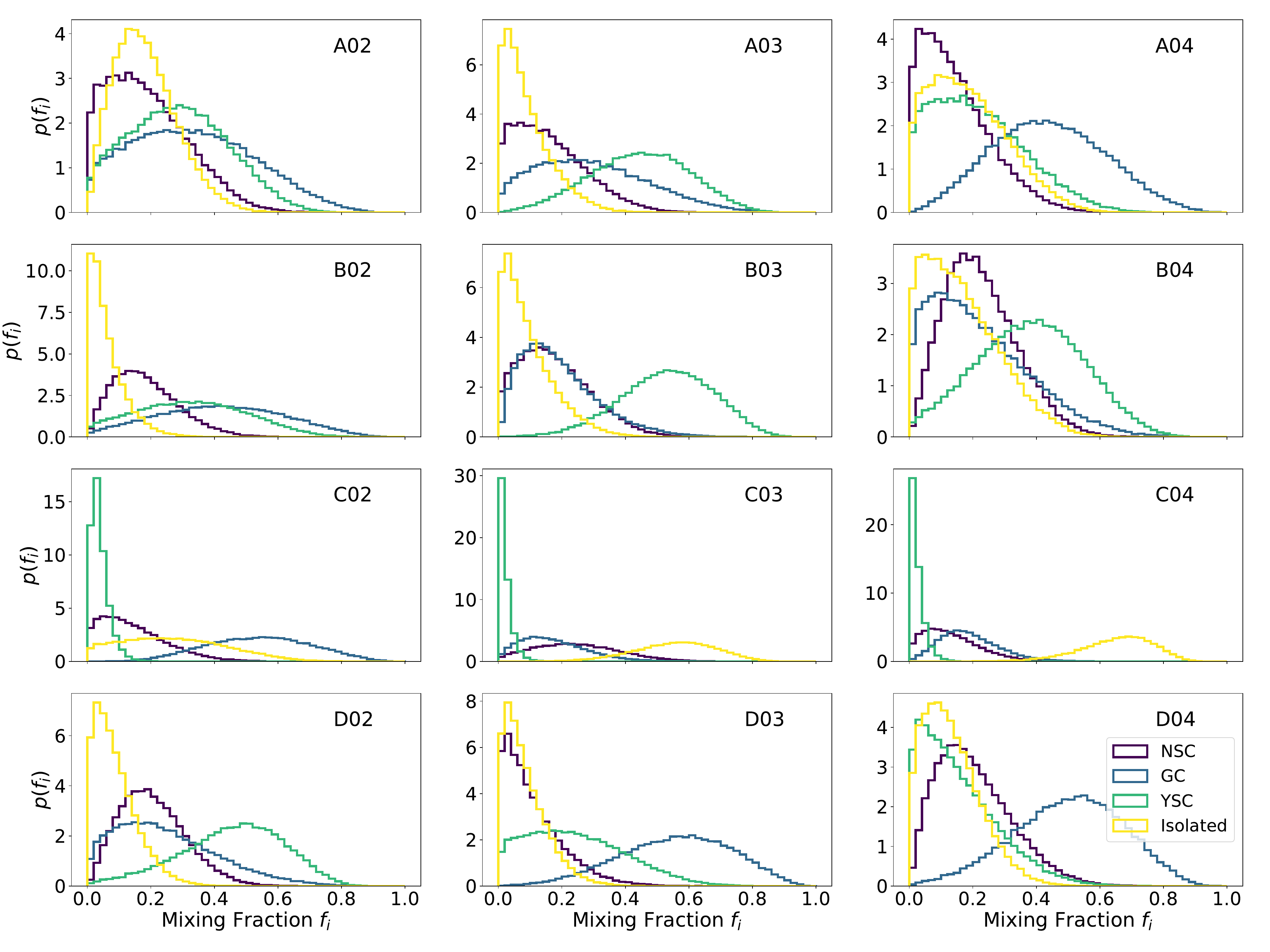}
    \end{center}
  \caption{Posterior probability distribution of the mixing fractions $f_i$ (with $i=$ iso, YSC, GC and NSC, eq.~\ref{eq:mixfrac}) for all our models. The order of the panels is the same as in Fig.~\ref{fig:rate}. Yellow line: isolated BBHs; light-blue line: BBHs in YSCs; blue line: BBHs in GCs; dark-blue line: BBHs in NSCs. To produce this Figure we used the posterior samples from GWTC-2 \citep{abbottO3a}. % and we took into account both detection efficiency and rates.
  \label{fig:mixing_frac}}
\end{figure*}
%%%%%%%%%%%%%%%%%%%%%%%%%%%%%%%%%%%%%%%%%%%%%%%%%%%%%%%%%%%%%%%%%%%%%%%%%%%%%%%%%%%%%%

Figure~\ref{fig:mixing_frac} shows the posterior distribution of the mixing fractions $f_i$ (with $i=$ iso, YSC, GC and NSC), defined in eq.~\ref{eq:mixfrac}. These values are obtained taking into account the detection efficiency (eq.~\ref{eq:beta}) 
%and the number of sources predicted by each channel (eq.~\ref{eq:approx_integral_likeli}). 
and marginalizing eq.~\ref{eq:approx_integral_likeli} over $N_\lambda$ (eq.~\ref{eq:post_hier_model_marg}). 
Table~\ref{tab:table3} shows the median and 90\% credible interval of the mixing fractions. The mixing fractions wildly depend on the details of each model: small differences between one model and another result in large differences in terms of $f_i$. There are still too many uncertainties about astrophysical models to claim we know the relative impact of each channel onto the global BBH population. 

However, there is a common feature of all our models: GWTC-2 data moderately support the co-existence of multiple channels: in each of our models, the mixing fraction is significantly larger than zero for at least two of the four considered channels. Hence, multiple formation channels likely are at work, to  produce the population of BBH mergers we observe with GWs. In particular,  the contribution of either isolated BBHs or BBHs in YSCs is needed to explain the low-mass portion of the BH mass function (Figs.~\ref{fig:masses} and \ref{fig:mix_mass}), while the contribution of BBHs in GCs or NSCs is fundamental to reproduce the high-mass tail ($m_1\geq{}50$ M$_\odot$). 

Isolated BBHs are associated with higher mixing fractions in models with very low spins (C02--C04) possibly because the observed population does not favour a strongly asymmetric $\chi_{\rm eff}$ distribution with support for large positive values \citep{abbottO3popandrate}. %The side effect is that the mixing fraction of YSCs is lower in the low spin models than elsewhere: 

The metallicity spread also has a large effect on the mixing fractions. Fig.~\ref{fig:mixing_frac} does not account for the predicted number of detections. Hence, the impact of $\sigma_{\rm Z}$ on our mixing fractions is rather connected with BH mass and redshift distribution than with rates. A larger metallicity spread increases the percentage of isolated BBHs born from metal-poor stars that merge in the low-redshift Universe. Since these tend to be more massive than BBHs from metal-rich stars, the mass function of isolated BBHs tends to be more top-heavy when $\sigma_{\rm Z}$ is large, hence more similar to the one of dynamical BBHs. As a consequence, $f_{\rm iso}$ tends to be larger.
%have a slight preference for positive $\chi_{\rm eff}$ which is supported by GWTC-2 population (e.g., Figure~11 of \citealt{abbottO3popandrate}).

\begin{figure}
  \begin{center}
    \includegraphics[width = 0.45 \textwidth]{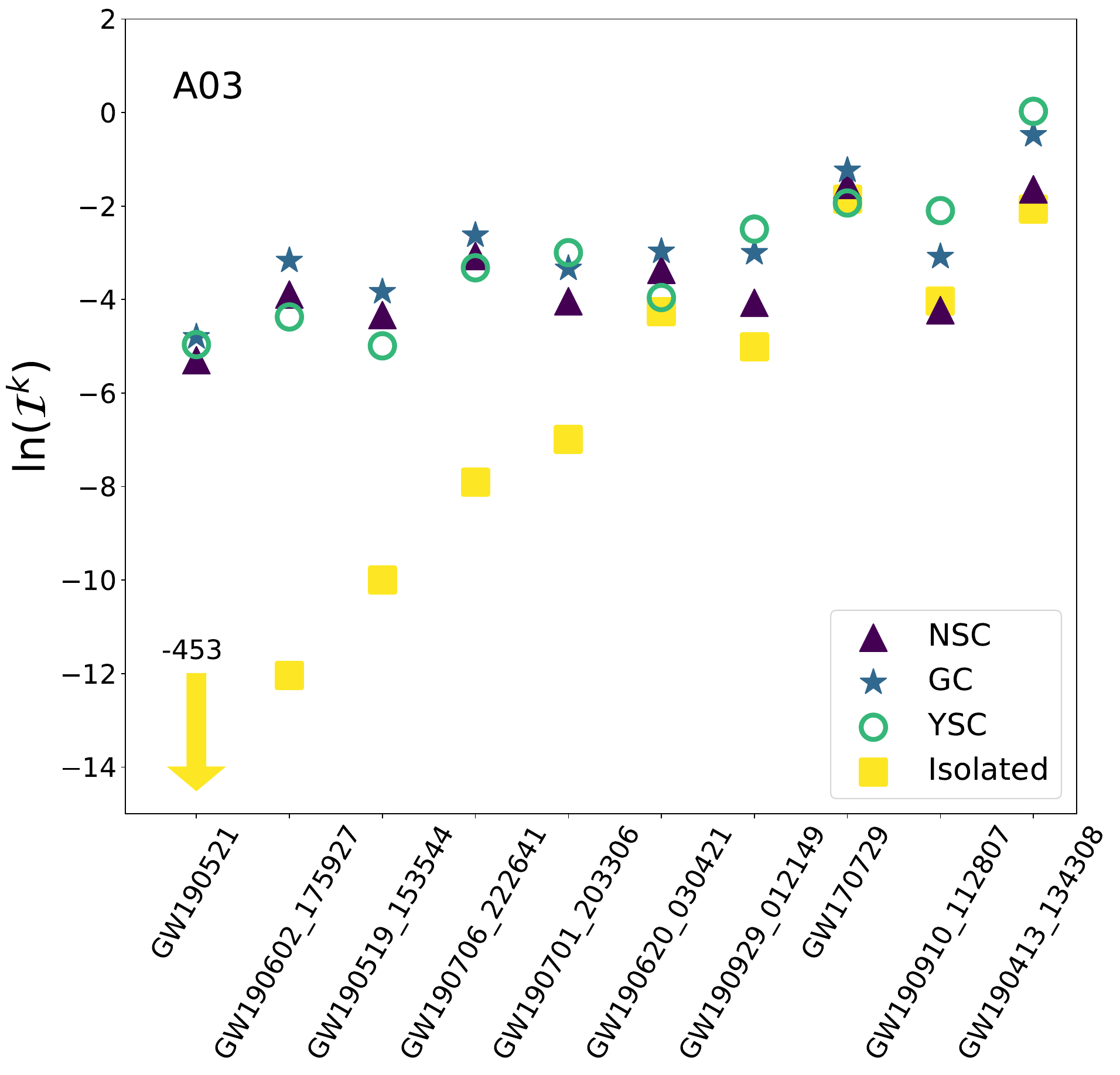}
    \end{center}
  \caption{Value of $\mathcal{I}^{k}$ (eq.~\ref{eq:approx_integral_likeli}) for the 10 GWTC-2 events with the largest chirp mass $\mathcal{M}$  and for model A03. The GW events on the $x$ axis are ordered by decreasing median value of $\mathcal{M}$. Yellow squares (and yellow arrow for GW190521): isolated BBHs; light-blue open circles: BBHs in YSCs; blue stars: BBHs in GCs; dark-blue  triangles: BBHs in NSCs. 
  %Value of $\mathcal{I}^{k}$ for GWTC-2 events where large discrepancies where found between formation channels models, i.e. events $k$ satisfying $| \ln(\mathcal{I}^{k}_{i}) - \ln(\mathcal{I}^{k}_{j})| \geq 3.0 \text{ for } i,j \in \left[\text{iso, YSC, GC, NSC}\right]$ for at least one couple $(i,j)$, for model A03 (upper panel) and B03 (lower panel). For readability, we restricted the range of the y-axis and indicated values outside of the range with arrows and their associated values.
  \label{fig:match_model}}
\end{figure}

 Figure~\ref{fig:match_model}  shows the values of $\mathcal{I}^{k}$ (defined in eq.~\ref{eq:approx_integral_likeli}) for model A03. %We do not show the other models, because they 
The other models yield similar results. The integral $\mathcal{I}^{k}$ %is useful to compare
%gives an idea of
 quantifies how well our  models are able to %explain 
match the posterior distributions of a  GW event. Figure \ref{fig:match_model} only  shows  the 10 BBHs with the largest chirp mass from GWTC-2 \citep{abbottO3a}. The isolated channel struggles to explain the five most massive events, which have $\mathcal{M}\ge{}40$~M$_\odot$. In the case of  GW190521, we find $\ln (\mathcal{I}^{k} )\approx{-453}$, even if we do not include $\chi_{\rm p}$ among the considered parameters. While a  strongly negative value of $\mathcal{I}^{k}$ for a single event does not significantly affect the mixing fractions, significantly negative values for at least five BBHs (over the 45 events we included in our sample) have some impact on $f_{\rm i}$. In model A03, the mixing fraction of the isolated channel increases from $f_{\rm iso}=0.07_{-0.07}^{+0.17}$ to $0.10_{-0.09}^{+0.20}$ if we recalculate it after removing the five events with the largest chirp mass (the reported uncertainty is the 90\% credible interval). Correspondingly, the mixing fraction of GCs decreases from  $f_{\rm GC}=0.28_{-0.23}^{+0.33}$ to $0.21_{-0.18}^{+0.32}$ when we remove these five events, while $f_{\rm YSC}$ and $f_{\rm NSC}$ remain nearly unchanged.

%\yb{First, when looking at GW190521 we see that the isolated model has an extremely low value of both for A03 ($-422$) and B03 ($-533$), indicating almost a zero match for this event. This originates from the model's sharp cutoff at masses close to $50 M_{\odot}$, which cannot describe the high values of masses of this event.}

%\yb{Another interesting event is GW190814 that exhibits large variations of $\ln (\mathcal{I}^{k} )$ between our various formation channels. For A03, the young star cluster distribution is the one that best fits the event's parameters.}

%%%%%%%%%%%%%%%%%%%%%%%%%%%%FIGURE%%%%%%%%%%%%%%%%%%%%%%%%%%%%%%%%%%%%%%%%%%%%%%%%%%%
%%%%%%%%%%%%%%%%%%%%%%%%%%%%%%%%%%%%%%%%%%%%%%%%%%%%%%%%%%%%%%%%%%%%%%%%%%%%%%%%%%%%%%

%%%%%%%%%%%%%%%%%%%%%%%%%%%%%%%%TABLE%%%%%%%%%%%%%%%%%%%%%%%%%%%%%%%%%%%
\begin{table}
	\begin{center}
	\caption{Median values of the mixing fractions.\label{tab:table3}}
	\begin{tabular}{lcccc}
          		\toprule
  Model & $f_{\rm iso}$ & $f_{\rm YSC}$ &$ f_{\rm GC}$ & $f_{\rm NSC}$\\
  \midrule
  A02 & $0.17_{-0.13}^{+0.19}$ & $0.28_{-0.23}^{+0.26}$ & $0.32_{-0.27}^{+0.34}$ & $0.17_{-0.15}^{+0.25}$ \vspace{0.1cm}\\ 
  A03 & $0.07_{-0.07}^{+0.17}$ & $0.45_{-0.27}^{+0.25}$ & $0.28_{-0.23}^{+0.33}$ & $0.15_{-0.13}^{+0.23}$ \vspace{0.1cm}\\ 
  A04 & $0.17_{-0.15}^{+0.24}$ & $0.20_{-0.18}^{+0.29}$ & $0.43_{-0.28}^{+0.30}$ & $0.13_{-0.11}^{+0.23}$ \vspace{0.1cm}\\ 
  B02 & $0.05_{-0.04}^{+0.13}$ & $0.33_{-0.27}^{+0.29}$ & $0.42_{-0.32}^{+0.32}$ & $0.17_{-0.13}^{+0.20}$ \vspace{0.1cm}\\ 
  B03 & $0.08_{-0.07}^{+0.18}$ & $0.54_{-0.26}^{+0.22}$ & $0.17_{-0.13}^{+0.26}$ & $0.16_{-0.14}^{+0.22}$ \vspace{0.1cm}\\ 
  B04 & $0.15_{-0.13}^{+0.24}$ & $0.38_{-0.28}^{+0.27}$ & $0.19_{-0.17}^{+0.33}$ & $0.21_{-0.15}^{+0.22}$ \vspace{0.1cm}\\ 
  C02 & $0.26_{-0.23}^{+0.30}$ & $0.03_{-0.03}^{+0.06}$ & $0.54_{-0.27}^{+0.26}$ & $0.13_{-0.11}^{+0.22}$ \vspace{0.1cm}\\ 
  C03 & $0.56_{-0.23}^{+0.19}$ & $0.02_{-0.01}^{+0.04}$ & $0.16_{-0.13}^{+0.21}$ & $0.24_{-0.19}^{+0.24}$ \vspace{0.1cm}\\ 
  C04 & $0.66_{-0.22}^{+0.15}$ & $0.02_{-0.02}^{+0.05}$ & $0.18_{-0.12}^{+0.19}$ & $0.12_{-0.10}^{+0.18}$ \vspace{0.1cm}\\ 
  D02 & $0.08_{-0.07}^{+0.15}$ & $0.47_{-0.31}^{+0.23}$ & $0.23_{-0.19}^{+0.33}$ & $0.20_{-0.14}^{+0.20}$ \vspace{0.1cm}\\ 
  D03 & $0.07_{-0.06}^{+0.15}$ & $0.23_{-0.20}^{+0.29}$ & $0.57_{-0.31}^{+0.26}$ & $0.09_{-0.08}^{+0.21}$ \vspace{0.1cm}\\ 
  D04 & $0.12_{-0.10}^{+0.18}$ & $0.13_{-0.12}^{+0.26}$ & $0.51_{-0.31}^{+0.26}$ & $0.19_{-0.14}^{+0.24}$ \\ 
		\bottomrule
	\end{tabular}
	\end{center}
%	\shiftleft{
	\footnotesize{This Table reports the median values and 90\% intervals of the mixing fractions shown in Fig.~\ref{fig:mixing_frac}.} %we obtain when we take into account both detection efficiency (eq.~\ref{eq:beta}) and the predicted number of detections by each channel (eq.~\ref{eq:approx_integral_likeli}). }
\end{table}

\section{Discussion: main sources of uncertainty and further caveats}\label{sec:uncertainties}

\begin{figure}
  \begin{center}
    \includegraphics[width=0.45 \textwidth]{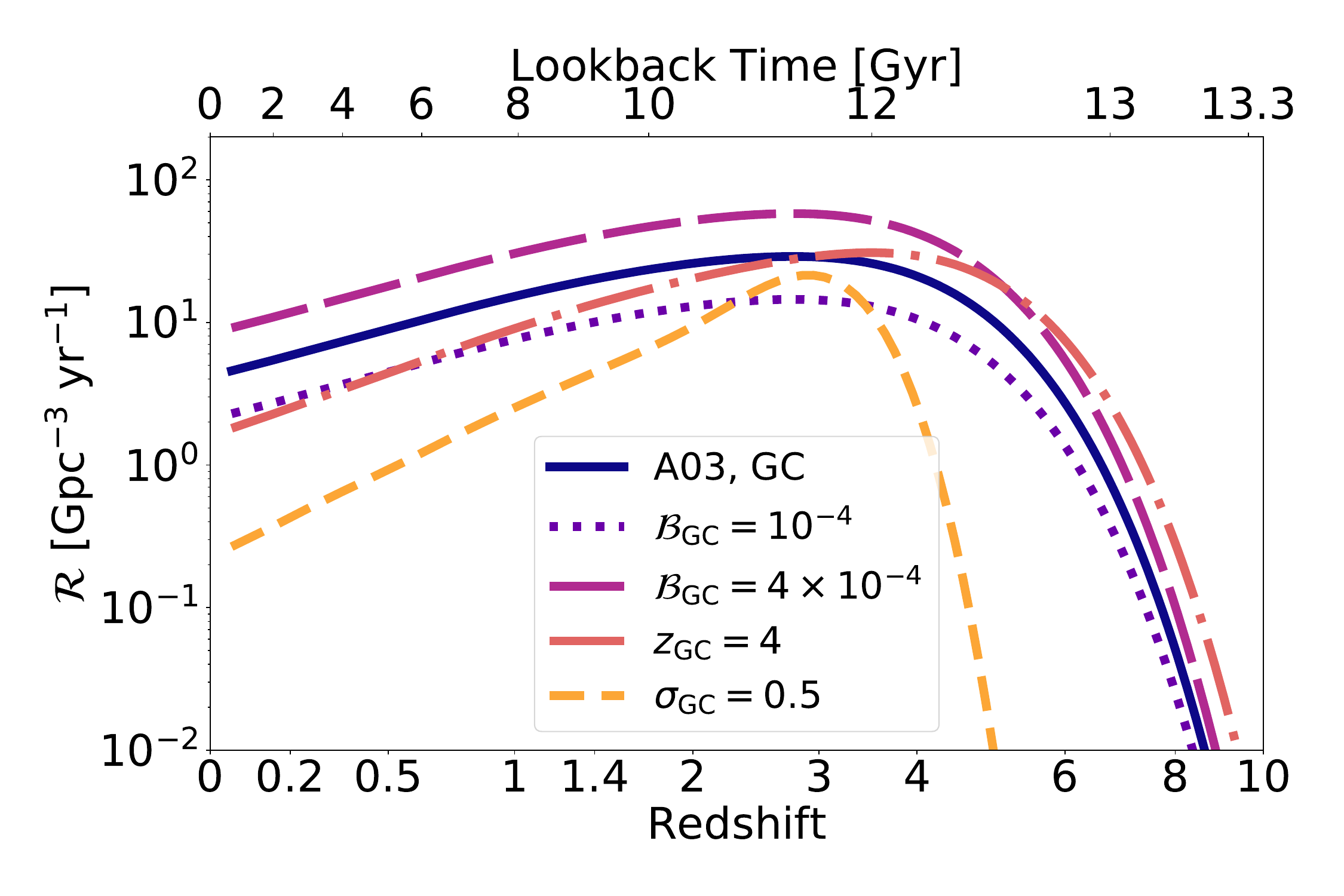}
    \end{center}
  \caption{Merger rate density of BBHs in GCs, as a function of redshift. Different lines show the uncertainties connected with the formation rate history of GCs. Blue solid line: model A03 for GCs. Violet dotted line: same  as A03 but with normalization $\mathcal{B}_{\rm GC}=10^{-4}$ in units of M$_\odot$
 Mpc$^{-3}$ yr$^{-1}$. Magenta long-dashed  line:  same as A03 but with normalization $\mathcal{B}_{\rm GC}=4\times{}10^{-4}$ M$_\odot$
 Mpc$^{-3}$ yr$^{-1}$. Pink dash-dotted  line: same as A03 but with peak redshift $z_{\rm GC}=4$. Orange short-dashed  line: same as A03 but with standard deviation $\sigma_{\rm GC}=0.5$.  \label{fig:GCrate}}
\end{figure}

The formation rate density of star clusters is extremely uncertain. Here, we discuss what happens if we consider different assumptions within the observational uncertainties. For GCs, we %consider 
start from model A03 %our fiducial model 
and change the normalization $\mathcal{B}_{\rm GC}$, the position of the peak $z_{\rm GC}$ and the spread of the distribution $\sigma_{\rm GC}$. A change of the normalization of the GC formation rate causes the same change of the value of the BBH merger rate density: if we increase (reduce) the normalization by a factor of two from $\mathcal{B}_{\rm GC}=2\times{}10^{-4}$  to $4\times{}10^{-4}$ M$_\odot$ Mpc$^{-3}$ yr$^{-2}$ ($10^{-4}$~M$_\odot$ Mpc$^{-3}$ yr$^{-2}$), we obtain a factor of two higher (lower) merger rate density at each redshift, as shown in Fig.~\ref{fig:GCrate}. 

If we change the peak redshift  from $z_{\rm GC}=3.2$, as inferred from Galactic GCs, to $z_{\rm GC}=4$, as suggested by the models of \cite{el-badry2019}, the BBH merger rate density also shifts: the maximum value of $\mathcal{R}(z)$ is at redshift $z=3.55$ ($z=2.75$) when   $z_{\rm GC}=4$  ($z_{\rm GC}=3.2$). This shift of the peak has a strong impact on the local merger rate density, which decreases from $\mathcal{R}(0)\approx{5}$ Gpc$^{-3}$ yr$^{-1}$ to  $\mathcal{R}(0)\approx{2}$ Gpc$^{-3}$ yr$^{-1}$ if we change $z_{\rm GC}$ from 3.2 to 4. Finally, 
%if we change 
the standard deviation $\sigma_{\rm GC}$ %we have 
has an even larger impact on the local merger rate: $\mathcal{R}(0)$ drops from $\approx{5}$ Gpc$^{-3}$ yr$^{-1}$ to  $\approx{0.3}$ Gpc$^{-3}$ yr$^{-1}$ if we change $\sigma_{\rm GC}$ from 1.5 to 0.5. However, $\sigma_{\rm GC}=0.5$ is an extreme value when compared with other models (e.g., \citealt{el-badry2019,reina-campos2019}).

\begin{figure}
  \begin{center}
    \includegraphics[width=0.45 \textwidth]{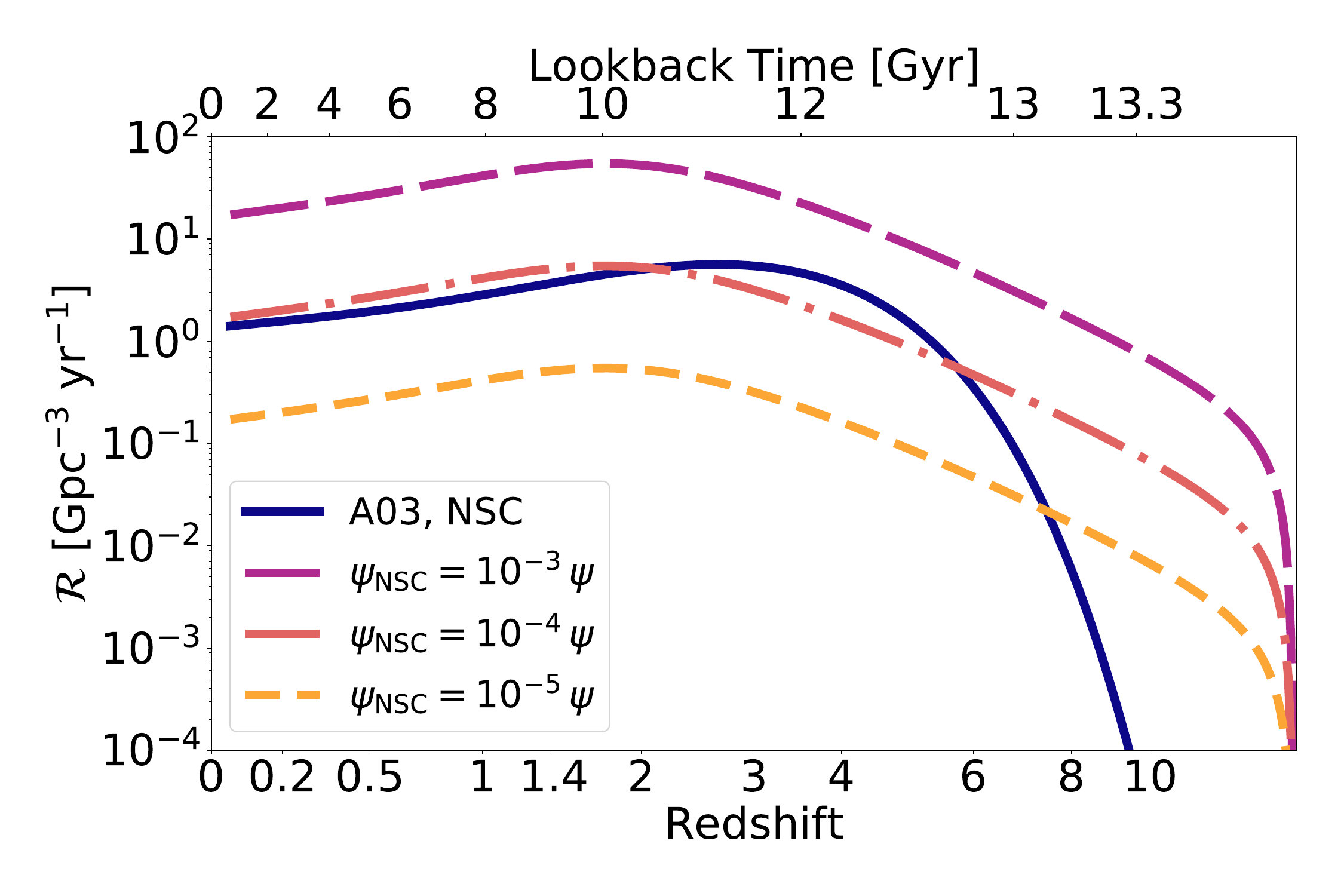}
    \end{center}
  \caption{  Merger rate density of BBHs in NSCs, as a function of redshift. Different lines show the uncertainties connected with the formation rate history of NSCs. Blue solid  line: model A03 for  NSCs. Magenta long-dashed  line: same  as A03 but with NSC formation rate density $\psi_{\rm NSC}(z)=10^{-3}\,{}\psi{}(z)$ ($\psi{}(z)$ is defined in eq.~\ref{eq:madau}). Pink dash-dotted line: same as A03 but with $\psi_{\rm NSC}(z)=10^{-4}\,{}\psi{}(z)$. Orange short-dashed line: same as A03 but with standard deviation but with $\psi_{\rm NSC}(z)=10^{-5}\,{}\psi{}(z)$.   
    \label{fig:NSCrate}}
\end{figure}

The formation history of NSCs is even more uncertain. We assumed that $\psi{}_{\rm NSC}(z)$ is a Gaussian function for analogy with GCs, but the shape of NSC formation history is essentially unconstrained \citep{neumayer2020}. In Figure~\ref{fig:NSCrate}, we assume that $\psi{}_{\rm NSC}(z)$  scales with the global star formation rate density  $\psi{}(z)$ \citep{madau2017} as %:
%\begin{equation}
$\psi{}_{\rm NSC}(z)=\mathcal{B}_{\rm NSC}\,{}\,{}\psi(z)$, 
%\end{equation}
where $\mathcal{B}_{\rm NSC}=10^{-5},$ 10$^{-4}$ and 10$^{-3}$ in the three cases shown in Fig.~\ref{fig:NSCrate}. The case with $\mathcal{B}_{\rm NSC}=10^{-4}$ has a similar behaviour to our fiducial model A03 at redshift $z<2$. The   models with $\mathcal{B}_{\rm NSC}=10^{-3}$  and $\mathcal{B}_{\rm NSC}=10^{-5}$ give a local merger rate density a factor of 10 higher and a factor of 10 lower than model A03, respectively. The case with $\mathcal{B}_{\rm NSC}=10^{-3}$ is a strong upper limit, because it gives a local density of NSCs $n_{\rm NSC}\sim{0.6}$ Mpc$^{-3}$, i.e. larger than the number of galaxies which can host such NSCs.

%%%%%%%%%%%%%%%%%%%%%%%%%%%%FIGURE%%%%%%%%%%%%%%%%%%%%%%%%%%%%%%%%%%%%%%%%%%%%%%%%%%%
%%%%%%%%%%%%%%%%%%%%%%%%%%%%%%%%%%%%%%%%%%%%%%%%%%%%%%%%%%%%%%%%%%%%%%%%%%%%%%%%%%%%%%

%%%%%%%%%%%%%%%%%%%%%%%%%%%%FIGURE%%%%%%%%%%%%%%%%%%%%%%%%%%%%%%%%%%%%%%%%%%%%%%%%%%%
%%%%%%%%%%%%%%%%%%%%%%%%%%%%%%%%%%%%%%%%%%%%%%%%%%%%%%%%%%%%%%%%%%%%%%%%%%%%%%%%%%%%%%

%%%%%%%%%%%%%%%%%%%%%%%%%%%%FIGURE%%%%%%%%%%%%%%%%%%%%%%%%%%%%%%%%%%%%%%%%%%%%%%%%%%%
\begin{figure}
  \begin{center}
    \includegraphics[width=0.45 \textwidth]{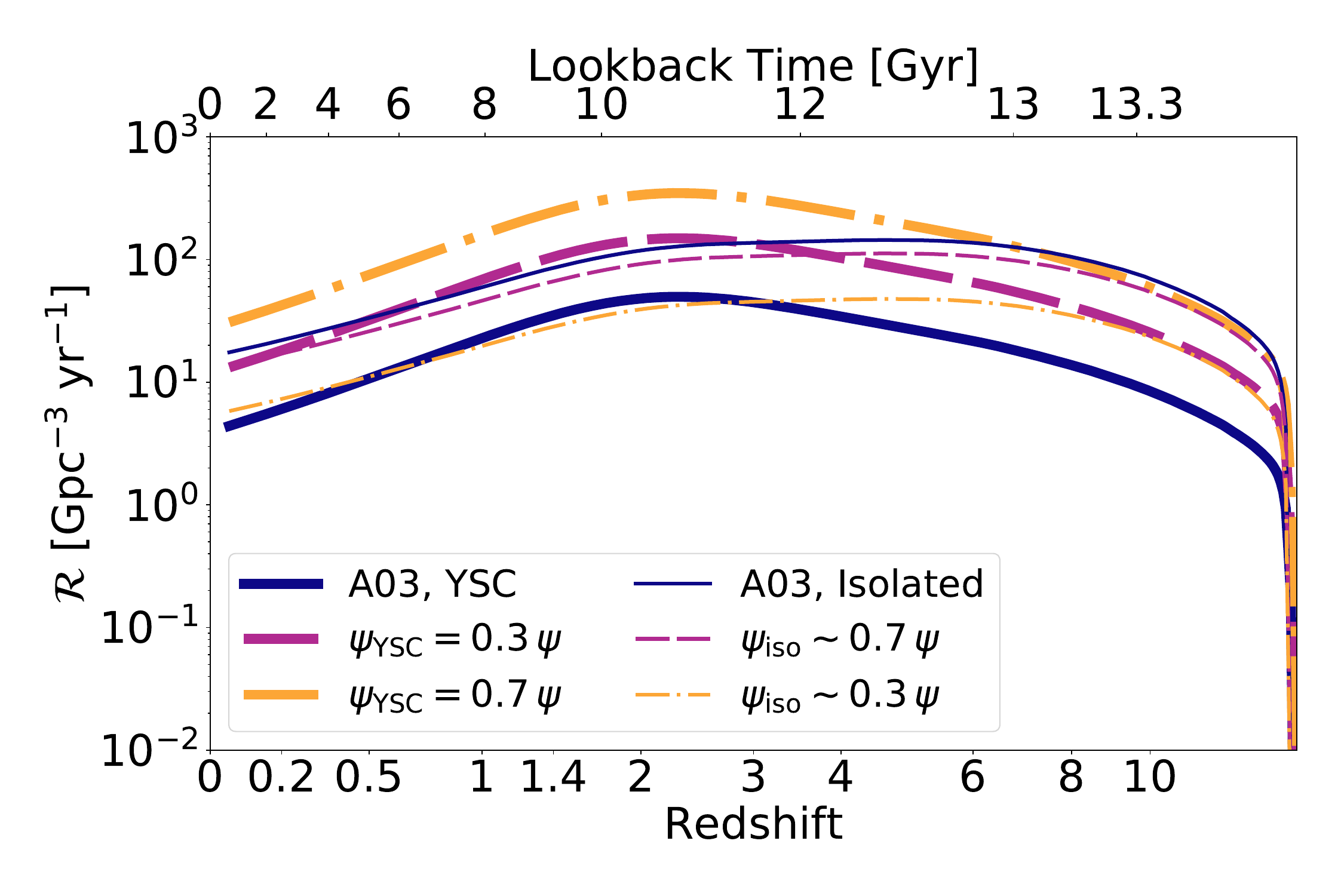}
    \end{center}
  \caption{Merger rate density of BBHs in YSCs (thick lines) and in the field (thin lines), as a function of redshift. Different lines show the uncertainties connected with the formation rate history of YSCs.  Blue thick (thin) solid line: model A03 for YSCs (isolated BBHs). Magenta long-dashed  line: same  as A03 but with YSC formation rate density $\psi_{\rm YSC}(z)=0.3\,{}\psi{}(z)$. Orange dash-dotted  line: same as A03 but with $\psi_{\rm YSC}(z)=0.7\,{}\psi{}(z)$. 
    \label{fig:YSCrate}}
\end{figure}
%%%%%%%%%%%%%%%%%%%%%%%%%%%%%%%%%%%%%%%%%%%%%%%%%%%%%%%%%%%%%%%%%%%%%%%%%%%%%%%%%%%%%%

In the case of YSCs, the main uncertainty concerns which fraction of the cosmic star formation rate happens in YSCs versus the field. In our fiducial model, we adopted a conservative assumption that only $\sim{10}$\% of the cosmic star formation rate happens in YSCs, as suggested by recent studies (e.g., \citealt{kruijssen2014,ward2020}). In Figure~\ref{fig:YSCrate}, we consider two more optimistic assumptions in which $\sim{30}$\% and $\sim{70}$\% of the cosmic star formation rate happen in YSCs \citep{lada2003,portegieszwart2010}. As expected, the merger rate density of BBHs in YSCs scales accordingly, while the merger rate density of isolated BBHs decreases by the corresponding amount.

\begin{figure}
  \begin{center}
    \includegraphics[width=0.45 \textwidth]{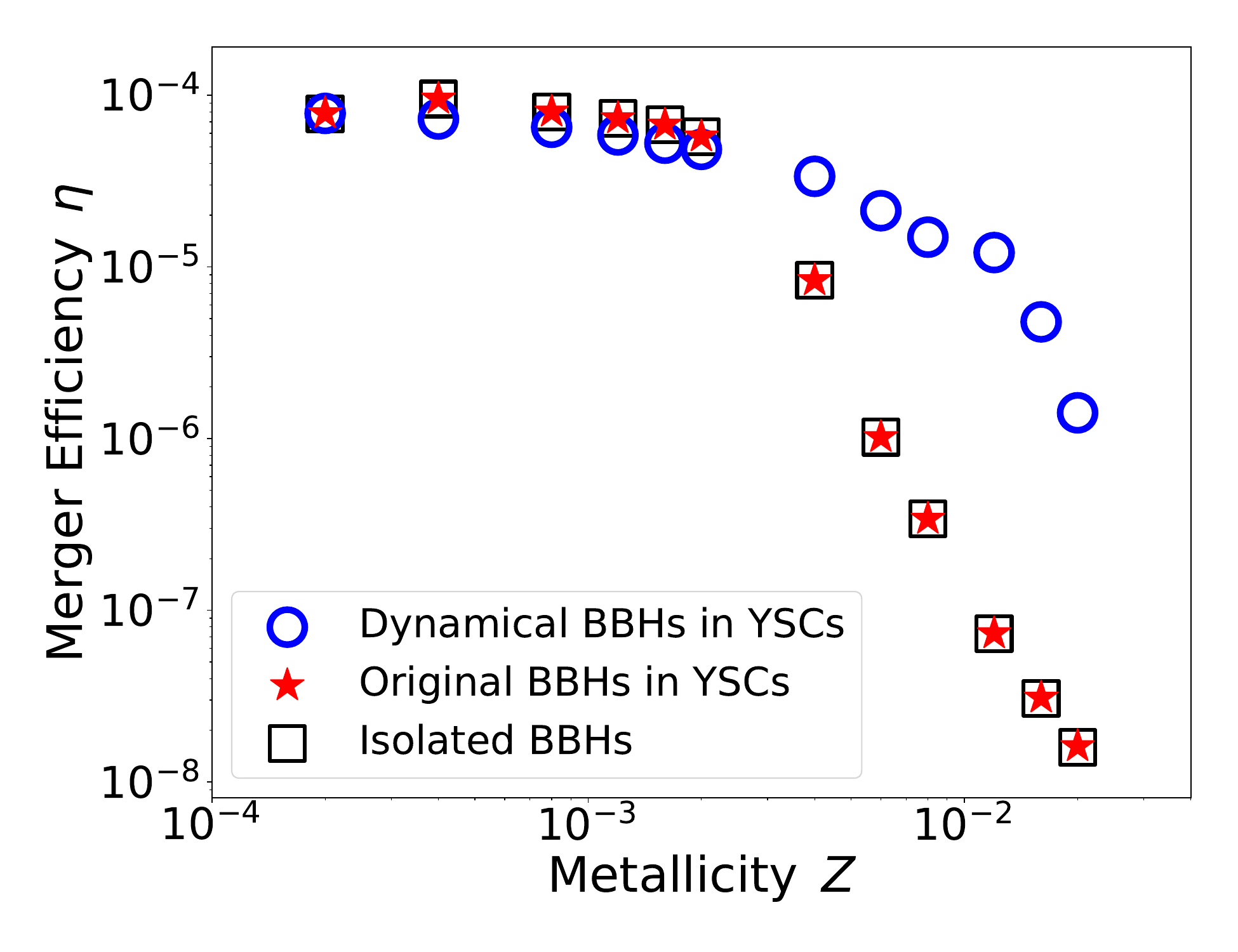}
    \end{center}
  \caption{Merger efficiency $\eta$ of BBHs as a function of metallicity $Z$, in model A03. Blue open circles: dynamical BBHs in YSCs. Red stars: original BBHs in YSCs. Black open squares: isolated BBHs.
    \label{fig:YSCeta}}
\end{figure}

Another source of uncertainty is the fraction of original versus dynamical BBHs. While the fraction of original BBHs is deemed to be very low in both GCs and NSCs  (hence their population properties are mostly driven by dynamical BBHs, \citealt{antonini2016,rodriguez2016}), the percentage of original BBHs in YSCs is more uncertain. Here, we have assumed they are 60\% of all BBH mergers, based on the results of \cite{rastello2021}. However, \cite{rastello2021} also show that the percentage of original BBHs strongly fluctuates from a cluster to another and possibly depends on both YSC mass and metallicity. Figure~\ref{fig:YSCeta} shows the merger efficiency in the field and in YSCs (defined in eqs. \ref{eq:eta_field} and \ref{eq:eta_dyn}). The merger efficiency of original BBHs in YSCs is very similar to the one of isolated BBHs, while the merger efficiency of dynamical BBHs in YSCs has a %milder 
much less steep dependence on metallicity. Hence, if we assume a higher percentage of dynamical BBHs in YSCs, we end up with a higher local merger rate density in YSCs and with a milder dependence of the YSC merger rate on metallicity spread.

One of the main approximations of our approach is that we do not model stellar and binary evolution together with dynamics. This approximation is well motivated for GCs and NSCs, which have two-body relaxation timescales of several Gyrs \citep{binney1987}. %, but is less good for YSCs, which have two-body relaxation timescales of several ten Myrs.
 In contrast, YSCs have two-body relaxation timescales of several ten Myrs. \cite{dicarlo2020b} showed that most dynamical exchanges leading to the formation of merging BBHs involve their stellar progenitors, before they collapse to BHs. Moreover, hierarchical BBH mergers are rare in YSCs, but runaway collisions seem to be more efficient in producing massive BHs in these environments \citep{mapelli2016,rizzuto2020,dicarlo2021}. Thus, our results likely underestimate the presence of massive BBHs ($m_1+m_2>100$ M$_\odot$) in YSCs. We will include a %combined 
treatment of stellar/binary evolution %and dynamics 
in {\sc fastcluster} in future work.
%The four different channels we consider in our analysis
%\section{Discussion}\label{sec:discussion}

The properties of our star clusters do not evolve with time. On the one hand, star clusters lose mass by stellar evolution and dynamical ejection and expand by two-body relaxation. This leads to lower star cluster mass and density, possibly quenching the formation of hierarchical mergers \citep[e.g.,][]{antonini2020a, antonini2020b}. On the other hand, by assuming no evolution with time, we do not account for core collapse episodes and gravothermal oscillations, which lead to a dramatic temporary increase of the central density, possibly boosting BBH formation and hierarchical mergers \citep[e.g.,][]{breen2013}. NSCs might even acquire mass during their life by fresh star formation \citep[e.g.,][]{mapelli2012,toyouchi2021,generozov2021} and by accreting GCs \citep[e.g.,][]{capuzzo2008,antonini2012}. These processes might lead to a higher efficiency of hierarchical mergers in NSCs. The overall effect of including star cluster evolution in our model is thus quite difficult to predict and might be significantly different for YSCs, GCs and NSCs. We will add a formalism for star cluster evolution in a follow-up study. Furthermore, we neglect the impact of additional formation channels, such as BBHs in AGN discs and field triples. %However, 
The approach of {\sc fastcluster} is very flexible, and we can add more channels in the future.

Comparing to previous studies, which use more  sophisticated and computationally expensive simulations, we find similar results. For example, our local merger rate density  in GCs is consistent with the one found by \cite{rodriguezloeb2018}, even if our values $\left[\mathcal{R}(0)\approx{4-8}\,{}{\rm Gpc}^{-3}\,{}{\rm yr}^{-1}\right]$ are rather on the lower side of their range $\left[\mathcal{R}(0)\approx{4-18}\,{} {\rm Gpc}^{-3}\,{} {\rm yr}^{-1}\right]$. The difference is easily  explained by the fact that we do not model GW captures, which require direct N-body integration with post-Newtonian terms \citep{samsing2018,zevin2019,kremer2020b}. Moreover, we use a different mass function and spin distribution. To confirm the good performance of {\sc fastcluster}, we also find that the maximum merger rate density (at $z\sim{2.8}$) is about six times higher than the local merger rate density, in perfect agreement with \cite{rodriguezloeb2018}.  Finally, our %percentages of Ng to 1g BBH mergers
percentages of Ng with respect to 1g BBH mergers in GCs are comparable to the ones derived by several authors with different approaches \citep{rodriguez2019,zevin2019,kimball2020a,kimball2020,doctor2020}. For more details on this comparison, see the Discussion in \cite{mapelli2021}.

The main result of our mixing fraction analysis is that at least two formation channels need to be at work to produce the population of GWTC-2. This result is in agreement with previous work \citep{abbottO3popandrate,zevin2021,bouffanais2021,wong2021}. Taking advantage of {\sc fastcluster} flexibility and speed, we probed a larger parameter space than previously done (including different metallicity spreads, different core-collapse SN models and a large number of stellar metallicities). This analysis shows that the mixing fraction of each channel varies wildly from one model to another, being extremely sensitive to the metallicity spread $\sigma_{\rm Z}$, spin parameter $\sigma_{\chi}$, common envelope parameter and  core-collapse SN model. Furthermore, despite the large number of models we ran and uncertainties we considered, the presented model selection analysis is not including all model uncertainties nor all proposed formation channels. Hence, we must be very cautious when drawing conclusions from a mixing-fraction analysis: the relevant parameter space and the uncertainties of current models are still utterly large.

%* we neglect AGN disks: will be included in future.

%* we neglect triples: will be included in future (?)

%* we do not have stellar and binary evolution (compare with Di Carlo - Santoliquido)

%* clusters do not evolve in density

%* compare to Rodriguez \& Loeb consistent especially peak redshift and factor of 6 between redshift 0 and peak redshift but our rates are on the lower side of theirs 

%%%%%%%%%%%%%%%%%%%%%%%%%%%%FIGURE%%%%%%%%%%%%%%%%%%%%%%%%%%%%%%%%%%%%%%%%%%%%%%%%%%%
%%%%%%%%%%%%%%%%%%%%%%%%%%%%%%%%%%%%%%%%%%%%%%%%%%%%%%%%%%%%%%%%%%%%%%%%%%%%%%%%%%%%%%

\section{Conclusions}\label{sec:summary}

We interfaced our semi-analytic codes {\sc fastcluster} \citep{mapelli2021} and {\sc cosmo$\mathcal{R}$ate} \citep{santoliquido2021}. {\sc fastcluster} dynamically pairs up binary black holes (BBHs) in dense star clusters,
%by taking into account dynamical friction, three-body captures and dynamical exchanges, 
and integrates their orbital evolution via three-body hardening and gravitational-wave (GW) decay. With  {\sc fastcluster} we can study the dynamical formation of BBHs in very different star clusters, from the least massive young star clusters (YSCs) to the most massive globular clusters (GCs) and nuclear star clusters (NSCs). Furthermore, {\sc fastcluster} includes a treatment for hierarchical mergers. 
% blah blah, while 
 {\sc cosmo$\mathcal{R}$ate} calculates the BBH merger rate evolution,  by using catalogs of BBH mergers simulated with {\sc fastcluster} and by  coupling them with the cosmic star formation rate and metallicity evolution. Here, we included the mass formation rate of NSCs, GCs and YSCs in {\sc cosmo$\mathcal{R}$ate}. 

We use {\sc fastcluster} + {\sc cosmo$\mathcal{R}$ate} to study four BBH formation channels: isolated BBHs and dynamical BBHs in NSCs, GCs and YSCs. This technique allows us to model different BBH formation channels with the same code, starting from the same %stellar/binary evolution and
 BH mass function. Our approach prevents any systematic bias which arises from comparing outputs of different codes, that assume different stellar evolution models and BH mass function. We consider a large range of progenitor's metallicities (twelve values of $Z\in[0.0002,\,{}0.02]$), three values of the metallicity spread ($\sigma_{\rm Z}=0.2,$ 0.3 and 0.4), two models of core-collapse SN (delayed and rapid), two values of the common envelope parameter ($\alpha=1$, 5) and two models for the dimensionless spin $\chi$ (two truncated Maxwellian distributions with $\sigma_{\chi}=0.01$ and 0.1).
 %connected with the comparison of outputs obtained with different codes, assuming different stellar evolution and BH mass function. 

 We find a local BBH merger rate density $\mathcal{R}(0)\sim{4-8}$ Gpc$^{-3}$ yr$^{-1}$ in GCs. The BBH merger rate density in GCs increases up to redshift $z\sim{2.5-2.8}$, reaching values $\sim{6}$ times higher than the local merger rate density. The local merger rate density of BBHs in NSCs spans $\mathcal{R}(0)\sim{1-2}$ Gpc$^{-3}$ yr$^{-1}$. The rate associated with NSCs also peaks at $z\sim{2.5-2.8}$, reaching values $\sim{4-5}$ times higher than at $z=0$ (Fig.~\ref{fig:rate}).
 
The merger rate density of BBHs in both GCs and NSCs is very mildly affected by stellar metallicity, while the merger rate of isolated BBHs changes wildly with the metallicity spread $\sigma{}_{\rm Z}$. BBHs in YSCs behave in an intermediate way between isolated BBHs and dynamical BBHs in GCs/NSCs. Enforcing or not the lower BH mass gap affects the merger rate density of all channels, from isolated BBHs to dynamical BBHs:  the rapid core-collapse SN model (which prevents the formation of BHs with mass $<5$ M$_\odot$)  produces a higher merger rate by $\sim{40-60}\%$ with respect to the delayed model (where we can have BHs with mass $3-5$ M$_\odot$). This happens because a higher minimum BH mass results in shorter delay times.

%common envelope parameter does not affect BBH rates in star clusters.
Our star cluster models grow a population of Nth generation (Ng) mergers. The local merger rate density of Ng BBHs is  $\sim{0.8-2.5}$,  $\sim{0.5-0.8}$, and  $\sim{0.1-0.9}$ Gpc$^{-3}$ yr$^{-1}$ in GCs, NSCs and YSCs, respectively (Fig.~\ref{fig:rate_ng}).  The total merger rate density of Ng BBHs in the local Universe, obtained by summing up these three channels, ranges from $\sim{1}$ to $\sim{4}$ Gpc$^{-3}$ yr$^{-1}$ and is mostly sensitive to the spin parameter: we find higher (lower) values  of the merger rate  for our low-spin model with $\sigma_\chi=0.01$ (fiducial model with $\sigma_\chi=0.1$).

The primary BH mass function has a high-mass tail, extending up to several hundred M$_\odot$ in the three dynamical channels, because of hierarchical mergers. The primary BH mass function evolves with redshift in both GCs and NSCs: lower mass BH mergers become less and less common at high redshift ($z\geq{}1$), because they are associated with longer delay times (Fig.~\ref{fig:masses}).  In contrast, the primary BH mass function does not significantly evolve with redshift in isolated  BBHs, in agreement with  previous studies \citep{mapelli2019,santoliquido2020}. This happens because  binary evolution processes (e.g., common envelope) generate tight systems of low-mass BHs with short delay time \citep[see, e.g.,][]{mapelli2019}. This difference has exciting implications for third-generation ground-based GW detectors: %if Einstein Telescope and Cosmic Explorer  find
%finding that the BH mass function shifts to larger masses at higher redshift, will
if Einstein Telescope and Cosmic Explorer  will find a heavier BH mass function at higher redshift, this will be a signature that most BBH mergers have a dynamical origin; %with respect to isolated formation 
vice versa, isolated BBHs dominate the observed population if the mass function does not evolve with redshift. %Important scientific case for 3G detectors % possibile progetto x filippo???

The resulting primary BH mass function we obtain by combining our four channels according to their merger rate is similar to the {\sc power law + peak} model used by the LIGO--Virgo--KAGRA collaboration \citep{abbottO3popandrate}. The main difference is that our models predict more BHs with mass $\sim{20}$ M$_\odot$ with respect to the {\sc power law + peak} model (Fig.~\ref{fig:mix_mass}). In our mass function, low-mass BHs are 
 mostly  given by isolated BBHs, YSCS and NSCs, while the high-mass tail ($\geq{}50$ M$_\odot$) is mostly due to Ng BHs in GCs and NSCs.

The distribution of effective ($\chi_{\rm eff}$) and precessing ($\chi_{\rm p}$) spins we obtain by combining our four channels %according to their merger rates 
strongly depend on $\sigma_\chi$. For $\sigma_\chi=0.01$ (low-spin models), 1g BBHs have vanishingly small values of $\chi_{\rm eff}$. Hence, the effective spin distribution has a sharp peak at  zero,  surrounded by two symmetric broad wings due to Ng BBHs. The distribution of $\chi_{\rm p}$ has two peaks: a primary peak, very narrow, at $\chi_{\rm p}=0$ and a secondary peak at $\chi_{\rm p}\approx{0.7}$, because of Ng BBHs. In contrast, for $\sigma_\chi=0.1$, the distribution of effective spins becomes asymmetric: it peaks at $\chi_{\rm eff}\approx{0.2}$, because of isolated BBHs. In this case, the distribution of precessing spins has three peaks: a sharp primary peak at $\chi_{\rm p}=0$ because of isolated BBHs, a broader secondary peak at  $\chi_{\rm p}=0.1-0.2$, because of 1g dynamical BBHs, and a third, lower peak at $\chi_{\rm p}\sim{0.7}$ because of Ng BBHs (Fig.~\ref{fig:mix_spin}).

We calculated the posterior probability distribution of the mixing fractions associated with our four channels, by running a Bayesian hierarchical analysis  with posterior samples from GWTC-2 \citep{abbottO3a}. The resulting mixing fractions indicate that at least two formation channels %need to be 
are likely at work to produce the observed BBH population (Fig.~\ref{fig:mixing_frac}).  However, the mixing fraction of each channel varies wildly from a model to another, being extremely sensitive to the metallicity spread $\sigma_{\rm Z}$, the spin parameter $\sigma_{\chi}$, the common envelope parameter $\alpha$ and the core-collapse SN model. Furthermore, our analysis still does does not include all the proposed formation channels and does not consider all possible sources of uncertainty. 
  %For example, for $\sigma_{\rm Z}=0.3$ isolated BBHs, GC BBHs and YSC BBHs are associated with the highest mixing fractions. 
Hence, our  models still suffer from large uncertainties (e.g. on the formation history of star clusters), but {\sc fastcluster} and {\sc cosmo$\mathcal{R}$ate} are extremely flexible and fast tool, and we can use them to probe the relevant  parameter space.

\section*{Acknowledgements}

MM, YB and FS acknowledge financial support from the European Research Council for the ERC Consolidator grant DEMOBLACK, under contract no. 770017. MCA and MM acknowledge financial support from the Austrian National Science Foundation through FWF stand-alone grant P31154-N27. MAS acknowledges financial support from the Alexander von Humboldt Foundation for the research program ``The evolution of black holes from stellar to galactic scales'', the Volkswagen Foundation Trilateral Partnership through project No. I/97778, and the Deutsche Forschungsgemeinschaft (DFG) -- Project-ID 138713538 -- SFB 881. We thank the anonymous referee for their comments, which helped us improve this work.

%%%%%%%%%% DATA %%%%%%%%%%%%%%%%
\section*{Data availability}

The data underlying this article will be shared on reasonable request to the corresponding authors. The latest public version of {\sc fastcluster} can be downloaded from \href{https://gitlab.com/micmap/fastcluster_open}{this repository}.

\appendix

\bibliographystyle{mnras}
\bibliography{bibliography}

\begin{thebibliography}{}
\makeatletter
\relax
\def\mn@urlcharsother{\let\do\@makeother \do\$\do\&\do\#\do\^\do\_\do\%\do\~}
\def\mn@doi{\begingroup\mn@urlcharsother \@ifnextchar [ {\mn@doi@}
  {\mn@doi@[]}}
\def\mn@doi@[#1]#2{\def\@tempa{#1}\ifx\@tempa\@empty \href
  {http://dx.doi.org/#2} {doi:#2}\else \href {http://dx.doi.org/#2} {#1}\fi
  \endgroup}
\def\mn@eprint#1#2{\mn@eprint@#1:#2::\@nil}
\def\mn@eprint@arXiv#1{\href {http://arxiv.org/abs/#1} {{\tt arXiv:#1}}}
\def\mn@eprint@dblp#1{\href {http://dblp.uni-trier.de/rec/bibtex/#1.xml}
  {dblp:#1}}
\def\mn@eprint@#1:#2:#3:#4\@nil{\def\@tempa {#1}\def\@tempb {#2}\def\@tempc
  {#3}\ifx \@tempc \@empty \let \@tempc \@tempb \let \@tempb \@tempa \fi \ifx
  \@tempb \@empty \def\@tempb {arXiv}\fi \@ifundefined
  {mn@eprint@\@tempb}{\@tempb:\@tempc}{\expandafter \expandafter \csname
  mn@eprint@\@tempb\endcsname \expandafter{\@tempc}}}

\bibitem[\protect\citeauthoryear{Aasi et~al.,}{Aasi
  et~al.}{2015}]{LIGOdetector}
Aasi J.,  et~al., 2015, Classical and Quantum Gravity, 32, 074001

\bibitem[\protect\citeauthoryear{{Abbott} et~al.,}{{Abbott}
  et~al.}{2020a}]{abbottGW190521}
{Abbott} R.,  et~al., 2020a, \mn@doi [\prl] {10.1103/PhysRevLett.125.101102},
  \href {https://ui.adsabs.harvard.edu/abs/2020PhRvL.125j1102A} {125, 101102}

\bibitem[\protect\citeauthoryear{{Abbott} et~al.,}{{Abbott}
  et~al.}{2020b}]{abbottGW190521astro}
{Abbott} R.,  et~al., 2020b, \mn@doi [\apjl] {10.3847/2041-8213/aba493}, \href
  {https://ui.adsabs.harvard.edu/abs/2020ApJ...900L..13A} {900, L13}

\bibitem[\protect\citeauthoryear{{Abbott} et~al.,}{{Abbott}
  et~al.}{2021a}]{abbottGWTC-2.1}
{Abbott} R.,  et~al., 2021a, arXiv e-prints, \href
  {https://ui.adsabs.harvard.edu/abs/2021arXiv210801045T} {p. arXiv:2108.01045}

\bibitem[\protect\citeauthoryear{{Abbott} et~al.,}{{Abbott}
  et~al.}{2021b}]{abbottO3a}
{Abbott} R.,  et~al., 2021b, \mn@doi [Physical Review X]
  {10.1103/PhysRevX.11.021053}, \href
  {https://ui.adsabs.harvard.edu/abs/2021PhRvX..11b1053A} {11, 021053}

\bibitem[\protect\citeauthoryear{{Abbott} et~al.,}{{Abbott}
  et~al.}{2021c}]{abbottO3popandrate}
{Abbott} R.,  et~al., 2021c, \mn@doi [\apjl] {10.3847/2041-8213/abe949}, \href
  {https://ui.adsabs.harvard.edu/abs/2021ApJ...913L...7A} {913, L7}

\bibitem[\protect\citeauthoryear{{Acernese} et~al.,}{{Acernese}
  et~al.}{2015}]{VIRGOdetector}
{Acernese} F.,  et~al., 2015, \mn@doi [Classical and Quantum Gravity]
  {10.1088/0264-9381/32/2/024001}, \href
  {http://adsabs.harvard.edu/abs/2015CQGra..32b4001A} {32, 024001}

\bibitem[\protect\citeauthoryear{{Ade}, {Aghanim}  \& {Zonca}}{{Ade}
  et~al.}{2016}]{planck2016}
{Ade} P.~A.~R.,  {Aghanim} N.,   {Zonca} A. e.~a.,  2016, \mn@doi [\aap]
  {10.1051/0004-6361/201525830}, \href
  {https://ui.adsabs.harvard.edu/abs/2016A&A...594A..13P} {594, A13}

\bibitem[\protect\citeauthoryear{{Ali-Ha{\"\i}moud}, {Kovetz}  \&
  {Kamionkowski}}{{Ali-Ha{\"\i}moud} et~al.}{2017}]{alihaimoud2017}
{Ali-Ha{\"\i}moud} Y.,  {Kovetz} E.~D.,   {Kamionkowski} M.,  2017, \mn@doi
  [\prd] {10.1103/PhysRevD.96.123523}, \href
  {https://ui.adsabs.harvard.edu/abs/2017PhRvD..96l3523A} {96, 123523}

\bibitem[\protect\citeauthoryear{{Anagnostou}, {Trenti}  \&
  {Melatos}}{{Anagnostou} et~al.}{2020}]{anagnostou2020}
{Anagnostou} O.,  {Trenti} M.,   {Melatos} A.,  2020, arXiv e-prints, \href
  {https://ui.adsabs.harvard.edu/abs/2020arXiv201006161A} {p. arXiv:2010.06161}

\bibitem[\protect\citeauthoryear{{Antonini} \& {Gieles}}{{Antonini} \&
  {Gieles}}{2020a}]{antonini2020b}
{Antonini} F.,  {Gieles} M.,  2020a, \mn@doi [\prd]
  {10.1103/PhysRevD.102.123016}, \href
  {https://ui.adsabs.harvard.edu/abs/2020PhRvD.102l3016A} {102, 123016}

\bibitem[\protect\citeauthoryear{{Antonini} \& {Gieles}}{{Antonini} \&
  {Gieles}}{2020b}]{antonini2020a}
{Antonini} F.,  {Gieles} M.,  2020b, \mn@doi [\mnras] {10.1093/mnras/stz3584},
  \href {https://ui.adsabs.harvard.edu/abs/2020MNRAS.492.2936A} {492, 2936}

\bibitem[\protect\citeauthoryear{{Antonini} \& {Rasio}}{{Antonini} \&
  {Rasio}}{2016}]{antonini2016}
{Antonini} F.,  {Rasio} F.~A.,  2016, \mn@doi [\apj]
  {10.3847/0004-637X/831/2/187}, \href
  {http://adsabs.harvard.edu/abs/2016ApJ...831..187A} {831, 187}

\bibitem[\protect\citeauthoryear{{Antonini}, {Capuzzo-Dolcetta},
  {Mastrobuono-Battisti}  \& {Merritt}}{{Antonini} et~al.}{2012}]{antonini2012}
{Antonini} F.,  {Capuzzo-Dolcetta} R.,  {Mastrobuono-Battisti} A.,   {Merritt}
  D.,  2012, \mn@doi [\apj] {10.1088/0004-637X/750/2/111}, \href
  {https://ui.adsabs.harvard.edu/abs/2012ApJ...750..111A} {750, 111}

\bibitem[\protect\citeauthoryear{{Antonini}, {Toonen}  \& {Hamers}}{{Antonini}
  et~al.}{2017}]{antonini2017}
{Antonini} F.,  {Toonen} S.,   {Hamers} A.~S.,  2017, \mn@doi [\apj]
  {10.3847/1538-4357/aa6f5e}, \href
  {http://adsabs.harvard.edu/abs/2017ApJ...841...77A} {841, 77}

\bibitem[\protect\citeauthoryear{{Antonini}, {Gieles}  \&
  {Gualandris}}{{Antonini} et~al.}{2019}]{antonini2019}
{Antonini} F.,  {Gieles} M.,   {Gualandris} A.,  2019, \mn@doi [\mnras]
  {10.1093/mnras/stz1149}, \href
  {https://ui.adsabs.harvard.edu/abs/2019MNRAS.486.5008A} {486, 5008}

\bibitem[\protect\citeauthoryear{{Arca Sedda}}{{Arca
  Sedda}}{2020}]{arcasedda2020b}
{Arca Sedda} M.,  2020, \mn@doi [\apj] {10.3847/1538-4357/ab723b}, \href
  {https://ui.adsabs.harvard.edu/abs/2020ApJ...891...47A} {891, 47}

\bibitem[\protect\citeauthoryear{{Arca Sedda}, {Mapelli}, {Spera},
  {Benacquista}  \& {Giacobbo}}{{Arca Sedda} et~al.}{2020}]{arcasedda2020}
{Arca Sedda} M.,  {Mapelli} M.,  {Spera} M.,  {Benacquista} M.,   {Giacobbo}
  N.,  2020, \mn@doi [\apj] {10.3847/1538-4357/ab88b2}, \href
  {https://ui.adsabs.harvard.edu/abs/2020ApJ...894..133A} {894, 133}

\bibitem[\protect\citeauthoryear{{Arca Sedda}, {Li}  \& {Kocsis}}{{Arca Sedda}
  et~al.}{2021a}]{arcasedda2018b}
{Arca Sedda} M.,  {Li} G.,   {Kocsis} B.,  2021a, \mn@doi [\aap]
  {10.1051/0004-6361/202038795}, \href
  {https://ui.adsabs.harvard.edu/abs/2021A&A...650A.189A} {650, A189}

\bibitem[\protect\citeauthoryear{{Arca Sedda}, {Amaro Seoane}  \& {Chen}}{{Arca
  Sedda} et~al.}{2021b}]{arcasedda2021b}
{Arca Sedda} M.,  {Amaro Seoane} P.,   {Chen} X.,  2021b, \mn@doi [\aap]
  {10.1051/0004-6361/202037785}, \href
  {https://ui.adsabs.harvard.edu/abs/2021A&A...652A..54A} {652, A54}

\bibitem[\protect\citeauthoryear{{Arca-Sedda}, {Rizzuto}, {Naab}, {Ostriker},
  {Giersz}  \& {Spurzem}}{{Arca-Sedda} et~al.}{2021c}]{arcasedda2021}
{Arca-Sedda} M.,  {Rizzuto} F.~P.,  {Naab} T.,  {Ostriker} J.,  {Giersz} M.,
  {Spurzem} R.,  2021c, \mn@doi [\apj] {10.3847/1538-4357/ac1419}, \href
  {https://ui.adsabs.harvard.edu/abs/2021ApJ...920..128A} {920, 128}

\bibitem[\protect\citeauthoryear{{Askar}, {Szkudlarek}, {Gondek-Rosi{\'n}ska},
  {Giersz}  \& {Bulik}}{{Askar} et~al.}{2017}]{askar2017}
{Askar} A.,  {Szkudlarek} M.,  {Gondek-Rosi{\'n}ska} D.,  {Giersz} M.,
  {Bulik} T.,  2017, \mn@doi [\mnras] {10.1093/mnrasl/slw177}, \href
  {http://adsabs.harvard.edu/abs/2017MNRAS.464L..36A} {464, L36}

\bibitem[\protect\citeauthoryear{{Baibhav}, {Gerosa}, {Berti}, {Wong}, {Helfer}
   \& {Mould}}{{Baibhav} et~al.}{2020}]{baibhav2020}
{Baibhav} V.,  {Gerosa} D.,  {Berti} E.,  {Wong} K. W.~K.,  {Helfer} T.,
  {Mould} M.,  2020, \mn@doi [\prd] {10.1103/PhysRevD.102.043002}, \href
  {https://ui.adsabs.harvard.edu/abs/2020PhRvD.102d3002B} {102, 043002}

\bibitem[\protect\citeauthoryear{{Banerjee}}{{Banerjee}}{2017}]{banerjee2017}
{Banerjee} S.,  2017, \mn@doi [\mnras] {10.1093/mnras/stw3392}, \href
  {http://adsabs.harvard.edu/abs/2017MNRAS.467..524B} {467, 524}

\bibitem[\protect\citeauthoryear{{Banerjee}}{{Banerjee}}{2021}]{banerjee2020}
{Banerjee} S.,  2021, \mn@doi [\mnras] {10.1093/mnras/staa2392}, \href
  {https://ui.adsabs.harvard.edu/abs/2021MNRAS.500.3002B} {500, 3002}

\bibitem[\protect\citeauthoryear{{Banerjee}, {Baumgardt}  \&
  {Kroupa}}{{Banerjee} et~al.}{2010}]{banerjee2010}
{Banerjee} S.,  {Baumgardt} H.,   {Kroupa} P.,  2010, \mn@doi [\mnras]
  {10.1111/j.1365-2966.2009.15880.x}, \href
  {http://adsabs.harvard.edu/abs/2010MNRAS.402..371B} {402, 371}

\bibitem[\protect\citeauthoryear{{Bartos}, {Kocsis}, {Haiman}  \&
  {M{\'a}rka}}{{Bartos} et~al.}{2017}]{bartos2017}
{Bartos} I.,  {Kocsis} B.,  {Haiman} Z.,   {M{\'a}rka} S.,  2017, \mn@doi
  [\apj] {10.3847/1538-4357/835/2/165}, \href
  {https://ui.adsabs.harvard.edu/abs/2017ApJ...835..165B} {835, 165}

\bibitem[\protect\citeauthoryear{{Bavera} et~al.,}{{Bavera}
  et~al.}{2020}]{bavera2020b}
{Bavera} S.~S.,  et~al., 2020, \mn@doi [\aap] {10.1051/0004-6361/201936204},
  \href {https://ui.adsabs.harvard.edu/abs/2020A&A...635A..97B} {635, A97}

\bibitem[\protect\citeauthoryear{{Bavera} et~al.,}{{Bavera}
  et~al.}{2021}]{bavera2020}
{Bavera} S.~S.,  et~al., 2021, \mn@doi [\aap] {10.1051/0004-6361/202039804},
  \href {https://ui.adsabs.harvard.edu/abs/2021A&A...647A.153B} {647, A153}

\bibitem[\protect\citeauthoryear{{Belczynski}}{{Belczynski}}{2020}]{belczynski2020b}
{Belczynski} K.,  2020, \mn@doi [\apjl] {10.3847/2041-8213/abcbf1}, \href
  {https://ui.adsabs.harvard.edu/abs/2020ApJ...905L..15B} {905, L15}

\bibitem[\protect\citeauthoryear{{Belczynski}, {Kalogera}  \&
  {Bulik}}{{Belczynski} et~al.}{2002}]{belczynski2002}
{Belczynski} K.,  {Kalogera} V.,   {Bulik} T.,  2002, \mn@doi [\apj]
  {10.1086/340304}, \href {http://adsabs.harvard.edu/abs/2002ApJ...572..407B}
  {572, 407}

\bibitem[\protect\citeauthoryear{{Belczynski}, {Kalogera}, {Rasio}, {Taam},
  {Zezas}, {Bulik}, {Maccarone}  \& {Ivanova}}{{Belczynski}
  et~al.}{2008}]{belczynski2008}
{Belczynski} K.,  {Kalogera} V.,  {Rasio} F.~A.,  {Taam} R.~E.,  {Zezas} A.,
  {Bulik} T.,  {Maccarone} T.~J.,   {Ivanova} N.,  2008, \mn@doi [\apjs]
  {10.1086/521026}, \href {http://adsabs.harvard.edu/abs/2008ApJS..174..223B}
  {174, 223}

\bibitem[\protect\citeauthoryear{{Belczynski}, {Holz}, {Bulik}  \&
  {O'Shaughnessy}}{{Belczynski} et~al.}{2016a}]{belczynski2016}
{Belczynski} K.,  {Holz} D.~E.,  {Bulik} T.,   {O'Shaughnessy} R.,  2016a,
  \mn@doi [\nat] {10.1038/nature18322}, \href
  {http://adsabs.harvard.edu/abs/2016Natur.534..512B} {534, 512}

\bibitem[\protect\citeauthoryear{{Belczynski} et~al.,}{{Belczynski}
  et~al.}{2016b}]{belczynski2016pair}
{Belczynski} K.,  et~al., 2016b, \mn@doi [\aap] {10.1051/0004-6361/201628980},
  \href {http://adsabs.harvard.edu/abs/2016A%26A...594A..97B} {594, A97}

\bibitem[\protect\citeauthoryear{{Belczynski}, {Ryu}, {Perna}, {Berti},
  {Tanaka}  \& {Bulik}}{{Belczynski} et~al.}{2017}]{belczynski2017}
{Belczynski} K.,  {Ryu} T.,  {Perna} R.,  {Berti} E.,  {Tanaka} T.~L.,
  {Bulik} T.,  2017, \mn@doi [\mnras] {10.1093/mnras/stx1759}, \href
  {https://ui.adsabs.harvard.edu/abs/2017MNRAS.471.4702B} {471, 4702}

\bibitem[\protect\citeauthoryear{{Belczynski} et~al.,}{{Belczynski}
  et~al.}{2020}]{belczynski2020}
{Belczynski} K.,  et~al., 2020, \mn@doi [\aap] {10.1051/0004-6361/201936528},
  \href {https://ui.adsabs.harvard.edu/abs/2020A&A...636A.104B} {636, A104}

\bibitem[\protect\citeauthoryear{{Bethe} \& {Brown}}{{Bethe} \&
  {Brown}}{1998}]{bethe1998}
{Bethe} H.~A.,  {Brown} G.~E.,  1998, \mn@doi [\apj] {10.1086/306265}, \href
  {http://adsabs.harvard.edu/abs/1998ApJ...506..780B} {506, 780}

\bibitem[\protect\citeauthoryear{{Binney} \& {Tremaine}}{{Binney} \&
  {Tremaine}}{1987}]{binney1987}
{Binney} J.,  {Tremaine} S.,  1987, {Galactic dynamics}.
Princeton University Press

\bibitem[\protect\citeauthoryear{{Bouffanais}, {Mapelli}, {Gerosa}, {Di Carlo},
  {Giacobbo}, {Berti}  \& {Baibhav}}{{Bouffanais}
  et~al.}{2019}]{bouffanais2019}
{Bouffanais} Y.,  {Mapelli} M.,  {Gerosa} D.,  {Di Carlo} U.~N.,  {Giacobbo}
  N.,  {Berti} E.,   {Baibhav} V.,  2019, \mn@doi [\apj]
  {10.3847/1538-4357/ab4a79}, \href
  {https://ui.adsabs.harvard.edu/abs/2019ApJ...886...25B} {886, 25}

\bibitem[\protect\citeauthoryear{{Bouffanais}, {Mapelli}, {Santoliquido},
  {Giacobbo}, {Iorio}  \& {Costa}}{{Bouffanais}
  et~al.}{2021a}]{bouffanais2021b}
{Bouffanais} Y.,  {Mapelli} M.,  {Santoliquido} F.,  {Giacobbo} N.,  {Iorio}
  G.,   {Costa} G.,  2021a, \mn@doi [\mnras] {10.1093/mnras/stab1589}, \href
  {https://ui.adsabs.harvard.edu/abs/2021MNRAS.505.3873B} {505, 3873}

\bibitem[\protect\citeauthoryear{{Bouffanais}, {Mapelli}, {Santoliquido},
  {Giacobbo}, {Di Carlo}, {Rastello}, {Artale}  \& {Iorio}}{{Bouffanais}
  et~al.}{2021b}]{bouffanais2021}
{Bouffanais} Y.,  {Mapelli} M.,  {Santoliquido} F.,  {Giacobbo} N.,  {Di Carlo}
  U.~N.,  {Rastello} S.,  {Artale} M.~C.,   {Iorio} G.,  2021b, \mn@doi
  [\mnras] {10.1093/mnras/stab2438}, \href
  {https://ui.adsabs.harvard.edu/abs/2021MNRAS.507.5224B} {507, 5224}

\bibitem[\protect\citeauthoryear{{Breen} \& {Heggie}}{{Breen} \&
  {Heggie}}{2013}]{breen2013}
{Breen} P.~G.,  {Heggie} D.~C.,  2013, \mn@doi [\mnras] {10.1093/mnras/stt628},
  \href {https://ui.adsabs.harvard.edu/abs/2013MNRAS.432.2779B} {432, 2779}

\bibitem[\protect\citeauthoryear{{Callister}, {Farr}  \& {Renzo}}{{Callister}
  et~al.}{2021}]{callister2021}
{Callister} T.~A.,  {Farr} W.~M.,   {Renzo} M.,  2021, \mn@doi [\apj]
  {10.3847/1538-4357/ac1347}, \href
  {https://ui.adsabs.harvard.edu/abs/2021ApJ...920..157C} {920, 157}

\bibitem[\protect\citeauthoryear{{Campanelli}, {Lousto}, {Zlochower}  \&
  {Merritt}}{{Campanelli} et~al.}{2007}]{campanelli2007}
{Campanelli} M.,  {Lousto} C.,  {Zlochower} Y.,   {Merritt} D.,  2007, \mn@doi
  [\apjl] {10.1086/516712}, \href
  {https://ui.adsabs.harvard.edu/abs/2007ApJ...659L...5C} {659, L5}

\bibitem[\protect\citeauthoryear{{Capuzzo-Dolcetta}}{{Capuzzo-Dolcetta}}{1993}]{capuzzo1993}
{Capuzzo-Dolcetta} R.,  1993, \mn@doi [\apj] {10.1086/173189}, \href
  {https://ui.adsabs.harvard.edu/abs/1993ApJ...415..616C} {415, 616}

\bibitem[\protect\citeauthoryear{{Capuzzo-Dolcetta} \&
  {Miocchi}}{{Capuzzo-Dolcetta} \& {Miocchi}}{2008}]{capuzzo2008}
{Capuzzo-Dolcetta} R.,  {Miocchi} P.,  2008, \mn@doi [\mnras]
  {10.1111/j.1745-3933.2008.00501.x}, \href
  {https://ui.adsabs.harvard.edu/abs/2008MNRAS.388L..69C} {388, L69}

\bibitem[\protect\citeauthoryear{{Carr} \& {Hawking}}{{Carr} \&
  {Hawking}}{1974}]{carr1974}
{Carr} B.~J.,  {Hawking} S.~W.,  1974, \mn@doi [\mnras]
  {10.1093/mnras/168.2.399}, \href
  {https://ui.adsabs.harvard.edu/abs/1974MNRAS.168..399C} {168, 399}

\bibitem[\protect\citeauthoryear{{Carr}, {K{\"u}hnel}  \& {Sandstad}}{{Carr}
  et~al.}{2016}]{carr2016}
{Carr} B.,  {K{\"u}hnel} F.,   {Sandstad} M.,  2016, \mn@doi [\prd]
  {10.1103/PhysRevD.94.083504}, \href
  {http://adsabs.harvard.edu/abs/2016PhRvD..94h3504C} {94, 083504}

\bibitem[\protect\citeauthoryear{{Chandrasekhar}}{{Chandrasekhar}}{1943}]{chandrasekhar1943}
{Chandrasekhar} S.,  1943, \mn@doi [\apj] {10.1086/144517}, \href
  {https://ui.adsabs.harvard.edu/abs/1943ApJ....97..255C} {97, 255}

\bibitem[\protect\citeauthoryear{{Chen}, {Bressan}, {Girardi}, {Marigo}, {Kong}
   \& {Lanza}}{{Chen} et~al.}{2015}]{chen2015}
{Chen} Y.,  {Bressan} A.,  {Girardi} L.,  {Marigo} P.,  {Kong} X.,   {Lanza}
  A.,  2015, \mn@doi [\mnras] {10.1093/mnras/stv1281}, \href
  {http://adsabs.harvard.edu/abs/2015MNRAS.452.1068C} {452, 1068}

\bibitem[\protect\citeauthoryear{{Choksi}, {Volonteri}, {Colpi}, {Gnedin}  \&
  {Li}}{{Choksi} et~al.}{2019}]{choksi2019}
{Choksi} N.,  {Volonteri} M.,  {Colpi} M.,  {Gnedin} O.~Y.,   {Li} H.,  2019,
  \mn@doi [\apj] {10.3847/1538-4357/aaffde}, \href
  {https://ui.adsabs.harvard.edu/abs/2019ApJ...873..100C} {873, 100}

\bibitem[\protect\citeauthoryear{{Chruslinska}, {Nelemans}  \&
  {Belczynski}}{{Chruslinska} et~al.}{2019}]{chruslinska2019}
{Chruslinska} M.,  {Nelemans} G.,   {Belczynski} K.,  2019, \mn@doi [\mnras]
  {10.1093/mnras/sty3087}, \href
  {https://ui.adsabs.harvard.edu/abs/2019MNRAS.482.5012C} {482, 5012}

\bibitem[\protect\citeauthoryear{{Clesse} \& {Garc{\'\i}a-Bellido}}{{Clesse} \&
  {Garc{\'\i}a-Bellido}}{2017}]{clesse2017}
{Clesse} S.,  {Garc{\'\i}a-Bellido} J.,  2017, \mn@doi [Physics of the Dark
  Universe] {10.1016/j.dark.2016.10.002}, \href
  {https://ui.adsabs.harvard.edu/abs/2017PDU....15..142C} {15, 142}

\bibitem[\protect\citeauthoryear{{Conselice}, {Wilkinson}, {Duncan}  \&
  {Mortlock}}{{Conselice} et~al.}{2016}]{conselice2016}
{Conselice} C.~J.,  {Wilkinson} A.,  {Duncan} K.,   {Mortlock} A.,  2016,
  \mn@doi [\apj] {10.3847/0004-637X/830/2/83}, \href
  {https://ui.adsabs.harvard.edu/abs/2016ApJ...830...83C} {830, 83}

\bibitem[\protect\citeauthoryear{{Costa}, {Bressan}, {Mapelli}, {Marigo},
  {Iorio}  \& {Spera}}{{Costa} et~al.}{2021}]{costa2021}
{Costa} G.,  {Bressan} A.,  {Mapelli} M.,  {Marigo} P.,  {Iorio} G.,   {Spera}
  M.,  2021, \mn@doi [\mnras] {10.1093/mnras/staa3916}, \href
  {https://ui.adsabs.harvard.edu/abs/2021MNRAS.501.4514C} {501, 4514}

\bibitem[\protect\citeauthoryear{{De Luca}, {Franciolini}, {Pani}  \&
  {Riotto}}{{De Luca} et~al.}{2021}]{deluca2021}
{De Luca} V.,  {Franciolini} G.,  {Pani} P.,   {Riotto} A.,  2021, \mn@doi
  [\jcap] {10.1088/1475-7516/2021/05/003}, \href
  {https://ui.adsabs.harvard.edu/abs/2021JCAP...05..003D} {2021, 003}

\bibitem[\protect\citeauthoryear{{Di Carlo}, {Giacobbo}, {Mapelli}, {Pasquato},
  {Spera}, {Wang}  \& {Haardt}}{{Di Carlo} et~al.}{2019}]{dicarlo2019}
{Di Carlo} U.~N.,  {Giacobbo} N.,  {Mapelli} M.,  {Pasquato} M.,  {Spera} M.,
  {Wang} L.,   {Haardt} F.,  2019, \mn@doi [\mnras] {10.1093/mnras/stz1453},
  \href {https://ui.adsabs.harvard.edu/abs/2019MNRAS.487.2947D} {487, 2947}

\bibitem[\protect\citeauthoryear{{Di Carlo}, {Mapelli}, {Bouffanais},
  {Giacobbo}, {Santoliquido}, {Bressan}, {Spera}  \& {Haardt}}{{Di Carlo}
  et~al.}{2020a}]{dicarlo2020a}
{Di Carlo} U.~N.,  {Mapelli} M.,  {Bouffanais} Y.,  {Giacobbo} N.,
  {Santoliquido} F.,  {Bressan} A.,  {Spera} M.,   {Haardt} F.,  2020a, \mn@doi
  [\mnras] {10.1093/mnras/staa1997}, \href
  {https://ui.adsabs.harvard.edu/abs/2020MNRAS.497.1043D} {497, 1043}

\bibitem[\protect\citeauthoryear{{Di Carlo} et~al.,}{{Di Carlo}
  et~al.}{2020b}]{dicarlo2020b}
{Di Carlo} U.~N.,  et~al., 2020b, \mn@doi [\mnras] {10.1093/mnras/staa2286},
  \href {https://ui.adsabs.harvard.edu/abs/2020MNRAS.498..495D} {498, 495}

\bibitem[\protect\citeauthoryear{{Di Carlo} et~al.,}{{Di Carlo}
  et~al.}{2021}]{dicarlo2021}
{Di Carlo} U.~N.,  et~al., 2021, \mn@doi [\mnras] {10.1093/mnras/stab2390},
  \href {https://ui.adsabs.harvard.edu/abs/2021MNRAS.tmp.2155D} {}

\bibitem[\protect\citeauthoryear{{Doctor}, {Wysocki}, {O'Shaughnessy}, {Holz}
  \& {Farr}}{{Doctor} et~al.}{2020}]{doctor2020}
{Doctor} Z.,  {Wysocki} D.,  {O'Shaughnessy} R.,  {Holz} D.~E.,   {Farr} B.,
  2020, \mn@doi [\apj] {10.3847/1538-4357/ab7fac}, \href
  {https://ui.adsabs.harvard.edu/abs/2020ApJ...893...35D} {893, 35}

\bibitem[\protect\citeauthoryear{{Dvorkin}, {Vangioni}, {Silk}, {Uzan}  \&
  {Olive}}{{Dvorkin} et~al.}{2016}]{dvorkin2016}
{Dvorkin} I.,  {Vangioni} E.,  {Silk} J.,  {Uzan} J.-P.,   {Olive} K.~A.,
  2016, \mn@doi [\mnras] {10.1093/mnras/stw1477}, \href
  {http://adsabs.harvard.edu/abs/2016MNRAS.461.3877D} {461, 3877}

\bibitem[\protect\citeauthoryear{{Dvorkin}, {Uzan}, {Vangioni}  \&
  {Silk}}{{Dvorkin} et~al.}{2018}]{dvorkin2018}
{Dvorkin} I.,  {Uzan} J.-P.,  {Vangioni} E.,   {Silk} J.,  2018, \mn@doi
  [\mnras] {10.1093/mnras/sty1414}, \href
  {https://ui.adsabs.harvard.edu/abs/2018MNRAS.479..121D} {479, 121}

\bibitem[\protect\citeauthoryear{{El-Badry}, {Quataert}, {Weisz}, {Choksi}  \&
  {Boylan-Kolchin}}{{El-Badry} et~al.}{2019}]{el-badry2019}
{El-Badry} K.,  {Quataert} E.,  {Weisz} D.~R.,  {Choksi} N.,   {Boylan-Kolchin}
  M.,  2019, \mn@doi [\mnras] {10.1093/mnras/sty3007}, \href
  {https://ui.adsabs.harvard.edu/abs/2019MNRAS.482.4528E} {482, 4528}

\bibitem[\protect\citeauthoryear{{Eldridge} \& {Stanway}}{{Eldridge} \&
  {Stanway}}{2016}]{eldridge2016}
{Eldridge} J.~J.,  {Stanway} E.~R.,  2016, \mn@doi [\mnras]
  {10.1093/mnras/stw1772}, \href
  {https://ui.adsabs.harvard.edu/abs/2016MNRAS.462.3302E} {462, 3302}

\bibitem[\protect\citeauthoryear{{Ertl}, {Woosley}, {Sukhbold}  \&
  {Janka}}{{Ertl} et~al.}{2020}]{ertl2020}
{Ertl} T.,  {Woosley} S.~E.,  {Sukhbold} T.,   {Janka} H.~T.,  2020, \mn@doi
  [\apj] {10.3847/1538-4357/ab6458}, \href
  {https://ui.adsabs.harvard.edu/abs/2020ApJ...890...51E} {890, 51}

\bibitem[\protect\citeauthoryear{{Farmer}, {Renzo}, {de Mink}, {Marchant}  \&
  {Justham}}{{Farmer} et~al.}{2019}]{farmer2019}
{Farmer} R.,  {Renzo} M.,  {de Mink} S.~E.,  {Marchant} P.,   {Justham} S.,
  2019, \mn@doi [\apj] {10.3847/1538-4357/ab518b}, \href
  {https://ui.adsabs.harvard.edu/abs/2019ApJ...887...53F} {887, 53}

\bibitem[\protect\citeauthoryear{{Farmer}, {Renzo}, {de Mink}, {Fishbach}  \&
  {Justham}}{{Farmer} et~al.}{2020}]{farmer2020}
{Farmer} R.,  {Renzo} M.,  {de Mink} S.~E.,  {Fishbach} M.,   {Justham} S.,
  2020, \mn@doi [\apjl] {10.3847/2041-8213/abbadd}, \href
  {https://ui.adsabs.harvard.edu/abs/2020ApJ...902L..36F} {902, L36}

\bibitem[\protect\citeauthoryear{{Farrell}, {Groh}, {Hirschi}, {Murphy},
  {Kaiser}, {Ekstr{\"o}m}, {Georgy}  \& {Meynet}}{{Farrell}
  et~al.}{2021}]{farrell2020}
{Farrell} E.,  {Groh} J.~H.,  {Hirschi} R.,  {Murphy} L.,  {Kaiser} E.,
  {Ekstr{\"o}m} S.,  {Georgy} C.,   {Meynet} G.,  2021, \mn@doi [\mnras]
  {10.1093/mnrasl/slaa196}, \href
  {https://ui.adsabs.harvard.edu/abs/2021MNRAS.502L..40F} {502, L40}

\bibitem[\protect\citeauthoryear{{Favata}, {Hughes}  \& {Holz}}{{Favata}
  et~al.}{2004}]{favata2004}
{Favata} M.,  {Hughes} S.~A.,   {Holz} D.~E.,  2004, \mn@doi [\apjl]
  {10.1086/421552}, \href
  {https://ui.adsabs.harvard.edu/abs/2004ApJ...607L...5F} {607, L5}

\bibitem[\protect\citeauthoryear{{Fishbach} \& {Kalogera}}{{Fishbach} \&
  {Kalogera}}{2021}]{fishbach2021b}
{Fishbach} M.,  {Kalogera} V.,  2021, \mn@doi [\apjl]
  {10.3847/2041-8213/ac05c4}, \href
  {https://ui.adsabs.harvard.edu/abs/2021ApJ...914L..30F} {914, L30}

\bibitem[\protect\citeauthoryear{{Fishbach}, {Holz}  \& {Farr}}{{Fishbach}
  et~al.}{2017}]{fishbach2017}
{Fishbach} M.,  {Holz} D.~E.,   {Farr} B.,  2017, \mn@doi [\apjl]
  {10.3847/2041-8213/aa7045}, \href
  {https://ui.adsabs.harvard.edu/abs/2017ApJ...840L..24F} {840, L24}

\bibitem[\protect\citeauthoryear{Fishbach, Holz  \& Farr}{Fishbach
  et~al.}{2018}]{fishbach2018}
Fishbach M.,  Holz D.~E.,   Farr W.~M.,  2018, \mn@doi [\apjl]
  {10.3847/2041-8213/aad800}, 863, L41

\bibitem[\protect\citeauthoryear{{Fishbach} et~al.,}{{Fishbach}
  et~al.}{2021}]{fishbach2021a}
{Fishbach} M.,  et~al., 2021, \mn@doi [\apj] {10.3847/1538-4357/abee11}, \href
  {https://ui.adsabs.harvard.edu/abs/2021ApJ...912...98F} {912, 98}

\bibitem[\protect\citeauthoryear{{Fitchett}}{{Fitchett}}{1983}]{fitchett1983}
{Fitchett} M.~J.,  1983, \mn@doi [\mnras] {10.1093/mnras/203.4.1049}, \href
  {https://ui.adsabs.harvard.edu/abs/1983MNRAS.203.1049F} {203, 1049}

\bibitem[\protect\citeauthoryear{{Fragione} \& {Kocsis}}{{Fragione} \&
  {Kocsis}}{2018}]{fragionekocsis2018}
{Fragione} G.,  {Kocsis} B.,  2018, \mn@doi [\prl]
  {10.1103/PhysRevLett.121.161103}, \href
  {https://ui.adsabs.harvard.edu/abs/2018PhRvL.121p1103F} {121, 161103}

\bibitem[\protect\citeauthoryear{{Fragione} \& {Kocsis}}{{Fragione} \&
  {Kocsis}}{2019}]{fragione2019b}
{Fragione} G.,  {Kocsis} B.,  2019, \mn@doi [\mnras] {10.1093/mnras/stz1175},
  \href {https://ui.adsabs.harvard.edu/abs/2019MNRAS.486.4781F} {486, 4781}

\bibitem[\protect\citeauthoryear{{Fragione} \& {Silk}}{{Fragione} \&
  {Silk}}{2020}]{fragione2020}
{Fragione} G.,  {Silk} J.,  2020, \mn@doi [\mnras] {10.1093/mnras/staa2629},
  \href {https://ui.adsabs.harvard.edu/abs/2020MNRAS.498.4591F} {498, 4591}

\bibitem[\protect\citeauthoryear{{Fragione}, {Loeb}  \& {Rasio}}{{Fragione}
  et~al.}{2020}]{fragione2020b}
{Fragione} G.,  {Loeb} A.,   {Rasio} F.~A.,  2020, \mn@doi [\apjl]
  {10.3847/2041-8213/abbc0a}, \href
  {https://ui.adsabs.harvard.edu/abs/2020ApJ...902L..26F} {902, L26}

\bibitem[\protect\citeauthoryear{{Fryer}, {Belczynski}, {Wiktorowicz},
  {Dominik}, {Kalogera}  \& {Holz}}{{Fryer} et~al.}{2012}]{fryer2012}
{Fryer} C.~L.,  {Belczynski} K.,  {Wiktorowicz} G.,  {Dominik} M.,  {Kalogera}
  V.,   {Holz} D.~E.,  2012, \mn@doi [\apj] {10.1088/0004-637X/749/1/91}, \href
  {http://adsabs.harvard.edu/abs/2012ApJ...749...91F} {749, 91}

\bibitem[\protect\citeauthoryear{{Fujii} \& {Portegies Zwart}}{{Fujii} \&
  {Portegies Zwart}}{2014}]{fujii2014}
{Fujii} M.~S.,  {Portegies Zwart} S.,  2014, \mn@doi [\mnras]
  {10.1093/mnras/stu015}, \href
  {http://adsabs.harvard.edu/abs/2014MNRAS.439.1003F} {439, 1003}

\bibitem[\protect\citeauthoryear{{Fuller} \& {Ma}}{{Fuller} \&
  {Ma}}{2019}]{fullerma2019}
{Fuller} J.,  {Ma} L.,  2019, \mn@doi [\apjl] {10.3847/2041-8213/ab339b}, \href
  {https://ui.adsabs.harvard.edu/abs/2019ApJ...881L...1F} {881, L1}

\bibitem[\protect\citeauthoryear{{Fuller}, {Piro}  \& {Jermyn}}{{Fuller}
  et~al.}{2019}]{fuller2019}
{Fuller} J.,  {Piro} A.~L.,   {Jermyn} A.~S.,  2019, \mn@doi [\mnras]
  {10.1093/mnras/stz514}, \href
  {https://ui.adsabs.harvard.edu/abs/2019MNRAS.485.3661F} {485, 3661}

\bibitem[\protect\citeauthoryear{{Gallegos-Garcia}, {Berry}, {Marchant}  \&
  {Kalogera}}{{Gallegos-Garcia} et~al.}{2021}]{gallegosgarcia2021}
{Gallegos-Garcia} M.,  {Berry} C. P.~L.,  {Marchant} P.,   {Kalogera} V.,
  2021, \mn@doi [\apj] {10.3847/1538-4357/ac2610}, \href
  {https://ui.adsabs.harvard.edu/abs/2021ApJ...922..110G} {922, 110}

\bibitem[\protect\citeauthoryear{{Generozov}, {Nayakshin}  \&
  {Madigan}}{{Generozov} et~al.}{2021}]{generozov2021}
{Generozov} A.,  {Nayakshin} S.,   {Madigan} A.~M.,  2021, arXiv e-prints,
  \href {https://ui.adsabs.harvard.edu/abs/2021arXiv211112744G} {p.
  arXiv:2111.12744}

\bibitem[\protect\citeauthoryear{{Georgiev}, {Puzia}, {Hilker}  \&
  {Goudfrooij}}{{Georgiev} et~al.}{2009a}]{georgiev2009a}
{Georgiev} I.~Y.,  {Puzia} T.~H.,  {Hilker} M.,   {Goudfrooij} P.,  2009a,
  \mn@doi [\mnras] {10.1111/j.1365-2966.2008.14104.x}, \href
  {https://ui.adsabs.harvard.edu/abs/2009MNRAS.392..879G} {392, 879}

\bibitem[\protect\citeauthoryear{{Georgiev}, {Hilker}, {Puzia}, {Goudfrooij}
  \& {Baumgardt}}{{Georgiev} et~al.}{2009b}]{georgiev2009b}
{Georgiev} I.~Y.,  {Hilker} M.,  {Puzia} T.~H.,  {Goudfrooij} P.,   {Baumgardt}
  H.,  2009b, \mn@doi [\mnras] {10.1111/j.1365-2966.2009.14776.x}, \href
  {https://ui.adsabs.harvard.edu/abs/2009MNRAS.396.1075G} {396, 1075}

\bibitem[\protect\citeauthoryear{{Georgiev}, {B{\"o}ker}, {Leigh},
  {L{\"u}tzgendorf}  \& {Neumayer}}{{Georgiev} et~al.}{2016}]{georgiev2016}
{Georgiev} I.~Y.,  {B{\"o}ker} T.,  {Leigh} N.,  {L{\"u}tzgendorf} N.,
  {Neumayer} N.,  2016, \mn@doi [\mnras] {10.1093/mnras/stw093}, \href
  {https://ui.adsabs.harvard.edu/abs/2016MNRAS.457.2122G} {457, 2122}

\bibitem[\protect\citeauthoryear{{Gerosa} \& {Berti}}{{Gerosa} \&
  {Berti}}{2017}]{gerosa2017}
{Gerosa} D.,  {Berti} E.,  2017, \mn@doi [\prd] {10.1103/PhysRevD.95.124046},
  \href {http://adsabs.harvard.edu/abs/2017PhRvD..95l4046G} {95, 124046}

\bibitem[\protect\citeauthoryear{{Gerosa} \& {Fishbach}}{{Gerosa} \&
  {Fishbach}}{2021}]{gerosa2021review}
{Gerosa} D.,  {Fishbach} M.,  2021, \mn@doi [Nature Astronomy]
  {10.1038/s41550-021-01398-w}, \href
  {https://ui.adsabs.harvard.edu/abs/2021NatAs.tmp..136G} {}

\bibitem[\protect\citeauthoryear{{Gerosa}, {Giacobbo}  \& {Vecchio}}{{Gerosa}
  et~al.}{2021}]{gerosa2021}
{Gerosa} D.,  {Giacobbo} N.,   {Vecchio} A.,  2021, \mn@doi [\apj]
  {10.3847/1538-4357/ac00bb}, \href
  {https://ui.adsabs.harvard.edu/abs/2021ApJ...915...56G} {915, 56}

\bibitem[\protect\citeauthoryear{{Giacobbo} \& {Mapelli}}{{Giacobbo} \&
  {Mapelli}}{2018}]{giacobbo2018b}
{Giacobbo} N.,  {Mapelli} M.,  2018, \mn@doi [\mnras] {10.1093/mnras/sty1999},
  \href {http://adsabs.harvard.edu/abs/2018MNRAS.480.2011G} {480, 2011}

\bibitem[\protect\citeauthoryear{{Giacobbo} \& {Mapelli}}{{Giacobbo} \&
  {Mapelli}}{2020}]{giacobbo2020}
{Giacobbo} N.,  {Mapelli} M.,  2020, \mn@doi [\apj] {10.3847/1538-4357/ab7335},
  \href {https://ui.adsabs.harvard.edu/abs/2020ApJ...891..141G} {891, 141}

\bibitem[\protect\citeauthoryear{{Giacobbo}, {Mapelli}  \& {Spera}}{{Giacobbo}
  et~al.}{2018}]{giacobbo2018}
{Giacobbo} N.,  {Mapelli} M.,   {Spera} M.,  2018, \mn@doi [\mnras]
  {10.1093/mnras/stx2933}, \href
  {http://adsabs.harvard.edu/abs/2018MNRAS.474.2959G} {474, 2959}

\bibitem[\protect\citeauthoryear{{Giersz}, {Leigh}, {Hypki}, {L{\"u}tzgendorf}
  \& {Askar}}{{Giersz} et~al.}{2015}]{giersz2015}
{Giersz} M.,  {Leigh} N.,  {Hypki} A.,  {L{\"u}tzgendorf} N.,   {Askar} A.,
  2015, \mn@doi [\mnras] {10.1093/mnras/stv2162}, \href
  {http://adsabs.harvard.edu/abs/2015MNRAS.454.3150G} {454, 3150}

\bibitem[\protect\citeauthoryear{{Gonz{\'a}lez}, {Kremer}, {Chatterjee},
  {Fragione}, {Rodriguez}, {Weatherford}, {Ye}  \& {Rasio}}{{Gonz{\'a}lez}
  et~al.}{2021}]{gonzalez2021}
{Gonz{\'a}lez} E.,  {Kremer} K.,  {Chatterjee} S.,  {Fragione} G.,  {Rodriguez}
  C.~L.,  {Weatherford} N.~C.,  {Ye} C.~S.,   {Rasio} F.~A.,  2021, \mn@doi
  [\apjl] {10.3847/2041-8213/abdf5b}, \href
  {https://ui.adsabs.harvard.edu/abs/2021ApJ...908L..29G} {908, L29}

\bibitem[\protect\citeauthoryear{{Goodman} \& {Hut}}{{Goodman} \&
  {Hut}}{1993}]{goodman1993}
{Goodman} J.,  {Hut} P.,  1993, \mn@doi [\apj] {10.1086/172200}, \href
  {http://adsabs.harvard.edu/abs/1993ApJ...403..271G} {403, 271}

\bibitem[\protect\citeauthoryear{{Gr{\"a}fener} \& {Hamann}}{{Gr{\"a}fener} \&
  {Hamann}}{2008}]{graefener2008}
{Gr{\"a}fener} G.,  {Hamann} W.-R.,  2008, \mn@doi [\aap]
  {10.1051/0004-6361:20066176}, \href
  {http://adsabs.harvard.edu/abs/2008A%26A...482..945G} {482, 945}

\bibitem[\protect\citeauthoryear{{Gratton}, {Fusi Pecci}, {Carretta},
  {Clementini}, {Corsi}  \& {Lattanzi}}{{Gratton} et~al.}{1997}]{gratton1997}
{Gratton} R.~G.,  {Fusi Pecci} F.,  {Carretta} E.,  {Clementini} G.,  {Corsi}
  C.~E.,   {Lattanzi} M.,  1997, \mn@doi [\apj] {10.1086/304987}, \href
  {https://ui.adsabs.harvard.edu/abs/1997ApJ...491..749G} {491, 749}

\bibitem[\protect\citeauthoryear{{Gratton}, {Bragaglia}, {Carretta},
  {Clementini}, {Desidera}, {Grundahl}  \& {Lucatello}}{{Gratton}
  et~al.}{2003}]{gratton2003}
{Gratton} R.~G.,  {Bragaglia} A.,  {Carretta} E.,  {Clementini} G.,  {Desidera}
  S.,  {Grundahl} F.,   {Lucatello} S.,  2003, \mn@doi [\aap]
  {10.1051/0004-6361:20031003}, \href
  {https://ui.adsabs.harvard.edu/abs/2003A&A...408..529G} {408, 529}

\bibitem[\protect\citeauthoryear{{Hamers} \& {Safarzadeh}}{{Hamers} \&
  {Safarzadeh}}{2020}]{hamers2020}
{Hamers} A.~S.,  {Safarzadeh} M.,  2020, \mn@doi [\apj]
  {10.3847/1538-4357/ab9b27}, \href
  {https://ui.adsabs.harvard.edu/abs/2020ApJ...898...99H} {898, 99}

\bibitem[\protect\citeauthoryear{{Harris}}{{Harris}}{1996}]{harris1996}
{Harris} W.~E.,  1996, \mn@doi [\aj] {10.1086/118116}, \href
  {https://ui.adsabs.harvard.edu/abs/1996AJ....112.1487H} {112, 1487}

\bibitem[\protect\citeauthoryear{{Hartwig}, {Volonteri}, {Bromm}, {Klessen},
  {Barausse}, {Magg}  \& {Stacy}}{{Hartwig} et~al.}{2016}]{hartwig2016}
{Hartwig} T.,  {Volonteri} M.,  {Bromm} V.,  {Klessen} R.~S.,  {Barausse} E.,
  {Magg} M.,   {Stacy} A.,  2016, \mn@doi [\mnras] {10.1093/mnrasl/slw074},
  \href {http://adsabs.harvard.edu/abs/2016MNRAS.460L..74H} {460, L74}

\bibitem[\protect\citeauthoryear{{Heger}, {Fryer}, {Woosley}, {Langer}  \&
  {Hartmann}}{{Heger} et~al.}{2003}]{heger2003}
{Heger} A.,  {Fryer} C.~L.,  {Woosley} S.~E.,  {Langer} N.,   {Hartmann} D.~H.,
   2003, \mn@doi [\apj] {10.1086/375341}, \href
  {http://adsabs.harvard.edu/abs/2003ApJ...591..288H} {591, 288}

\bibitem[\protect\citeauthoryear{{Heggie}}{{Heggie}}{1975}]{heggie1975}
{Heggie} D.~C.,  1975, \mn@doi [\mnras] {10.1093/mnras/173.3.729}, \href
  {http://adsabs.harvard.edu/abs/1975MNRAS.173..729H} {173, 729}

\bibitem[\protect\citeauthoryear{{Hills}}{{Hills}}{1983}]{hills1983}
{Hills} J.~G.,  1983, \mn@doi [\aj] {10.1086/113418}, \href
  {https://ui.adsabs.harvard.edu/abs/1983AJ.....88.1269H} {88, 1269}

\bibitem[\protect\citeauthoryear{{Holley-Bockelmann}, {G{\"u}ltekin},
  {Shoemaker}  \& {Yunes}}{{Holley-Bockelmann}
  et~al.}{2008}]{holley-bockelmann2008}
{Holley-Bockelmann} K.,  {G{\"u}ltekin} K.,  {Shoemaker} D.,   {Yunes} N.,
  2008, \mn@doi [\apj] {10.1086/591218}, \href
  {https://ui.adsabs.harvard.edu/abs/2008ApJ...686..829H} {686, 829}

\bibitem[\protect\citeauthoryear{{Hong}, {Vesperini}, {Askar}, {Giersz},
  {Szkudlarek}  \& {Bulik}}{{Hong} et~al.}{2018}]{hong2018}
{Hong} J.,  {Vesperini} E.,  {Askar} A.,  {Giersz} M.,  {Szkudlarek} M.,
  {Bulik} T.,  2018, \mn@doi [\mnras] {10.1093/mnras/sty2211}, \href
  {https://ui.adsabs.harvard.edu/abs/2018MNRAS.480.5645H} {480, 5645}

\bibitem[\protect\citeauthoryear{{Hurley}, {Tout}  \& {Pols}}{{Hurley}
  et~al.}{2002}]{hurley2002}
{Hurley} J.~R.,  {Tout} C.~A.,   {Pols} O.~R.,  2002, \mn@doi [\mnras]
  {10.1046/j.1365-8711.2002.05038.x}, \href
  {http://adsabs.harvard.edu/abs/2002MNRAS.329..897H} {329, 897}

\bibitem[\protect\citeauthoryear{{Ishibashi} \& {Gr{\"o}bner}}{{Ishibashi} \&
  {Gr{\"o}bner}}{2020}]{ishibashi2020}
{Ishibashi} W.,  {Gr{\"o}bner} M.,  2020, \mn@doi [\aap]
  {10.1051/0004-6361/202037799}, \href
  {https://ui.adsabs.harvard.edu/abs/2020A&A...639A.108I} {639, A108}

\bibitem[\protect\citeauthoryear{{Ji} \& {Bregman}}{{Ji} \&
  {Bregman}}{2015}]{jibregman2015}
{Ji} J.,  {Bregman} J.~N.,  2015, \mn@doi [\apj] {10.1088/0004-637X/807/1/32},
  \href {https://ui.adsabs.harvard.edu/abs/2015ApJ...807...32J} {807, 32}

\bibitem[\protect\citeauthoryear{{Jim{\'e}nez-Forteza}, {Keitel}, {Husa},
  {Hannam}, {Khan}  \& {P{\"u}rrer}}{{Jim{\'e}nez-Forteza}
  et~al.}{2017}]{jimenez-forteza2017}
{Jim{\'e}nez-Forteza} X.,  {Keitel} D.,  {Husa} S.,  {Hannam} M.,  {Khan} S.,
  {P{\"u}rrer} M.,  2017, \mn@doi [\prd] {10.1103/PhysRevD.95.064024}, \href
  {https://ui.adsabs.harvard.edu/abs/2017PhRvD..95f4024J} {95, 064024}

\bibitem[\protect\citeauthoryear{{Kamlah} et~al.,}{{Kamlah}
  et~al.}{2021}]{kamlah2021}
{Kamlah} A.~W.~H.,  et~al., 2021, \mn@doi [\mnras] {10.1093/mnras/stab3748},
  \href {https://ui.adsabs.harvard.edu/abs/2021MNRAS.tmp.3435K} {}

\bibitem[\protect\citeauthoryear{{Kimball}, {Talbot}, {Berry}, {Carney},
  {Zevin}, {Thrane}  \& {Kalogera}}{{Kimball} et~al.}{2020}]{kimball2020}
{Kimball} C.,  {Talbot} C.,  {Berry} C. P.~L.,  {Carney} M.,  {Zevin} M.,
  {Thrane} E.,   {Kalogera} V.,  2020, \mn@doi [\apj]
  {10.3847/1538-4357/aba518}, \href
  {https://ui.adsabs.harvard.edu/abs/2020ApJ...900..177K} {900, 177}

\bibitem[\protect\citeauthoryear{{Kimball} et~al.,}{{Kimball}
  et~al.}{2021}]{kimball2020a}
{Kimball} C.,  et~al., 2021, \mn@doi [\apjl] {10.3847/2041-8213/ac0aef}, \href
  {https://ui.adsabs.harvard.edu/abs/2021ApJ...915L..35K} {915, L35}

\bibitem[\protect\citeauthoryear{{Kinugawa}, {Miyamoto}, {Kanda}  \&
  {Nakamura}}{{Kinugawa} et~al.}{2016}]{kinugawa2016}
{Kinugawa} T.,  {Miyamoto} A.,  {Kanda} N.,   {Nakamura} T.,  2016, \mn@doi
  [\mnras] {10.1093/mnras/stv2624}, \href
  {http://adsabs.harvard.edu/abs/2016MNRAS.456.1093K} {456, 1093}

\bibitem[\protect\citeauthoryear{{Klencki}, {Moe}, {Gladysz}, {Chruslinska},
  {Holz}  \& {Belczynski}}{{Klencki} et~al.}{2018}]{klencki2018}
{Klencki} J.,  {Moe} M.,  {Gladysz} W.,  {Chruslinska} M.,  {Holz} D.~E.,
  {Belczynski} K.,  2018, \mn@doi [\aap] {10.1051/0004-6361/201833025}, \href
  {https://ui.adsabs.harvard.edu/abs/2018A&A...619A..77K} {619, A77}

\bibitem[\protect\citeauthoryear{{Klencki}, {Nelemans}, {Istrate}  \&
  {Chruslinska}}{{Klencki} et~al.}{2021}]{klencki2021}
{Klencki} J.,  {Nelemans} G.,  {Istrate} A.~G.,   {Chruslinska} M.,  2021,
  \mn@doi [\aap] {10.1051/0004-6361/202038707}, \href
  {https://ui.adsabs.harvard.edu/abs/2021A&A...645A..54K} {645, A54}

\bibitem[\protect\citeauthoryear{{Kremer} et~al.,}{{Kremer}
  et~al.}{2020a}]{kremer2020b}
{Kremer} K.,  et~al., 2020a, \mn@doi [\apjs] {10.3847/1538-4365/ab7919}, \href
  {https://ui.adsabs.harvard.edu/abs/2020ApJS..247...48K} {247, 48}

\bibitem[\protect\citeauthoryear{{Kremer} et~al.,}{{Kremer}
  et~al.}{2020b}]{kremer2020}
{Kremer} K.,  et~al., 2020b, \mn@doi [\apj] {10.3847/1538-4357/abb945}, \href
  {https://ui.adsabs.harvard.edu/abs/2020ApJ...903...45K} {903, 45}

\bibitem[\protect\citeauthoryear{{Kroupa}}{{Kroupa}}{2001}]{kroupa2001}
{Kroupa} P.,  2001, \mn@doi [\mnras] {10.1046/j.1365-8711.2001.04022.x}, \href
  {http://adsabs.harvard.edu/abs/2001MNRAS.322..231K} {322, 231}

\bibitem[\protect\citeauthoryear{{Kruckow}, {Tauris}, {Langer}, {Kramer}  \&
  {Izzard}}{{Kruckow} et~al.}{2018}]{kruckow2018}
{Kruckow} M.~U.,  {Tauris} T.~M.,  {Langer} N.,  {Kramer} M.,   {Izzard} R.~G.,
   2018, \mn@doi [\mnras] {10.1093/mnras/sty2190}, \href
  {http://adsabs.harvard.edu/abs/2018MNRAS.481.1908K} {481, 1908}

\bibitem[\protect\citeauthoryear{{Kruijssen}}{{Kruijssen}}{2014}]{kruijssen2014}
{Kruijssen} J.~M.~D.,  2014, \mn@doi [Classical and Quantum Gravity]
  {10.1088/0264-9381/31/24/244006}, \href
  {https://ui.adsabs.harvard.edu/abs/2014CQGra..31x4006K} {31, 244006}

\bibitem[\protect\citeauthoryear{{Kumamoto}, {Fujii}  \& {Tanikawa}}{{Kumamoto}
  et~al.}{2019}]{kumamoto2019}
{Kumamoto} J.,  {Fujii} M.~S.,   {Tanikawa} A.,  2019, \mn@doi [\mnras]
  {10.1093/mnras/stz1068}, \href
  {https://ui.adsabs.harvard.edu/abs/2019MNRAS.486.3942K} {486, 3942}

\bibitem[\protect\citeauthoryear{{Kumamoto}, {Fujii}  \& {Tanikawa}}{{Kumamoto}
  et~al.}{2020}]{kumamoto2020}
{Kumamoto} J.,  {Fujii} M.~S.,   {Tanikawa} A.,  2020, \mn@doi [\mnras]
  {10.1093/mnras/staa1440}, \href
  {https://ui.adsabs.harvard.edu/abs/2020MNRAS.495.4268K} {495, 4268}

\bibitem[\protect\citeauthoryear{{Lada} \& {Lada}}{{Lada} \&
  {Lada}}{2003}]{lada2003}
{Lada} C.~J.,  {Lada} E.~A.,  2003, \mn@doi [\araa]
  {10.1146/annurev.astro.41.011802.094844}, \href
  {http://adsabs.harvard.edu/abs/2003ARA%26A..41...57L} {41, 57}

\bibitem[\protect\citeauthoryear{{Lee}}{{Lee}}{1995}]{lee1995}
{Lee} H.~M.,  1995, \mn@doi [\mnras] {10.1093/mnras/272.3.605}, \href
  {https://ui.adsabs.harvard.edu/abs/1995MNRAS.272..605L} {272, 605}

\bibitem[\protect\citeauthoryear{{Liu} \& {Bromm}}{{Liu} \&
  {Bromm}}{2020}]{liubromm2020}
{Liu} B.,  {Bromm} V.,  2020, \mn@doi [\mnras] {10.1093/mnras/staa1362}, \href
  {https://ui.adsabs.harvard.edu/abs/2020MNRAS.495.2475L} {495, 2475}

\bibitem[\protect\citeauthoryear{{Liu} \& {Lai}}{{Liu} \&
  {Lai}}{2019}]{liu2019}
{Liu} B.,  {Lai} D.,  2019, \mn@doi [\mnras] {10.1093/mnras/sty3432}, \href
  {https://ui.adsabs.harvard.edu/abs/2019MNRAS.483.4060L} {483, 4060}

\bibitem[\protect\citeauthoryear{{Liu} \& {Lai}}{{Liu} \&
  {Lai}}{2021}]{liu2021}
{Liu} B.,  {Lai} D.,  2021, \mn@doi [\mnras] {10.1093/mnras/stab178}, \href
  {https://ui.adsabs.harvard.edu/abs/2021MNRAS.502.2049L} {502, 2049}

\bibitem[\protect\citeauthoryear{Loredo}{Loredo}{2004}]{loredo2004}
Loredo T.~J.,  2004, \mn@doi [AIP Conf. Proc.] {10.1063/1.1835214}, 735, 195

\bibitem[\protect\citeauthoryear{{Lousto} \& {Zlochower}}{{Lousto} \&
  {Zlochower}}{2011}]{lousto2011}
{Lousto} C.~O.,  {Zlochower} Y.,  2011, \mn@doi [\prl]
  {10.1103/PhysRevLett.107.231102}, \href
  {https://ui.adsabs.harvard.edu/abs/2011PhRvL.107w1102L} {107, 231102}

\bibitem[\protect\citeauthoryear{{Lousto}, {Zlochower}, {Dotti}  \&
  {Volonteri}}{{Lousto} et~al.}{2012}]{lousto2012}
{Lousto} C.~O.,  {Zlochower} Y.,  {Dotti} M.,   {Volonteri} M.,  2012, \mn@doi
  [\prd] {10.1103/PhysRevD.85.084015}, \href
  {https://ui.adsabs.harvard.edu/abs/2012PhRvD..85h4015L} {85, 084015}

\bibitem[\protect\citeauthoryear{{Madau} \& {Fragos}}{{Madau} \&
  {Fragos}}{2017}]{madau2017}
{Madau} P.,  {Fragos} T.,  2017, \mn@doi [\apj] {10.3847/1538-4357/aa6af9},
  \href {https://ui.adsabs.harvard.edu/abs/2017ApJ...840...39M} {840, 39}

\bibitem[\protect\citeauthoryear{{Mandel} \& {Broekgaarden}}{{Mandel} \&
  {Broekgaarden}}{2021}]{mandel2021}
{Mandel} I.,  {Broekgaarden} F.~S.,  2021, arXiv e-prints, \href
  {https://ui.adsabs.harvard.edu/abs/2021arXiv210714239M} {p. arXiv:2107.14239}

\bibitem[\protect\citeauthoryear{{Mandel} \& {de Mink}}{{Mandel} \& {de
  Mink}}{2016}]{mandel2016}
{Mandel} I.,  {de Mink} S.~E.,  2016, \mn@doi [\mnras] {10.1093/mnras/stw379},
  \href {http://adsabs.harvard.edu/abs/2016MNRAS.458.2634M} {458, 2634}

\bibitem[\protect\citeauthoryear{{Mandel}, {Farr}  \& {Gair}}{{Mandel}
  et~al.}{2019}]{mandel2018}
{Mandel} I.,  {Farr} W.~M.,   {Gair} J.~R.,  2019, \mn@doi [\mnras]
  {10.1093/mnras/stz896}, \href
  {https://ui.adsabs.harvard.edu/abs/2019MNRAS.486.1086M} {486, 1086}

\bibitem[\protect\citeauthoryear{{Mapelli}}{{Mapelli}}{2016}]{mapelli2016}
{Mapelli} M.,  2016, \mn@doi [\mnras] {10.1093/mnras/stw869}, \href
  {http://adsabs.harvard.edu/abs/2016MNRAS.459.3432M} {459, 3432}

\bibitem[\protect\citeauthoryear{Mapelli}{Mapelli}{2021}]{mapelli2021review}
Mapelli M.,  2021, Formation Channels of Single and Binary Stellar-Mass Black
  Holes.
Springer Singapore, Singapore, pp 1--65,
  \mn@doi{10.1007/978-981-15-4702-7_16-1}, \url
  {https://doi.org/10.1007/978-981-15-4702-7_16-1}

\bibitem[\protect\citeauthoryear{{Mapelli}, {Hayfield}, {Mayer}  \&
  {Wadsley}}{{Mapelli} et~al.}{2012}]{mapelli2012}
{Mapelli} M.,  {Hayfield} T.,  {Mayer} L.,   {Wadsley} J.,  2012, \mn@doi
  [\apj] {10.1088/0004-637X/749/2/168}, \href
  {https://ui.adsabs.harvard.edu/abs/2012ApJ...749..168M} {749, 168}

\bibitem[\protect\citeauthoryear{{Mapelli}, {Giacobbo}, {Ripamonti}  \&
  {Spera}}{{Mapelli} et~al.}{2017}]{mapelli2017}
{Mapelli} M.,  {Giacobbo} N.,  {Ripamonti} E.,   {Spera} M.,  2017, \mn@doi
  [\mnras] {10.1093/mnras/stx2123}, \href
  {http://adsabs.harvard.edu/abs/2017MNRAS.472.2422M} {472, 2422}

\bibitem[\protect\citeauthoryear{{Mapelli}, {Giacobbo}, {Santoliquido}  \&
  {Artale}}{{Mapelli} et~al.}{2019}]{mapelli2019}
{Mapelli} M.,  {Giacobbo} N.,  {Santoliquido} F.,   {Artale} M.~C.,  2019,
  \mn@doi [\mnras] {10.1093/mnras/stz1150}, \href
  {http://adsabs.harvard.edu/abs/2019MNRAS.tmp.1108M} {}

\bibitem[\protect\citeauthoryear{{Mapelli}, {Spera}, {Montanari}, {Limongi},
  {Chieffi}, {Giacobbo}, {Bressan}  \& {Bouffanais}}{{Mapelli}
  et~al.}{2020}]{mapelli2020}
{Mapelli} M.,  {Spera} M.,  {Montanari} E.,  {Limongi} M.,  {Chieffi} A.,
  {Giacobbo} N.,  {Bressan} A.,   {Bouffanais} Y.,  2020, \mn@doi [\apj]
  {10.3847/1538-4357/ab584d}, \href
  {https://ui.adsabs.harvard.edu/abs/2020ApJ...888...76M} {888, 76}

\bibitem[\protect\citeauthoryear{{Mapelli} et~al.,}{{Mapelli}
  et~al.}{2021}]{mapelli2021}
{Mapelli} M.,  et~al., 2021, \mn@doi [\mnras] {10.1093/mnras/stab1334}, \href
  {https://ui.adsabs.harvard.edu/abs/2021MNRAS.505..339M} {505, 339}

\bibitem[\protect\citeauthoryear{{Marchant}, {Langer}, {Podsiadlowski},
  {Tauris}  \& {Moriya}}{{Marchant} et~al.}{2016}]{marchant2016}
{Marchant} P.,  {Langer} N.,  {Podsiadlowski} P.,  {Tauris} T.~M.,   {Moriya}
  T.~J.,  2016, \mn@doi [\aap] {10.1051/0004-6361/201628133}, \href
  {http://adsabs.harvard.edu/abs/2016A%26A...588A..50M} {588, A50}

\bibitem[\protect\citeauthoryear{{Marchant}, {Renzo}, {Farmer}, {Pappas},
  {Taam}, {de Mink}  \& {Kalogera}}{{Marchant} et~al.}{2019}]{marchant2019}
{Marchant} P.,  {Renzo} M.,  {Farmer} R.,  {Pappas} K. M.~W.,  {Taam} R.~E.,
  {de Mink} S.~E.,   {Kalogera} V.,  2019, \mn@doi [\apj]
  {10.3847/1538-4357/ab3426}, \href
  {https://ui.adsabs.harvard.edu/abs/2019ApJ...882...36M} {882, 36}

\bibitem[\protect\citeauthoryear{{McKernan} et~al.,}{{McKernan}
  et~al.}{2018}]{mckernan2018}
{McKernan} B.,  et~al., 2018, \mn@doi [\apj] {10.3847/1538-4357/aadae5}, \href
  {https://ui.adsabs.harvard.edu/abs/2018ApJ...866...66M} {866, 66}

\bibitem[\protect\citeauthoryear{{Miller} \& {Hamilton}}{{Miller} \&
  {Hamilton}}{2002}]{miller2002}
{Miller} M.~C.,  {Hamilton} D.~P.,  2002, \mn@doi [\mnras]
  {10.1046/j.1365-8711.2002.05112.x}, \href
  {http://adsabs.harvard.edu/abs/2002MNRAS.330..232C} {330, 232}

\bibitem[\protect\citeauthoryear{{Miller} \& {Lauburg}}{{Miller} \&
  {Lauburg}}{2009}]{millerlauburg2009}
{Miller} M.~C.,  {Lauburg} V.~M.,  2009, \mn@doi [\apj]
  {10.1088/0004-637X/692/1/917}, \href
  {http://adsabs.harvard.edu/abs/2009ApJ...692..917M} {692, 917}

\bibitem[\protect\citeauthoryear{{Milone} et~al.,}{{Milone}
  et~al.}{2012}]{milone2012}
{Milone} A.~P.,  et~al., 2012, \mn@doi [\aap] {10.1051/0004-6361/201016384},
  \href {https://ui.adsabs.harvard.edu/abs/2012A&A...540A..16M} {540, A16}

\bibitem[\protect\citeauthoryear{{Moody} \& {Sigurdsson}}{{Moody} \&
  {Sigurdsson}}{2009}]{moody2009}
{Moody} K.,  {Sigurdsson} S.,  2009, \mn@doi [\apj]
  {10.1088/0004-637X/690/2/1370}, \href
  {https://ui.adsabs.harvard.edu/abs/2009ApJ...690.1370M} {690, 1370}

\bibitem[\protect\citeauthoryear{{Neijssel} et~al.,}{{Neijssel}
  et~al.}{2019}]{neijssel2019}
{Neijssel} C.~J.,  et~al., 2019, \mn@doi [\mnras] {10.1093/mnras/stz2840},
  \href {https://ui.adsabs.harvard.edu/abs/2019MNRAS.490.3740N} {490, 3740}

\bibitem[\protect\citeauthoryear{{Neumayer}, {Seth}  \& {B{\"o}ker}}{{Neumayer}
  et~al.}{2020}]{neumayer2020}
{Neumayer} N.,  {Seth} A.,   {B{\"o}ker} T.,  2020, \mn@doi [\aapr]
  {10.1007/s00159-020-00125-0}, \href
  {https://ui.adsabs.harvard.edu/abs/2020A&ARv..28....4N} {28, 4}

\bibitem[\protect\citeauthoryear{{Ng}, {Vitale}, {Farr}  \& {Rodriguez}}{{Ng}
  et~al.}{2021}]{ng2021}
{Ng} K. K.~Y.,  {Vitale} S.,  {Farr} W.~M.,   {Rodriguez} C.~L.,  2021, \mn@doi
  [\apjl] {10.3847/2041-8213/abf8be}, \href
  {https://ui.adsabs.harvard.edu/abs/2021ApJ...913L...5N} {913, L5}

\bibitem[\protect\citeauthoryear{{O'Connor} \& {Ott}}{{O'Connor} \&
  {Ott}}{2011}]{oconnor2011}
{O'Connor} E.,  {Ott} C.~D.,  2011, \mn@doi [\apj]
  {10.1088/0004-637X/730/2/70}, \href
  {https://ui.adsabs.harvard.edu/abs/2011ApJ...730...70O} {730, 70}

\bibitem[\protect\citeauthoryear{{O'Leary}, {Meiron}  \& {Kocsis}}{{O'Leary}
  et~al.}{2016}]{oleary2016}
{O'Leary} R.~M.,  {Meiron} Y.,   {Kocsis} B.,  2016, \mn@doi [\apjl]
  {10.3847/2041-8205/824/1/L12}, \href
  {http://adsabs.harvard.edu/abs/2016ApJ...824L..12O} {824, L12}

\bibitem[\protect\citeauthoryear{{Olejak} \& {Belczynski}}{{Olejak} \&
  {Belczynski}}{2021}]{olejak2021b}
{Olejak} A.,  {Belczynski} K.,  2021, \mn@doi [\apjl]
  {10.3847/2041-8213/ac2f48}, \href
  {https://ui.adsabs.harvard.edu/abs/2021ApJ...921L...2O} {921, L2}

\bibitem[\protect\citeauthoryear{{Olejak}, {Belczynski}  \& {Ivanova}}{{Olejak}
  et~al.}{2021}]{olejak2021}
{Olejak} A.,  {Belczynski} K.,   {Ivanova} N.,  2021, \mn@doi [\aap]
  {10.1051/0004-6361/202140520}, \href
  {https://ui.adsabs.harvard.edu/abs/2021A&A...651A.100O} {651, A100}

\bibitem[\protect\citeauthoryear{{Patton} \& {Sukhbold}}{{Patton} \&
  {Sukhbold}}{2020}]{patton2020}
{Patton} R.~A.,  {Sukhbold} T.,  2020, \mn@doi [\mnras]
  {10.1093/mnras/staa3029}, \href
  {https://ui.adsabs.harvard.edu/abs/2020MNRAS.499.2803P} {499, 2803}

\bibitem[\protect\citeauthoryear{{Pejcha} \& {Thompson}}{{Pejcha} \&
  {Thompson}}{2015}]{pejcha2015}
{Pejcha} O.,  {Thompson} T.~A.,  2015, \mn@doi [\apj]
  {10.1088/0004-637X/801/2/90}, \href
  {https://ui.adsabs.harvard.edu/abs/2015ApJ...801...90P} {801, 90}

\bibitem[\protect\citeauthoryear{{Peters}}{{Peters}}{1964}]{peters1964}
{Peters} P.~C.,  1964, \mn@doi [Physical Review] {10.1103/PhysRev.136.B1224},
  \href {http://adsabs.harvard.edu/abs/1964PhRv..136.1224P} {136, 1224}

\bibitem[\protect\citeauthoryear{{Petrovich} \& {Antonini}}{{Petrovich} \&
  {Antonini}}{2017}]{petrovich2017}
{Petrovich} C.,  {Antonini} F.,  2017, \mn@doi [\apj]
  {10.3847/1538-4357/aa8628}, \href
  {https://ui.adsabs.harvard.edu/abs/2017ApJ...846..146P} {846, 146}

\bibitem[\protect\citeauthoryear{{Portegies Zwart} \& {McMillan}}{{Portegies
  Zwart} \& {McMillan}}{2000}]{portegieszwart2000}
{Portegies Zwart} S.~F.,  {McMillan} S.~L.~W.,  2000, \mn@doi [\apjl]
  {10.1086/312422}, \href {http://adsabs.harvard.edu/abs/2000ApJ...528L..17P}
  {528, L17}

\bibitem[\protect\citeauthoryear{{Portegies Zwart} \& {Yungelson}}{{Portegies
  Zwart} \& {Yungelson}}{1998}]{portegieszwart1998}
{Portegies Zwart} S.~F.,  {Yungelson} L.~R.,  1998, \aap, \href
  {http://adsabs.harvard.edu/abs/1998A%26A...332..173P} {332, 173}

\bibitem[\protect\citeauthoryear{{Portegies Zwart}, {McMillan}  \&
  {Gieles}}{{Portegies Zwart} et~al.}{2010}]{portegieszwart2010}
{Portegies Zwart} S.~F.,  {McMillan} S.~L.~W.,   {Gieles} M.,  2010, \mn@doi
  [\araa] {10.1146/annurev-astro-081309-130834}, \href
  {http://adsabs.harvard.edu/abs/2010ARA%26A..48..431P} {48, 431}

\bibitem[\protect\citeauthoryear{{Qin}, {Fragos}, {Meynet}, {Andrews},
  {S{\o}rensen}  \& {Song}}{{Qin} et~al.}{2018}]{qin2018}
{Qin} Y.,  {Fragos} T.,  {Meynet} G.,  {Andrews} J.,  {S{\o}rensen} M.,
  {Song} H.~F.,  2018, \mn@doi [\aap] {10.1051/0004-6361/201832839}, \href
  {https://ui.adsabs.harvard.edu/abs/2018A&A...616A..28Q} {616, A28}

\bibitem[\protect\citeauthoryear{{Qin}, {Marchant}, {Fragos}, {Meynet}  \&
  {Kalogera}}{{Qin} et~al.}{2019}]{qin2019}
{Qin} Y.,  {Marchant} P.,  {Fragos} T.,  {Meynet} G.,   {Kalogera} V.,  2019,
  \mn@doi [\apjl] {10.3847/2041-8213/aaf97b}, \href
  {https://ui.adsabs.harvard.edu/abs/2019ApJ...870L..18Q} {870, L18}

\bibitem[\protect\citeauthoryear{{Quinlan}}{{Quinlan}}{1996}]{quinlan1996}
{Quinlan} G.~D.,  1996, \mn@doi [\na] {10.1016/S1384-1076(96)00003-6}, \href
  {https://ui.adsabs.harvard.edu/abs/1996NewA....1...35Q} {1, 35}

\bibitem[\protect\citeauthoryear{{Rastello}, {Mapelli}, {Di Carlo}, {Iorio},
  {Ballone}, {Giacobbo}, {Santoliquido}  \& {Torniamenti}}{{Rastello}
  et~al.}{2021}]{rastello2021}
{Rastello} S.,  {Mapelli} M.,  {Di Carlo} U.~N.,  {Iorio} G.,  {Ballone} A.,
  {Giacobbo} N.,  {Santoliquido} F.,   {Torniamenti} S.,  2021, \mn@doi
  [\mnras] {10.1093/mnras/stab2355}, \href
  {https://ui.adsabs.harvard.edu/abs/2021MNRAS.507.3612R} {507, 3612}

\bibitem[\protect\citeauthoryear{{Reina-Campos}, {Kruijssen}, {Pfeffer},
  {Bastian}  \& {Crain}}{{Reina-Campos} et~al.}{2019}]{reina-campos2019}
{Reina-Campos} M.,  {Kruijssen} J.~M.~D.,  {Pfeffer} J.~L.,  {Bastian} N.,
  {Crain} R.~A.,  2019, \mn@doi [\mnras] {10.1093/mnras/stz1236}, \href
  {https://ui.adsabs.harvard.edu/abs/2019MNRAS.486.5838R} {486, 5838}

\bibitem[\protect\citeauthoryear{{Renzo}, {Cantiello}, {Metzger}  \&
  {Jiang}}{{Renzo} et~al.}{2020}]{renzo2020b}
{Renzo} M.,  {Cantiello} M.,  {Metzger} B.~D.,   {Jiang} Y.~F.,  2020, \mn@doi
  [\apjl] {10.3847/2041-8213/abc6a6}, \href
  {https://ui.adsabs.harvard.edu/abs/2020ApJ...904L..13R} {904, L13}

\bibitem[\protect\citeauthoryear{{Riley}, {Mandel}, {Marchant}, {Butler},
  {Nathaniel}, {Neijssel}, {Shortt}  \& {Vigna-G{\'o}mez}}{{Riley}
  et~al.}{2021}]{riley2021}
{Riley} J.,  {Mandel} I.,  {Marchant} P.,  {Butler} E.,  {Nathaniel} K.,
  {Neijssel} C.,  {Shortt} S.,   {Vigna-G{\'o}mez} A.,  2021, \mn@doi [\mnras]
  {10.1093/mnras/stab1291}, \href
  {https://ui.adsabs.harvard.edu/abs/2021MNRAS.505..663R} {505, 663}

\bibitem[\protect\citeauthoryear{{Rizzuto} et~al.,}{{Rizzuto}
  et~al.}{2021}]{rizzuto2020}
{Rizzuto} F.~P.,  et~al., 2021, \mn@doi [\mnras] {10.1093/mnras/staa3634},
  \href {https://ui.adsabs.harvard.edu/abs/2021MNRAS.501.5257R} {501, 5257}

\bibitem[\protect\citeauthoryear{{Rodriguez} \& {Loeb}}{{Rodriguez} \&
  {Loeb}}{2018}]{rodriguezloeb2018}
{Rodriguez} C.~L.,  {Loeb} A.,  2018, \mn@doi [\apjl]
  {10.3847/2041-8213/aae377}, \href
  {https://ui.adsabs.harvard.edu/abs/2018ApJ...866L...5R} {866, L5}

\bibitem[\protect\citeauthoryear{{Rodriguez}, {Chatterjee}  \&
  {Rasio}}{{Rodriguez} et~al.}{2016}]{rodriguez2016}
{Rodriguez} C.~L.,  {Chatterjee} S.,   {Rasio} F.~A.,  2016, \mn@doi [\prd]
  {10.1103/PhysRevD.93.084029}, \href
  {http://adsabs.harvard.edu/abs/2016PhRvD..93h4029R} {93, 084029}

\bibitem[\protect\citeauthoryear{{Rodriguez}, {Zevin}, {Amaro-Seoane},
  {Chatterjee}, {Kremer}, {Rasio}  \& {Ye}}{{Rodriguez}
  et~al.}{2019}]{rodriguez2019}
{Rodriguez} C.~L.,  {Zevin} M.,  {Amaro-Seoane} P.,  {Chatterjee} S.,  {Kremer}
  K.,  {Rasio} F.~A.,   {Ye} C.~S.,  2019, \mn@doi [\prd]
  {10.1103/PhysRevD.100.043027}, \href
  {https://ui.adsabs.harvard.edu/abs/2019PhRvD.100d3027R} {100, 043027}

\bibitem[\protect\citeauthoryear{{Samsing}}{{Samsing}}{2018}]{samsing2018}
{Samsing} J.,  2018, \mn@doi [\prd] {10.1103/PhysRevD.97.103014}, \href
  {http://adsabs.harvard.edu/abs/2018PhRvD..97j3014S} {97, 103014}

\bibitem[\protect\citeauthoryear{{Samsing}, {MacLeod}  \&
  {Ramirez-Ruiz}}{{Samsing} et~al.}{2014}]{samsing2014}
{Samsing} J.,  {MacLeod} M.,   {Ramirez-Ruiz} E.,  2014, \mn@doi [\apj]
  {10.1088/0004-637X/784/1/71}, \href
  {https://ui.adsabs.harvard.edu/abs/2014ApJ...784...71S} {784, 71}

\bibitem[\protect\citeauthoryear{{Sana} et~al.,}{{Sana}
  et~al.}{2012}]{sana2012}
{Sana} H.,  et~al., 2012, \mn@doi [Science] {10.1126/science.1223344}, \href
  {http://adsabs.harvard.edu/abs/2012Sci...337..444S} {337, 444}

\bibitem[\protect\citeauthoryear{{Santoliquido}, {Mapelli}, {Bouffanais},
  {Giacobbo}, {Di Carlo}, {Rastello}, {Artale}  \& {Ballone}}{{Santoliquido}
  et~al.}{2020}]{santoliquido2020}
{Santoliquido} F.,  {Mapelli} M.,  {Bouffanais} Y.,  {Giacobbo} N.,  {Di Carlo}
  U.~N.,  {Rastello} S.,  {Artale} M.~C.,   {Ballone} A.,  2020, \mn@doi [\apj]
  {10.3847/1538-4357/ab9b78}, \href
  {https://ui.adsabs.harvard.edu/abs/2020ApJ...898..152S} {898, 152}

\bibitem[\protect\citeauthoryear{{Santoliquido}, {Mapelli}, {Giacobbo},
  {Bouffanais}  \& {Artale}}{{Santoliquido} et~al.}{2021}]{santoliquido2021}
{Santoliquido} F.,  {Mapelli} M.,  {Giacobbo} N.,  {Bouffanais} Y.,   {Artale}
  M.~C.,  2021, \mn@doi [\mnras] {10.1093/mnras/stab280}, \href
  {https://ui.adsabs.harvard.edu/abs/2021MNRAS.502.4877S} {502, 4877}

\bibitem[\protect\citeauthoryear{{Sasaki}, {Suyama}, {Tanaka}  \&
  {Yokoyama}}{{Sasaki} et~al.}{2016}]{sasaki2016}
{Sasaki} M.,  {Suyama} T.,  {Tanaka} T.,   {Yokoyama} S.,  2016, \mn@doi
  [Physical Review Letters] {10.1103/PhysRevLett.117.061101}, \href
  {http://adsabs.harvard.edu/abs/2016PhRvL.117f1101S} {117, 061101}

\bibitem[\protect\citeauthoryear{{Sesana}, {Haardt}  \& {Madau}}{{Sesana}
  et~al.}{2006}]{sesana2006}
{Sesana} A.,  {Haardt} F.,   {Madau} P.,  2006, \mn@doi [\apj]
  {10.1086/507596}, \href
  {https://ui.adsabs.harvard.edu/abs/2006ApJ...651..392S} {651, 392}

\bibitem[\protect\citeauthoryear{{Shao} \& {Li}}{{Shao} \&
  {Li}}{2021}]{shao2021}
{Shao} Y.,  {Li} X.-D.,  2021, \mn@doi [\apj] {10.3847/1538-4357/ac173e}, \href
  {https://ui.adsabs.harvard.edu/abs/2021ApJ...920...81S} {920, 81}

\bibitem[\protect\citeauthoryear{{Silsbee} \& {Tremaine}}{{Silsbee} \&
  {Tremaine}}{2017}]{silsbee2017}
{Silsbee} K.,  {Tremaine} S.,  2017, \mn@doi [\apj] {10.3847/1538-4357/aa5729},
  \href {https://ui.adsabs.harvard.edu/abs/2017ApJ...836...39S} {836, 39}

\bibitem[\protect\citeauthoryear{{Sollima}, {Beccari}, {Ferraro}, {Fusi Pecci}
  \& {Sarajedini}}{{Sollima} et~al.}{2007}]{sollima2007}
{Sollima} A.,  {Beccari} G.,  {Ferraro} F.~R.,  {Fusi Pecci} F.,   {Sarajedini}
  A.,  2007, \mn@doi [\mnras] {10.1111/j.1365-2966.2007.12116.x}, \href
  {https://ui.adsabs.harvard.edu/abs/2007MNRAS.380..781S} {380, 781}

\bibitem[\protect\citeauthoryear{{Spera} \& {Mapelli}}{{Spera} \&
  {Mapelli}}{2017}]{spera2017}
{Spera} M.,  {Mapelli} M.,  2017, \mn@doi [\mnras] {10.1093/mnras/stx1576},
  \href {http://adsabs.harvard.edu/abs/2017MNRAS.470.4739S} {470, 4739}

\bibitem[\protect\citeauthoryear{{Spera}, {Mapelli}, {Giacobbo}, {Trani},
  {Bressan}  \& {Costa}}{{Spera} et~al.}{2019}]{spera2019}
{Spera} M.,  {Mapelli} M.,  {Giacobbo} N.,  {Trani} A.~A.,  {Bressan} A.,
  {Costa} G.,  2019, \mn@doi [\mnras] {10.1093/mnras/stz359}, \href
  {https://ui.adsabs.harvard.edu/abs/2019MNRAS.485..889S} {485, 889}

\bibitem[\protect\citeauthoryear{{Spitzer}}{{Spitzer}}{1969}]{spitzer1969}
{Spitzer} Jr. L.,  1969, \mn@doi [\apjl] {10.1086/180451}, \href
  {http://adsabs.harvard.edu/abs/1969ApJ...158L.139S} {158, L139}

\bibitem[\protect\citeauthoryear{{Spitzer}}{{Spitzer}}{1987}]{spitzer1987}
{Spitzer} L.,  1987, {Dynamical evolution of globular clusters}.
Princeton University Press

\bibitem[\protect\citeauthoryear{{Stevenson}, {Berry}  \& {Mandel}}{{Stevenson}
  et~al.}{2017}]{stevenson2017}
{Stevenson} S.,  {Berry} C. P.~L.,   {Mandel} I.,  2017, \mn@doi [\mnras]
  {10.1093/mnras/stx1764}, \href
  {https://ui.adsabs.harvard.edu/abs/2017MNRAS.471.2801S} {471, 2801}

\bibitem[\protect\citeauthoryear{{Stevenson}, {Sampson}, {Powell},
  {Vigna-G{\'o}mez}, {Neijssel}, {Sz{\'e}csi}  \& {Mandel}}{{Stevenson}
  et~al.}{2019}]{stevenson2019}
{Stevenson} S.,  {Sampson} M.,  {Powell} J.,  {Vigna-G{\'o}mez} A.,  {Neijssel}
  C.~J.,  {Sz{\'e}csi} D.,   {Mandel} I.,  2019, \mn@doi [\apj]
  {10.3847/1538-4357/ab3981}, \href
  {https://ui.adsabs.harvard.edu/abs/2019ApJ...882..121S} {882, 121}

\bibitem[\protect\citeauthoryear{{Stone}, {Metzger}  \& {Haiman}}{{Stone}
  et~al.}{2017}]{stone2017}
{Stone} N.~C.,  {Metzger} B.~D.,   {Haiman} Z.,  2017, \mn@doi [\mnras]
  {10.1093/mnras/stw2260}, \href
  {https://ui.adsabs.harvard.edu/abs/2017MNRAS.464..946S} {464, 946}

\bibitem[\protect\citeauthoryear{{Sukhbold}, {Ertl}, {Woosley}, {Brown}  \&
  {Janka}}{{Sukhbold} et~al.}{2016}]{sukhbold2016}
{Sukhbold} T.,  {Ertl} T.,  {Woosley} S.~E.,  {Brown} J.~M.,   {Janka} H.~T.,
  2016, \mn@doi [\apj] {10.3847/0004-637X/821/1/38}, \href
  {https://ui.adsabs.harvard.edu/abs/2016ApJ...821...38S} {821, 38}

\bibitem[\protect\citeauthoryear{{Tagawa}, {Haiman}  \& {Kocsis}}{{Tagawa}
  et~al.}{2020}]{tagawa2020}
{Tagawa} H.,  {Haiman} Z.,   {Kocsis} B.,  2020, \mn@doi [\apj]
  {10.3847/1538-4357/ab9b8c}, \href
  {https://ui.adsabs.harvard.edu/abs/2020ApJ...898...25T} {898, 25}

\bibitem[\protect\citeauthoryear{{Tanikawa}}{{Tanikawa}}{2013}]{tanikawa2013}
{Tanikawa} A.,  2013, \mn@doi [\mnras] {10.1093/mnras/stt1380}, \href
  {https://ui.adsabs.harvard.edu/abs/2013MNRAS.435.1358T} {435, 1358}

\bibitem[\protect\citeauthoryear{{Tanikawa}, {Kinugawa}, {Yoshida}, {Hijikawa}
  \& {Umeda}}{{Tanikawa} et~al.}{2021a}]{tanikawa2020}
{Tanikawa} A.,  {Kinugawa} T.,  {Yoshida} T.,  {Hijikawa} K.,   {Umeda} H.,
  2021a, \mn@doi [\mnras] {10.1093/mnras/stab1421}, \href
  {https://ui.adsabs.harvard.edu/abs/2021MNRAS.505.2170T} {505, 2170}

\bibitem[\protect\citeauthoryear{{Tanikawa}, {Susa}, {Yoshida}, {Trani}  \&
  {Kinugawa}}{{Tanikawa} et~al.}{2021b}]{tanikawa2021}
{Tanikawa} A.,  {Susa} H.,  {Yoshida} T.,  {Trani} A.~A.,   {Kinugawa} T.,
  2021b, \mn@doi [\apj] {10.3847/1538-4357/abe40d}, \href
  {https://ui.adsabs.harvard.edu/abs/2021ApJ...910...30T} {910, 30}

\bibitem[\protect\citeauthoryear{{Toyouchi}, {Inayoshi}, {Ishigaki}  \&
  {Tominaga}}{{Toyouchi} et~al.}{2021}]{toyouchi2021}
{Toyouchi} D.,  {Inayoshi} K.,  {Ishigaki} M.~N.,   {Tominaga} N.,  2021, arXiv
  e-prints, \href {https://ui.adsabs.harvard.edu/abs/2021arXiv211206151T} {p.
  arXiv:2112.06151}

\bibitem[\protect\citeauthoryear{{Tremaine}, {Ostriker}  \&
  {Spitzer}}{{Tremaine} et~al.}{1975}]{tremaine1975}
{Tremaine} S.~D.,  {Ostriker} J.~P.,   {Spitzer} L. J.,  1975, \mn@doi [\apj]
  {10.1086/153422}, \href
  {https://ui.adsabs.harvard.edu/abs/1975ApJ...196..407T} {196, 407}

\bibitem[\protect\citeauthoryear{{Tutukov} \& {Yungelson}}{{Tutukov} \&
  {Yungelson}}{1973}]{tutukov1973}
{Tutukov} A.,  {Yungelson} L.,  1973, Nauchnye Informatsii, \href
  {http://adsabs.harvard.edu/abs/1973NInfo..27...70T} {27, 70}

\bibitem[\protect\citeauthoryear{{Ugliano}, {Janka}, {Marek}  \&
  {Arcones}}{{Ugliano} et~al.}{2012}]{ugliano2012}
{Ugliano} M.,  {Janka} H.-T.,  {Marek} A.,   {Arcones} A.,  2012, \mn@doi
  [\apj] {10.1088/0004-637X/757/1/69}, \href
  {https://ui.adsabs.harvard.edu/abs/2012ApJ...757...69U} {757, 69}

\bibitem[\protect\citeauthoryear{{VandenBerg}, {Brogaard}, {Leaman}  \&
  {Casagrand e}}{{VandenBerg} et~al.}{2013}]{vandenberg2013}
{VandenBerg} D.~A.,  {Brogaard} K.,  {Leaman} R.,   {Casagrand e} L.,  2013,
  \mn@doi [\apj] {10.1088/0004-637X/775/2/134}, \href
  {https://ui.adsabs.harvard.edu/abs/2013ApJ...775..134V} {775, 134}

\bibitem[\protect\citeauthoryear{{Vigna-G{\'o}mez}, {Toonen}, {Ramirez-Ruiz},
  {Leigh}, {Riley}  \& {Haster}}{{Vigna-G{\'o}mez} et~al.}{2021}]{vigna2021}
{Vigna-G{\'o}mez} A.,  {Toonen} S.,  {Ramirez-Ruiz} E.,  {Leigh} N. W.~C.,
  {Riley} J.,   {Haster} C.-J.,  2021, \mn@doi [\apjl]
  {10.3847/2041-8213/abd5b7}, \href
  {https://ui.adsabs.harvard.edu/abs/2021ApJ...907L..19V} {907, L19}

\bibitem[\protect\citeauthoryear{{Vink}, {de Koter}  \& {Lamers}}{{Vink}
  et~al.}{2001}]{vink2001}
{Vink} J.~S.,  {de Koter} A.,   {Lamers} H.~J.~G.~L.~M.,  2001, \mn@doi [\aap]
  {10.1051/0004-6361:20010127}, \href
  {http://adsabs.harvard.edu/abs/2001A%26A...369..574V} {369, 574}

\bibitem[\protect\citeauthoryear{{Vink}, {Higgins}, {Sander}  \&
  {Sabhahit}}{{Vink} et~al.}{2021}]{vink2021}
{Vink} J.~S.,  {Higgins} E.~R.,  {Sander} A. A.~C.,   {Sabhahit} G.~N.,  2021,
  \mn@doi [\mnras] {10.1093/mnras/stab842}, \href
  {https://ui.adsabs.harvard.edu/abs/2021MNRAS.504..146V} {504, 146}

\bibitem[\protect\citeauthoryear{{Ward}, {Kruijssen}  \& {Rix}}{{Ward}
  et~al.}{2020}]{ward2020}
{Ward} J.~L.,  {Kruijssen} J.~M.~D.,   {Rix} H.-W.,  2020, \mn@doi [\mnras]
  {10.1093/mnras/staa1056}, \href
  {https://ui.adsabs.harvard.edu/abs/2020MNRAS.495..663W} {495, 663}

\bibitem[\protect\citeauthoryear{{Wong}, {Breivik}, {Kremer}  \&
  {Callister}}{{Wong} et~al.}{2021}]{wong2021}
{Wong} K. W.~K.,  {Breivik} K.,  {Kremer} K.,   {Callister} T.,  2021, \mn@doi
  [\prd] {10.1103/PhysRevD.103.083021}, \href
  {https://ui.adsabs.harvard.edu/abs/2021PhRvD.103h3021W} {103, 083021}

\bibitem[\protect\citeauthoryear{{Woosley}}{{Woosley}}{2017}]{woosley2017}
{Woosley} S.~E.,  2017, \mn@doi [\apj] {10.3847/1538-4357/836/2/244}, \href
  {http://adsabs.harvard.edu/abs/2017ApJ...836..244W} {836, 244}

\bibitem[\protect\citeauthoryear{{Yang}, {Bartos}, {Haiman}, {Kocsis},
  {M{\'a}rka}, {Stone}  \& {M{\'a}rka}}{{Yang} et~al.}{2019}]{yang2019}
{Yang} Y.,  {Bartos} I.,  {Haiman} Z.,  {Kocsis} B.,  {M{\'a}rka} Z.,  {Stone}
  N.~C.,   {M{\'a}rka} S.,  2019, \mn@doi [\apj] {10.3847/1538-4357/ab16e3},
  \href {https://ui.adsabs.harvard.edu/abs/2019ApJ...876..122Y} {876, 122}

\bibitem[\protect\citeauthoryear{{Zevin}, {Pankow}, {Rodriguez}, {Sampson},
  {Chase}, {Kalogera}  \& {Rasio}}{{Zevin} et~al.}{2017}]{zevin2017}
{Zevin} M.,  {Pankow} C.,  {Rodriguez} C.~L.,  {Sampson} L.,  {Chase} E.,
  {Kalogera} V.,   {Rasio} F.~A.,  2017, \mn@doi [\apj]
  {10.3847/1538-4357/aa8408}, \href
  {http://adsabs.harvard.edu/abs/2017ApJ...846...82Z} {846, 82}

\bibitem[\protect\citeauthoryear{{Zevin}, {Samsing}, {Rodriguez}, {Haster}  \&
  {Ramirez-Ruiz}}{{Zevin} et~al.}{2019}]{zevin2019}
{Zevin} M.,  {Samsing} J.,  {Rodriguez} C.,  {Haster} C.-J.,   {Ramirez-Ruiz}
  E.,  2019, \mn@doi [\apj] {10.3847/1538-4357/aaf6ec}, \href
  {https://ui.adsabs.harvard.edu/abs/2019ApJ...871...91Z} {871, 91}

\bibitem[\protect\citeauthoryear{{Zevin} et~al.,}{{Zevin}
  et~al.}{2021}]{zevin2021}
{Zevin} M.,  et~al., 2021, \mn@doi [\apj] {10.3847/1538-4357/abe40e}, \href
  {https://ui.adsabs.harvard.edu/abs/2021ApJ...910..152Z} {910, 152}

\bibitem[\protect\citeauthoryear{{Ziosi}, {Mapelli}, {Branchesi}  \&
  {Tormen}}{{Ziosi} et~al.}{2014}]{ziosi2014}
{Ziosi} B.~M.,  {Mapelli} M.,  {Branchesi} M.,   {Tormen} G.,  2014, \mn@doi
  [\mnras] {10.1093/mnras/stu824}, \href
  {http://adsabs.harvard.edu/abs/2014MNRAS.441.3703Z} {441, 3703}

\bibitem[\protect\citeauthoryear{{de Mink} \& {Mandel}}{{de Mink} \&
  {Mandel}}{2016}]{demink2016}
{de Mink} S.~E.,  {Mandel} I.,  2016, \mn@doi [\mnras] {10.1093/mnras/stw1219},
  \href {http://adsabs.harvard.edu/abs/2016MNRAS.460.3545D} {460, 3545}

\bibitem[\protect\citeauthoryear{{du Buisson} et~al.,}{{du Buisson}
  et~al.}{2020}]{dubuisson2020}
{du Buisson} L.,  et~al., 2020, \mn@doi [\mnras] {10.1093/mnras/staa3225},
  \href {https://ui.adsabs.harvard.edu/abs/2020MNRAS.499.5941D} {499, 5941}

\bibitem[\protect\citeauthoryear{{van Son} et~al.,}{{van Son}
  et~al.}{2021}]{vanson2021}
{van Son} L.~A.~C.,  et~al., 2021, arXiv e-prints, \href
  {https://ui.adsabs.harvard.edu/abs/2021arXiv211001634V} {p. arXiv:2110.01634}

\makeatother
\end{thebibliography}


\end{document}